\documentclass[numbook,natbib,final,runningheads]{svjour3}
\usepackage{graphicx,mathptmx} 
\usepackage{makeidx} 
\usepackage{amssymb}
\usepackage{url}
\usepackage{lscape}

\makeindex

\newcommand{\degr}{\ifmmode^\circ\else$^\circ$\fi}

\newcommand{\lapprox} {\, \lower3pt\hbox{$\sim$}\llap{\raise2pt\hbox{$<$}}\,}
\newcommand{\gapprox} {\, \lower3pt\hbox{$\sim$}\llap{\raise2pt\hbox{$>$}}\,}

\begin{document}

\title{An Observational Overview of Solar Flares}

 \author{L. {Fletcher}$^{1}$,
        B.~R.~{Dennis}$^{2}$,
        H.~S.~{Hudson}$^{3}$,
	S.~{Krucker}$^{3}$,
	K.~{Phillips}$^{4}$,
	A.~{Veronig}$^{5}$,
        M.~{Battaglia}$^{1}$,
        L.~{Bone}$^{4}$,
        A.~{Caspi}$^3$,
        Q.~{Chen}$^{7}$ ,
        P.  {Gallagher}$^{8}$,
        P.~T.~{Grigis}$^{9}$, \and
        H.~{Ji}$^{10,11}$,
        W.~{Liu}$^{2,12}$,
        R.~O.~{Milligan}$^{2}$, and
        M.~{Temmer}$^{5,13,14}$} 
 
 \institute{$^{1}$ Department of Physics and Astronomy, University of Glasgow, Glasgow G12 	8QQ, U. K.\\
                $^{2}$ NASA Goddard Space Flight Center, Greenbelt, MD USA\\ 
               $^{3}$ Space Sciences Laboratory, U.C. Berkeley, CA USA 94720-7450\\      
                              $^{4}$ Mullard Space Science Laboratory, Holmbury St. Mary, Dorking, RH5 CNT, U.~K.\\ 
               $^{5}$ Institute for Geophysics, Astrophysics and Meteorology, University of Graz, Graz A-8010,  Austria\\
               $^{7}$ Stanford University Physics Department, Varian Building, 382 Via Pueblo Mall, Stanford CA  USA  94305\\ 
              $^{8}$ School of Physics, Trinity College Dublin, Dublin 2, Ireland\\
              $^{9}$ Harvard-Smithsonian center for Astrophysics, 60 Garden Street, Cambridge MA USA 02138\\
             $^{10}$ Purple Mountain Observatory, 2 W. Beijing Rd, Nanjing, 210008, China\\
              $^{11}$ Big Bear Solar Observatory, New Jersey Institute of Technology, 40386 North Shore Lane,  Big Bear City, CA USA 92314 \\
              $^{12}$ Stanford-Lockheed Institute for Space Research, 466 Via Ortega, Cypress Hall, Stanford, CA USA 94305-4085\\ 
             $^{13}$ Hvar Observatory, Faculty of Geodesy, Kaciceva 26, 10000 Zagreb, Croatia\\
             $^{14}$ Space Research Institute, Austrian Academy of Sciences, Schmiedlstra{\ss}e 6, A-8042 Graz, Austria}

\date{}

\authorrunning{Fletcher et al.}
\titlerunning{Observational Overview}
\maketitle

\begin{abstract}
We present an overview of solar flares and associated phenomena,
drawing upon a wide range of observational data primarily from the
\textit{RHESSI} era.\index{eras!RHESSI@\textit{RHESSI}}  
Following an introductory discussion and overview of
the status of observational capabilities, the article is split into
topical sections which deal with different areas of  flare phenomena
(footpoints and ribbons, coronal sources, relationship to coronal
mass ejections) and their interconnections.  We also discuss flare
soft X-ray spectroscopy and the energetics of the process.  The
emphasis is  to describe the observations from multiple points of
view,  while bearing in mind the models that link them to each other
and to theory.  The present theoretical and observational understanding
of solar flares is far from complete, so we conclude with a brief
discussion of models, and a list of missing but important observations.
\end{abstract}


\tableofcontents
\section{The multi-wavelength flare}

Solar flares are the most powerful magnetic events in the solar
system. In tens of minutes they can release in excess of $10^{32}$
erg of energy.
They emit radiation across the entire electromagnetic
spectrum, from radio to $\gamma$-rays, and are also intimately
associated with the acceleration of particles into interplanetary
space and with coronal mass ejections.\index{spectrum!flare!broad-band nature of}
The flare results from the
rapid release of energy previously stored as the inductive magnetic
fields due to electrical currents flowing into the corona.  The
total flare energy is compatible with the amount of  magnetic ``free''
energy (usually defined as the energy stored in the magnetic field
that is over and above the energy of the potential magnetic field
defined by the same boundaries) inferred to be available in the
magnetic active regions (i.e., the coronal connections of a sunspot
group) where most flares take place.  \index{magnetic field!free
energy} The magnetic free energy is is hard to evaluate from
\index{sunspots!and free magnetic energy}
\index{free energy}
observations, depending as it does on the magnetic vector field,
but in the few cases in which this has been possible
\citep[e.g.,][]{1995ApJ...439..474M,2005ApJ...623L..53M,2008ApJ...675.1637S,2008ApJ...676L..81J},
and it is found that the free energy is comparable with that of
large flares. Furthermore, the energy budget is  difficult to explain
from other possible coronal or chromospheric energy sources
\citep[e.g.,][]{2007ASPC..368..365H}.  So we can conclude that
conversion of stored magnetic energy is at the heart of the flare
process.  \index{flares!energy content!magnetic}

The term ``flare'' is normally taken to refer specifically to the
electromagnetic radiation of this whole magnetically-driven event,
which embodies a significant fraction of the total energy liberated.
The total energy released varies from event to event, with many
more small events than large events. The distribution\index{distribution functions!flare occurrence} 
of the number
of flares as a function of their peak energy, or their total energy,
or their duration, is approximately a power law, the gradient of
which is a critical factor in understanding the contribution of
flare-like heating\index{coronal heating!in flare-like events} events to the overall energy budget of the solar
corona \citep[e.g.,][]{1993SoPh..143..275C,Chapter6}. 
The primary way of classifying the ``importance'' of a flare
is via its soft X-ray (SXR) flux at 1--8~\AA, as measured by 
\textit{GOES}\index{satellites!GOES@\textit{GOES}}
(the \textit{Geostationary Orbiting  Environmental Satellites}). 
Flares are classified\index{flares!GOES@\textit{GOES} classification}\index{GOES@\textit{GOES} classification}\index{flares!classification}\index{flare classification}
into X, M, C, B and A flares, with X corresponding to
\textit{GOES} flux in excess of $10^{-4}\ {\rm W\ m}^{-2}$ at Earth, and
successive classifications decreasing in decades. 
Table~\ref{tab:class} puts the X-ray class of a flare in the context
of the older classification based on H$\alpha$~area \citep[data from][]{1971SoPh...16..431T}.

\begin{table}
\caption{\textit{GOES} and H$\alpha$ classifications}
\begin{tabular}{l l l l}
\hline\noalign{\smallskip}
\textit{GOES} & EM$^a$ &H$\alpha$ & H$\alpha$ area  \\
class      & cm$^{-3}$      &            class                    & Sq. degrees \\
\noalign{\smallskip}\hline\noalign{\smallskip}
X10& 10$^{51}$ & 4 & 24.7\\
X   &10$^{50}$ & 3 & 12.4  \\
M  &10$^{49}$  & 2 & 5.1\\
C  &10$^{48}$ & 1 & 2.0   \\
B   & 10$^{47}$& S  & $<$2.0 \\
A   & 10$^{46}$& S & $<$2.0  \\
\noalign{\smallskip}\hline
\end{tabular}

\noindent $^a$Soft X-ray emission measure (approx.)
\end{table}
\label{tab:class}
\index{emission measure!by \textit{GOES} class}
\index{flares!classification!table}\index{flare classification!table}
\index{flares!classification!H$\alpha$}\index{flare classification!H$\alpha$}

The majority of the radiative flare energy emerges at visible and
ultraviolet wavelengths 
\citep{2006JGRA..11110S14W}.  
Where a bolometric measurement is possible, i.e.,
in the most energetic flares, we find that the radiated optical
luminosity is comparable to the kinetic energy of the coronal mass
ejection (see Section~\ref{sec:energetics}), and also to the the
energy of the accelerated electrons as inferred from the hard
X-radiation (HXR) \citep{2007ApJ...656.1187F} under the assumptions
of the collisional thick-target model \citep{1971SoPh...18..489B}.\index{thick-target model!collisional}
A lot of emphasis has been placed on hard X-rays (HXRs) in understanding
the flare energization process -- see \cite{Chapter3} --
despite the fact that energetically they represent only a small
fraction of the total radiation. However, as HXRs result mainly
from the well-understood bremsstrahlung radiation process, and the
sources are optically thin, it is relatively straightforward to
interpret them. The HXR emission is thus a powerful diagnostic for
flare electrons, compared to longer wavelength, optically thick
radiation, and the measurement of flare HXRs has been a primary
goal of the \textit{Reuven Ramaty High Energy Solar Spectroscopic 
Imager}\index{RHESSI@\textit{RHESSI}} \citep[\textit{RHESSI},][]{2002SoPh..210....3L}. 
However, hard X-rays alone give only a restricted view of the overall configuration, development
and energetics of a flare, and of its relationship to accompanying
dynamical processes. 
The aim of this article is therefore to set the HXRs in the context of the multi-wavelength flare\index{hard X-rays!multi-wavelength context}, to give an
up-to-date observational picture, and to provide context  for
subsequent articles in this volume. 
This article focuses on the radiative flare,
and discusses the coronal mass ejection and solar energetic particles
only in association. It is not intended as a comprehensive historical
review, although selected historical observations appear. The
following recent reviews of observations of flare and related
phenomena, including theory, are also
recommended for further reading:  \cite{2002A&ARv..10..313P};
\cite{2002SSRv..101....1A}; \cite{2008LRSP....5....1B};
\cite{2008A&ARv..16..155K}; \cite{2008SoPh..253..215V};
\cite{2009AdSpR..43..739S}.

\begin{figure}
\begin{center}
\includegraphics[width=\textwidth]{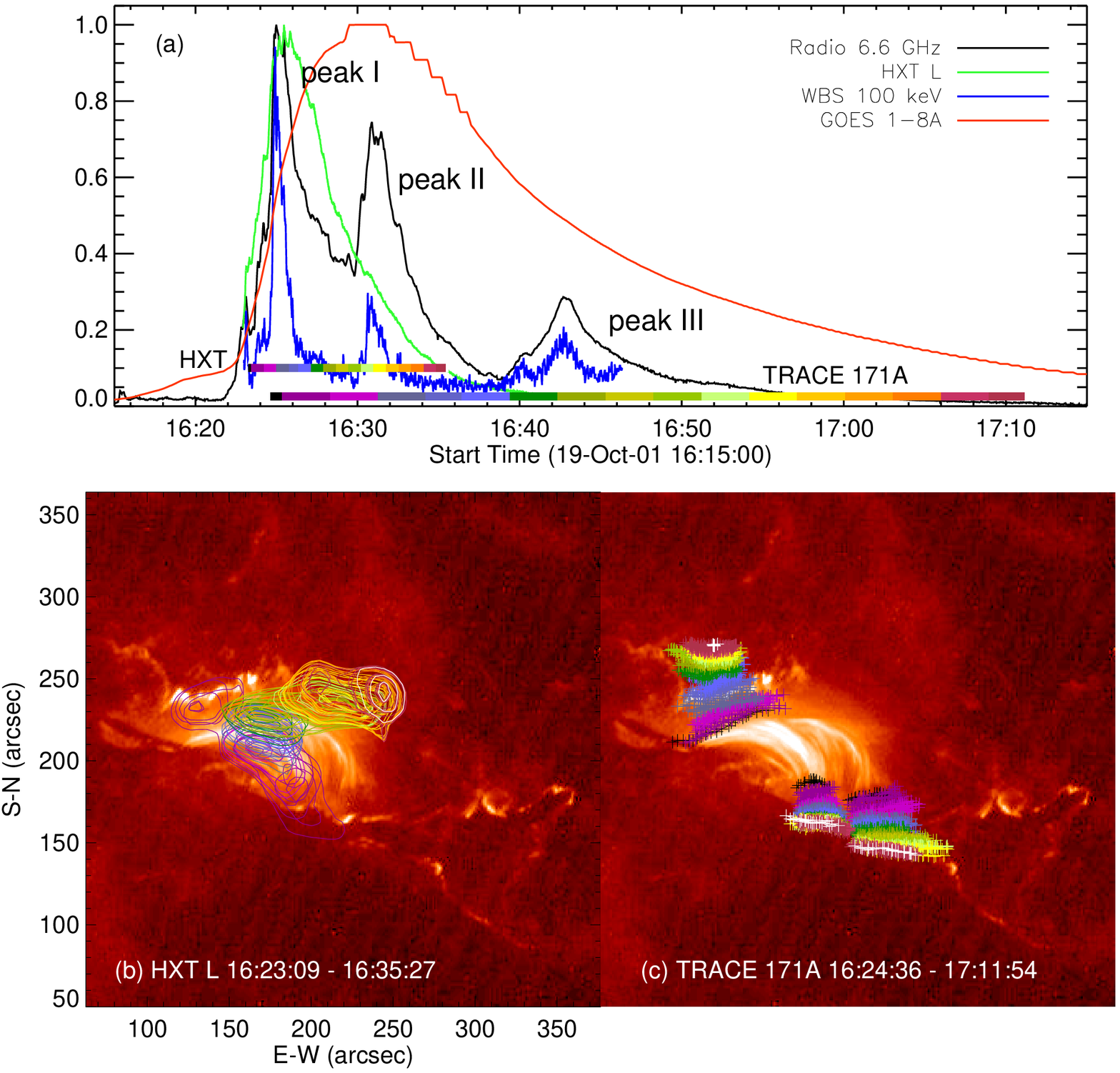}
\end{center}
\caption{\label{fig:timeprofile_overview} Time evolution of the flare
SOL2001-10-19T01:05 (X1.6)
in multiple X-ray, EUV, and radio wavelengths
\citep[from][]{2009SoPh..255..107Q}.  The \textit{impulsive phase}
is best characterized by the hard X-ray light curve (blue, in the
100~keV band of \textit{Yohkoh} Wide Band Spectrometer (WBS) or the shortest-wavelength
radio emissions (black, at 6.6~GHz), from the Owens Valley Solar Array.\index{Owens Valley Solar Array (OVSA)}
\index{satellites!Yohkoh@\textit{Yohkoh}}
The \textit{GOES} lightcurve (red)
shows the \textit{gradual phase} well. 
\index{satellites!GOES@\textit{GOES}} 
The two panels at the bottom
show \textit{TRACE} 171~\AA~images defining the flare arcade and its footpoints;
\textit{left,} with hard X-ray contours; \textit{right,} with  EUV
footpoint locations color-coded by time in the upper panel.
\index{arcade} The black line shows the 6.6~GHz microwave emission, at
the electron gyrofrequency for a 2,400~G magnetic field.\index{frequency!Larmor}
} 
\index{flares!overview!illustration} \index{arcade!illustration}\index{frequency!Larmor}
\index{gyrosynchrotron emission!illustration}
\index{flare (individual)!SOL2001-10-19T01:05 (X1.6)!illustration}

\end{figure}

As a preview of the introductory sections of this article,
Figure~\ref{fig:timeprofile_overview} \citep[from][]{2009SoPh..255..107Q}
sketches out the temporal and spatial evolution of a well-observed
major flare, SOL2001-10-19T01:05 (X1.6)\index{gradual phase}\index{impulsive phase}\index{soft X-rays!ribbons}\index{ribbons}.

\subsection{Flare development}

\begin{figure}
\begin{center}
\includegraphics[width=\textwidth]{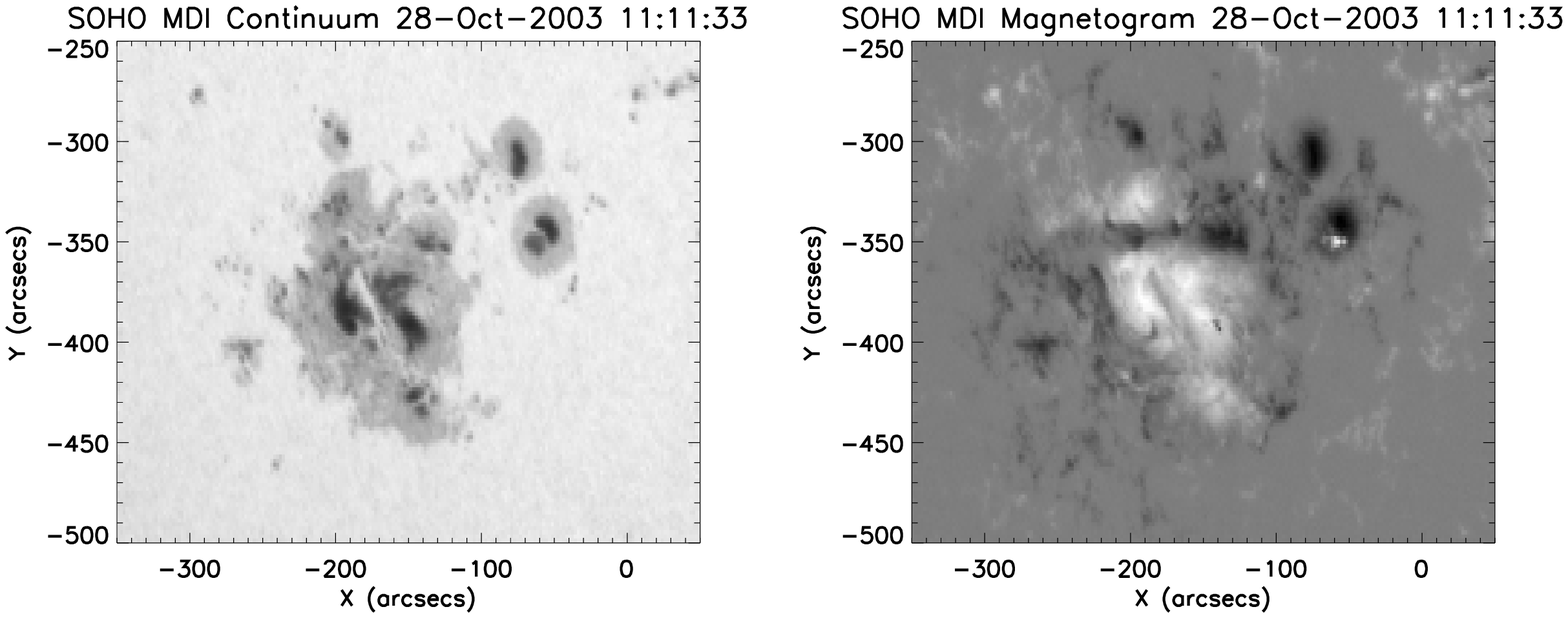}
\end{center}
\caption{\label{fig:AR10486} Continuum (\textit{left}) and magnetogram (\textit{right})
images of NOAA Active Region 10486, which was the seat of many major
flares in late October/early November 2003, including \index{flare
(individual)!SOL2003-10-29T20:49 (X10.0)} SOL2003-10-29T20:49 (X10.0),
\index{flare (individual)!SOL2003-11-03T09:55 (X3.9)} SOL2003-11-03T09:55 (X3.9), and \index{flare (individual)!SOL2003-11-04T19:53 (X17.4)} SOL2003-11-04T19:53 (X17.4), the most powerful \textit{GOES} flare on record. 
The $\delta$~configuration is visible in the positive polarity
region of this group, with three major umbrae (those for X~in the
range [-140,-200] and Y~in the range [--370, -410]) within one
penumbra.  Also visible in the magnetogram are oppositely-colored
inclusions within the main spot areas, e.g., at [-140, -390]. These
occur as a result of disruption of the magnetically-sensitive
spectral line because of the strong disturbance to the atmosphere
at the time of the flare. Such features are transient and unrelated
to actual magnetic changes. However, non-reversing magnetic changes
at the time of flares are observed; these are discussed in
Section~\ref{sec:fletcher_deep}.} 
\end{figure}

Flares -- of all sizes -- tend strongly to occur in magnetic
active regions\index{active regions!and flare occurrence} and are associated with strong magnetic fields in
the neighborhood of magnetic polarity inversion\index{magnetic structures!polarity inversion line}\index{magnetic structures!neutral line} lines (which is the
dividing line between regions of positive and negative vertical
component of the photospheric magnetic field, sometimes also called
a neutral line).  A small fraction of flares do occur in so-called
``spotless'' regions \citep{1970SoPh...13..401D,1980SoPh...68..217M}\index{flare types!spotless},
and large-scale filament eruptions with flare-like\index{flare types!filament eruption} properties can
happen anywhere on the quiet Sun \citep[e.g.,][]{1986stp..conf..198H}.
It is not yet possible to predict the time or location of a solar
flare. Several statistical studies have attempted to identify
active-region magnetic properties that are correlated with active
region flare productivity, or can even act as solar flare forecasters.
The best indicators of flare productivity are those known in the
``lore'' of flare observers regarding the size and complexity of  the sunspots
in the active region.
The largest flares occur in
\index{sunspots!delta spots@$\delta$ spots}
\index{magnetic structures!delta spot@$\delta$ spot}
``delta-spot'' regions which have two umbrae within a single penumbra
\citep{1987SoPh..113..267Z,1990SoPh..125..251M}.
They follow the rate of evolution of the active region \citep{2005ApJ...628..501S} and require the presence of strong magnetic gradients\index{magnetic structures!strong gradients} \citep{1984SoPh...91..115H}. 
An example of a flare-productive active region, AR10486, is shown in 
Figure~\ref{fig:AR10486}.  
The magnetic properties of flaring regions are
encapsulated in various studies of the photospheric field which
find higher flare probabilities in regions with high total photospheric
magnetic flux, excess magnetic energy, long polarity inversion lines\index{magnetic structures!long polarity inversion lines}
with a strong, highly variable distribution of shear\index{magnetic structures!variable shear} along their
length \citep{2006SoPh..237...45C,2007ApJ...656.1173L} and a high
fractal dimension\index{magnetic structures!fractal dimension} of the photospheric field  \citep{2005ApJ...631..628M}.
Incorporating information on the evolution of observed photospheric
parameters, the rate of change of the strongest photospheric magnetic
twists\index{magnetic structures!twists!and flare prediction} in the region, is the best predictor of a flare
\citep{2003ApJ...595.1277L}.  \index{flares!energy content!magnetic}
However, in general the photospheric properties alone appear to
offer poor predictive capabilities, and it appears likely that
parameters of the coronal magnetic configuration offer a better
prospect. 
For example, a high degree of complexity -- expressed in
parameters such as the number of topologically distinct regions
\citep[e.g.,][]{2006SoPh..237...45C,2006ApJ...646.1303B}, and the
``effective connected magnetic field''  \citep{2007ApJ...661L.109G} --
shows promise, as do estimators of the global non-potentiality\index{non-potential energy} 
of the magnetic field such as the flux-normalized field twist
\citep[e.g.,][]{2002ApJ...569.1016F}.  
Aside from those methods
based on magnetic field information, a Bayesian approach\index{Bayesian approach} using past
history of flare occurrence in an active region has also been
proposed \citep{2004ApJ...609.1134W}.

One final indicator of approaching flare activity that is worth
mentioning is filament activation\index{filaments!activation}. 
A filament\index{filaments} is a narrow concentration
of dense, cool material ($n_e \approx 10^{12}\ {\rm cm}^{-3}, T\approx
10^4\ {\rm K}$) which overlies and runs parallel to a magnetic neutral
line. 
The filament material is supported in the atmosphere by a
strongly-sheared magnetic field in the corona, primarily oriented
along the neutral line, and it exhibits substantial plasma flows parallel to this. 
Quiescent filaments are visible in absorption in H$\alpha$\index{Fraunhofer lines!H$\alpha$} against the disk (and emission at the limb, against the background sky), and also in absorption in the \textit{Transition Region and Coronal Explorer}\index{satellites!TRACE@\textit{TRACE}} (\textit{TRACE}) extreme ultraviolet (EUV) lines. 
Prior to a flare\index{flares!pre-flare filament motions}, such features are often observed to start
rising slowly. 
Just prior to their eruption, brightenings in part
or all of the filament may also be observed along its length, in
wavelengths from H$\alpha$ to EUV. 
Filaments and their activation/eruption
are interesting in the context of flare development as their
involvement in the earliest phase of the eruption, their low altitude
in the corona, and their strong concentration along the region of
sheared field all point towards the flare initiation, and possibly
a large part of the pre-flare energy storage\index{flares!energy storage} taking place within
1-2~$\times$~10$^4$~km above the photosphere.

\index{magnetic structures!delta spot@$\delta$ spot!illustration}

\bigskip
\noindent{\bf{Terminology:}} 
Over the decades, the task of describing the appearance and time
evolution of a flare led to a flowering of classifications and
related vocabulary, which evolved along with our understanding of
the range of flare and flare-related phenomena.  This included much
to-ing and fro-ing over the direction of causality\index{causality} between flare
and CME phenomena \citep[see][for a historical
overview]{1995SoPh..157..285C}.  
\index{coronal mass ejections (CMEs)}
The time evolution of a flare based
on observations is now normally characterized by two main phases,
these being the impulsive phase and the gradual phase.  
The rough
division of the time-history of emission into an abrupt and a more
slowly-varying phase has been recognized\index{flares!phases} through the history of
flare observations in different wavelengths, including
H$\alpha$~\citep{1946MNRAS.106..500E}, microwaves \citep{1963ARA&A...1..291W},
and HXRs \citep{1969ApJ...157L.139K,1971ApJ...164..151K}.  ``Spotless''
flares, mentioned above, tend to have less rapid rises and less
significant impulsive phases than flares in normal active regions.
Some authors have also classified events as either ``gradual flares"
or ``impulsive flares"  based on their time characteristics within
a single wavelength range, for example SXRs \citep{1977ApJ...216..108P}
or  HXRs \citep{1983SoPh...86..301O,1986ApJ...308..912B} emissions.
The notion of the ``impulsive flare,'' i.e., one with no gradual phase has more-or-less faded from parlance among flare physicists, though it persists in discussions of solar energetic particle events.  
However, exclusively ``gradual flares'' do occur; such events have essentially no impulsive component and show only a more slowly-varying X-ray component at relatively low peak temperatures
-- so low, indeed, as to be undetectable by \textit{GOES} \citep[see][for example]{1995JGR...100.3473H}. 
\index{soft X-rays!long-decay events (LDE)}
These are often associated with filament eruptions and have properties similar to ``long decay
events'' (LDEs)\index{flare types!long-decay events (LDEs)} that tend to be accompanied by CMEs
\citep{1977ApJ...214..891K,1983ApJ...272..349S}, but which typically do have impulsive phases.

\bigskip
\noindent{\bf{Time evolution:}}
The time evolution of a flare is characterized by different timescales
visible at different wavelengths. In the SXR range, following a
rise phase lasting a few minutes, evolution is slow.\index{soft X-rays!gradual time profiles}
The return
of this thermal emission to its pre-event levels is a smooth decay
that can last for hours. 
At the opposite extreme, the lower-frequency
end of the radio spectrum exhibits bursts known as type~III bursts,
with a high brightness temperature, rapid drift rate (frequency
decreasing with time) and duration as short as tens of milliseconds
\citep{1995ApJ...455..347A}.\index{radio emission!type III burst}
These are most normally seen in the decimetric regime and far below, but
can also be found in the microwaves.  
Observations of X-rays at
tens of keV and above, with the BATSE\index{satellites!CGRO@\textit{CGRO}!BATSE} detector (on the \textit{Compton Gamma Ray Observatory, CGRO}) 
exhibit pulses with widths of hundreds of milliseconds
to a second or two \citep[see][for details]{1995ApJ...447..923A}\index{hard X-rays!short time scales}.
This fast structure is sometimes apparently superposed on a slower,
large amplitude variation (though the slow component  could be the
superposition of many rapid variations).  
It has not proved possible to detect these short pulses with \textit{RHESSI}.  
\index{satellites!Yohkoh@\textit{Yohkoh}}
However with both
{\emph{Yohkoh}}/HXT and \textit{RHESSI}\index{RHESSI@\textit{RHESSI}!and QPP}, the phenomenon of HXR quasi-periodic
pulsations has become very clear
\citep{2005A&A...440L..59F,2006ApJ...644L.149O,2008SoPh..247...77L,2009A&A...493..259I}.
\index{quasi-periodic pulsations}
\index{waves!quasi-periodic pulsations}
These are modulations in the X-ray intensity, in the few to tens
of keV range with a modulation depth of up to 90\%,  and observed
periods of tens of seconds to a few minutes.  
The periods observed demand an explanation in terms of MHD timescales; for example, a
sausage mode\index{waves!sausage mode} could alter the magnetic field in the loop leading to
variations in trapped particle precipitation rates \citep{2009A&A...493..259I}, or perturbations to the accelerator itself \citep{2006ApJ...644L.149O}. 
Optical flare variations are
typically abrupt, tracking the HXR lightcurves well at energies of
tens of keV in the rise phase, but sometimes having a slower decay.
UV and EUV also show mixed impulsive and slow variations -- major
HXR spikes are reflected in the lightcurves, but superposed on a
slowly-varying background.

In the HXR impulsive phase, a well-known temporal pattern is the
so-called \textit{soft-hard-soft} spectral pattern -- namely that
the spectral index of the non-thermal part of the photon spectrum
becomes harder as the non-thermal flux increases.  
\index{hard X-rays!soft-hard-soft}\index{soft-hard-soft}
This is a pattern that holds universally across individual bursts
in the impulsive phase of a flare, both in the footpoint regions
and the looptop regions\index{looptop sources!impulsive phase}
(e.g., Battaglia \& Benz 2006; see also Figure~\ref{grigis_fpim1}),
\nocite{2006A&A...456..751B} 
and on timescales down to a few seconds.  
In many flares it also represents
the evolution of the entire event at lower time resolution, as the
HXR flux increases and then decreases again.  However there is a
separate class of events -- the \textit{soft-hard-harder} events --
in which the spectrum continues to become harder throughout the
duration of the event \citep{2000ApJ...545.1116S}.\index{flare types!long-duration}\index{soft-hard-harder}
This is a property of gradual HXR events\index{hard X-rays!gradual events}\index{hard X-rays!coronal sources}
\citep[e.g.,][]{1986ApJ...305..920C,2008ApJ...673.1169S}, and in at
least some events may reflect the presence of long-lived high energy
coronal sources \citep{2008ApJ...678L..63K}.  It has been interpreted
both as a consequence of energy-dependent particle losses from a
coronal trap (with low-energy particles being scattered out of the
trap before high-energy particles), and as evidence of continued
injection of high energy particles after the flare impulsive phase.

The final time-behavior mentioned here is the \textit{Neupert effect}\index{Neupert effect}
-- the phenomenon that in many flares the time integral of the
non-thermal emissions tracks the time profile of the thermal
emissions. The underlying reason for this is that the mostly thermal
coronal plasma has a much longer energy loss timescale than the
chromosphere, and ``integrates'' the energy deposited there, presumably
from evaporation caused by more impulsive lower-atmosphere energy
input.  
This was first discovered as a  delay between the peaks of
SXR and microwave emissions \citep{1968ApJ...153L..59N}, but  has
since become more commonly associated with the HXR/SXR time profiles;
essentially the HXRs show the (non-thermal) energy release as it
happens, and the SXRs show the part of it that winds up in the corona
as high-temperature thermal plasma. 
The fidelity with which the Neupert effect holds depends on the wavelength
ranges chosen to test it, with higher temperature thermal emissions
showing a better relationship \citep{1999ApJ...514..472M}. 
The
Neupert timing relationship was found to hold in 80\% of 66 large
events studied using the \textit{GOES} 1-8~\AA~ channel and the 
\textit{SMM}/HXRBS\index{satellites!SMM@\textit{SMM}!HXRBS}
26-41~keV channel \citep{1993SoPh..146..177D}. A statistical study
of the timing of the SXR peak compared to the HXR impulsive phase in
more than 1000 events, using \textit{GOES} 1-8~\AA~ and 
\textit{CGRO}/BATSE\index{satellites!CGRO@\textit{CGRO}!BATSE} HXR counts
at 25-50 and 50-100~keV, found that 50\% of events were consistent
with Neupert-like timing behavior, 25\% were inconsistent, with SXR
emission peaking substantially after the end of the HXR emission,
and the remaining 25\% were unclear \citep{2002A&A...392..699V}. 
In those flares consistent with Neupert timing there was also a strong linear
correlation between the HXR fluence (time-integrated counts) and
SXR peak flux, as expected. It has been speculated
\citep{2006ApJ...652L..61L} that the flares which do not show the
expected Neupert timing behavior may occur if the SXR peak only
when plasma ``evaporated'' from the chromosphere (see
Section~\ref{sec:fletcher_fpevap})\index{chromospheric evaporation} reaches the top of the flaring
loops  \citep{1995A&A...299..225R}, which introduces a longer time
lag for larger flare loops.  
But the level of disagreement has
prompted further investigations of individual flares, in which the
beam power inferred from HXR and the power required to explain the
SXR flux and spectrum were tested for consistency
\citep{2005ApJ...621..482V}.  
Under straightforward model assumptions
these two powers were found not to correlate well in time, which
may suggest that there is energy input other than by non-thermal
electrons during some phases of the flare 
\citep[see counter-examples from][]{2008SoPh..248...99N}. 
Of course, plasma flows may also lead to heating via compression that might not match
the non-thermal signatures so well \citep[e.g.][]{2010ApJ...725L.161C}.
Violating one
particular model assumption, that of a constant value low-energy
cutoff, could (if the cutoff varies through the flare in the right
way) lead to better agreement
\citep{2005ApJ...626.1102S,2005ApJ...621..482V}.  At the present
time, the discrepancies between observed and theoretical Neupert
effects are probably  within the observational limits, but hints
of different physics from the standard ideas that  hot thermal
emission is coronal and non-thermal emission mostly chromospheric
should of course not be ignored.

\index{acceleration!low-energy cutoff} 
\index{low-energy cutoff} 
Cross-correlation
analysis has indicated a delay of the SXR time derivative by
$\sim$10~s relative to the HXR flux \citep{2006ApJ...649.1124L}.
This delay can be interpreted as the hydrodynamic timescale for the
redistribution of energy deposited by non-thermal electrons, consistent
with the results of radiation hydrodynamic simulations
\citep[e.g.,][]{1993ApJ...417..313L,2008sfpa.book.....L,Chapter8}.\index{simulations!radiation hydrodynamic}

\bigskip
\noindent{\bf{Preflare evolution:}}
Preceding the impulsive phase, there may be initial signs of activity
termed the \textit{pre-flare phase}\index{flares!pre-flare phase}. This term covers both  pre-flare
activity, which refers to the very earliest stages of the flare
before the impulsive phase radiation is detectable, and the
\textit{flare precursor}\index{flares!precursor events} events, which are small-scale brightenings
in UV to SXR wavelengths happening some tens of minutes before the
flare.\index{flares!hard X-ray precursor}\index{precursor}
Spatially unresolved lightcurve data such as that obtained with
\textit{GOES} may be misleading in this respect, since apparent flare
precursors can originate from distant active regions, whereas actual
flare precursors may fail to be visible against the integrated
intensity from the disk. However with spatially resolved observations
it is apparent that flare precursors often do occur in the neighborhood
of, but not usually at exactly the same location as, the site in
which the majority of the flare radiation will subsequently occur
\citep{1996SoPh..165..169F,1998SoPh..183..339F,2001ApJ...560L..87W,2003A&A...399.1159F}.
Some authors have explicitly linked pre-flare brightenings with the
destabilization of the magnetic structure that will lead to a CME
\index{magnetic structures!CME formation}
\index{coronal mass ejections (CMEs)}
\citep[e.g.,][]{1985SoPh...97..387H,2005ApJ...630.1148S} or a
filament eruption
\citep{2003A&A...399.1159F,2007A&A...472..967C,2009ApJ...698..632L}
although coronagraphic studies of the former kind are plagued by
the lack of knowledge of the CME launch time, since substantial
parts of the event occur unseen below the coronagraph occulting
disk.  The onset of pre-flare activity, in the form of weak SXR
emission, precedes the onset of impulsive HXR emission by around 3
minutes in the vast majority of flares, regardless of their total
energy or duration \citep{2002SoPh..208..297V}.  Spectral line
broadening has been observed in the pre-flare phase
\citep{2001ApJ...549L.245H,2009ApJ...691L..99H} starting minutes
to hours before the impulsive phase, consistent with non-thermal
effects such as plasma turbulence\index{turbulence}\index{plasma turbulence}.

\subsection{The impulsive phase}\index{impulsive phase}
The primary energy release occurs during the impulsive phase\index{flares!impulsive phase}.  
This phase of flare activity lasts from tens of seconds to tens of minutes
and is characterized by HXRs, $\gamma$-ray, non-thermal (synchrotron)
microwaves and white-light continuum emission, indicating strong
acceleration of both electrons and ions.  These radiations are
accompanied also by strong enhancements in chromospheric line and
continuum emission, ultraviolet and extreme ultraviolet radiation,
and bulk plasma upflows\index{upflows!EUV}\index{upflows!SXR} 
in the EUV and SXRs at speeds on the order
of 100~km~s$^{-1}$  coupled with downflows in cooler lines such as
H$\alpha$~\citep{1988ApJ...324..582Z,2006ApJ...638L.117M}\index{chromosphere!flare-related flows}. 
The impulsive-phase radiation is
concentrated at the chromospheric endpoints of the magnetic field
involved in the flare; indeed prior to the availability of the EUV
and SXR imaging that has led to a shift of focus to the coronal
aspects of flares, the strong lower atmosphere signatures led to
the term ``chromospheric eruption'' or ``chromospheric flare'' being
used in early studies \citep{1948MNRAS.108..163G} even as the
relationship of the radiation increase to flows and to processes
higher in the atmosphere and at the Earth were becoming apparent.
Even now, it is clear that while the magnetic drama takes place in
the corona, the dominant radiative energy of a flare, from both
non-thermal and thermal particles,  is from the lower atmosphere
\citep[e.g.,][]{1973Natur.241..333C,1972SoPh...24..414H,
1979ApJ...234..669K,2004A&A...418..737L,1972SoPh...23..444M,1983ApJ...265..530W}.
So somehow the flare energy must be transported into the chromosphere,
there to be dissipated by radiation and flows.  There is unfortunately
only very sparse knowledge of the optical and UV properties of the
impulsive phase, and this is a substantial shortcoming and missed
opportunity. Rich diagnostic information, particularly in spectroscopic
data, informs our understanding of the quiet-Sun chromosphere, and
to have such data available also for flaring regions would substantially
improve our knowledge of the results of flare energy deposition in
these layers, and even give insights into the flare particle
acceleration problem.

The magnetic reconfiguration that allows the rapid release of stored
magnetic energy in a flare is generally agreed to occur somewhere
above the chromosphere in the (low-$\beta$) corona\index{magnetic structures!energy storage}. 
The main
theoretical argument for a coronal energy release is that the corona
provides adequate volume for storing the energy required for a
flare. Observationally, the coronal manifestations such as large
SXR and H$\alpha$ flare loops, HXR looptop sources\index{looptop sources}, and coronal
mass ejections have almost universally been interpreted in a framework
involving large-scale coronal magnetic reconnection\index{flares!coronal manifestations}\index{reconnection}. 
Non-linear force-free\index{magnetic structures!force-free} reconstructions of the magnetic field find that the energy is concentrated low in the corona in a newly-emerged active
region\index{flux emergence}, and can be sufficient for flaring activity
\citep[][]{2007A&A...468..701R,2008ApJ...675.1637S}  whereas in an
older decaying active region it is stored higher in the corona.  
It has proved marginally possible to detect differences between the free
magnetic energy before and after a flare or CME event
\citep{2002mwoc.conf..249M,2002A&A...395..685B,2008ApJ...675.1637S}.
More quantitative reconstructions of the coronal magnetic field which can track the
actual redistribution of coronal magnetic energy during a flare are
just beginning to appear \citep{2008ApJ...676L..81J,2008A&A...484..495T}\index{reconnection!impulsive phase}\index{flares!energy content!magnetic}\index{magnetic field!energy storage}.

\begin{figure}
\begin{center}
\includegraphics[width=0.9\textwidth,height=0.45\textwidth]{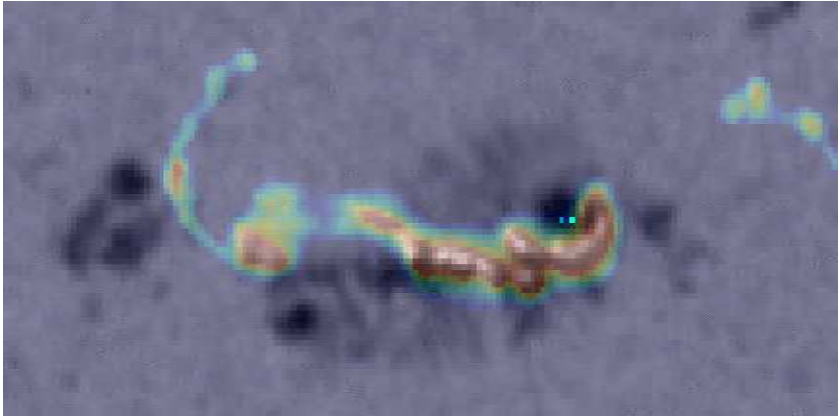}
\end{center}
\caption{\label{fig:ribbons_fps} SOL2002-10-04T05:38 (M4.0),
\index{flare (individual)!SOL2002-10-04T05:38 (M4.0)} 
with \textit{TRACE}~UV (1700~\AA) ribbons in color overlaid on \textit{TRACE} continuum
in black and white. The flare white-light sources are visible as
small white patches, within the UV~ribbons, and having a considerably
smaller area.  The field of view is 135$''$ by 65$''$, pixel size
0.5$''$.  Here the main neutral line lies between the concave-up
UV feature under the sunspot on the right, and the more linear
feature below it.} 
\index{sunspots!and white-light emission!illustration}
\index{TRACE@\textit{TRACE}!UV ribbons!illustration}
\index{flare types!white light!illustration}
\index{ribbons!illustration}
\end{figure}

The impulsive-phase flare signatures in the lower solar atmosphere
\index{hard X-rays!footpoints}
are termed ``footpoints'' (originally HXR) or ``ribbons'' (originally
H$\alpha$) and are now detectable in a wide range of wavelengths.\index{ribbons!vs. footpoints}\index{footpoints!vs. ribbons}
They
are interpreted as the chromospheric ends of the coronal magnetic
field structures involved with the flare energization at  a given
instant.  The impulsive-phase reconnection of the coronal magnetic
field is not visible in coronal signatures with current instruments,
though its effects certainly may be.  These effects include the EUV
and SXR flare loops and cusp-shaped structures\index{magnetic structures!cusps} 
that appear in the
gradual phase (see Section~\ref{sec:fletcher_gradualphase}) and
impulsive phase supra-arcade downflows
\citep[][]{2004ApJ...605L..77A,2007A&A...475..333K}, interpreted
as the ``dipolarization'' of newly-reconnected field. These correspond
extremely well in time with impulsive phase HXR bursts\index{downflows!supra-arcade}.
\index{dipolarization} 
Flare footpoints occur on either side of the
magnetic polarity inversion line, as illustrated in
Figure~\ref{fig:ribbons_fps}.  Early in the impulsive phase they
tend to be concentrated around this line, and move with respect to
it as the flare evolves.\index{ribbons!H$\alpha$}
In the later phase of a flare, when strong
H$\alpha$~and UV ribbons\index{ribbons!UV} are visible, the ribbons tend to move outward
from the polarity inversion line, but in the impulsive phase, both
ribbon and HXR footpoint motion is more complex, sometimes showing
parallel or approaching motions. This is discussed in more detail
in Section~\ref{sec:fletcher_fpmotions}.

The importance of flare ribbons and footpoints in marking regions
of changing magnetic connectivity is now well established
\citep[e.g.,][]{1991A&A...250..541M}. The large-scale reconnection
model in two dimensions  \citep[e.g.,][]{1976SoPh...50...85K} was
originally developed to explain the spreading H$\alpha$~ribbons and the
H$\alpha$~arcade that appears in the gradual phase of a flare.  
The outer
edges of the ribbons show the chromospheric projection of the
interface (magnetic separatrix surface) between the post-reconnection
(``post-flare'') arcade fields, and the field that is yet to be
reconnected\index{magnetic structures!separatrices}.\index{separatrix}\index{ribbons!and separatrices}
The importance to
flare energy release of the three-dimensional connectivity of the
\index{sunspots!connectivity}
\index{magnetic structures!sunspots!field topology}
\index{magnetic structures!three-dimensionality}
solar magnetic field around sunspots was known early on
\citep{1948MNRAS.108..163G,1969ARA&A...7..149S,1981ARA&A..19..163S} and
early observational associations were made between topological
structures and observable chromospheric H$\alpha$~features by e.g.,
\cite{1989SvA....33...57G}, \cite{1991A&A...250..541M}, and \cite{1992SoPh..139..105D}.
Interpreting the chromospheric features within this framework, it
becomes possible to establish some of the global properties of the
magnetic field and its evolution, such as the different magnetic
domains, the amount and rate of magnetic flux transfer during flare
events, and also -- under the assumption of two-dimensional
translational symmetry -- the convective electric field\index{electric fields!convective} ({\bf
v}~$\times$~{\bf B}) of the magnetic reconnection\index{reconnection!convective electric field}.

\subsection{The gradual phase}\label{sec:fletcher_gradualphase}
\index{gradual phase}
\index{reconnection!gradual phase}

\begin{figure}
\begin{center}
\hbox{
\includegraphics[width=0.5\textwidth]{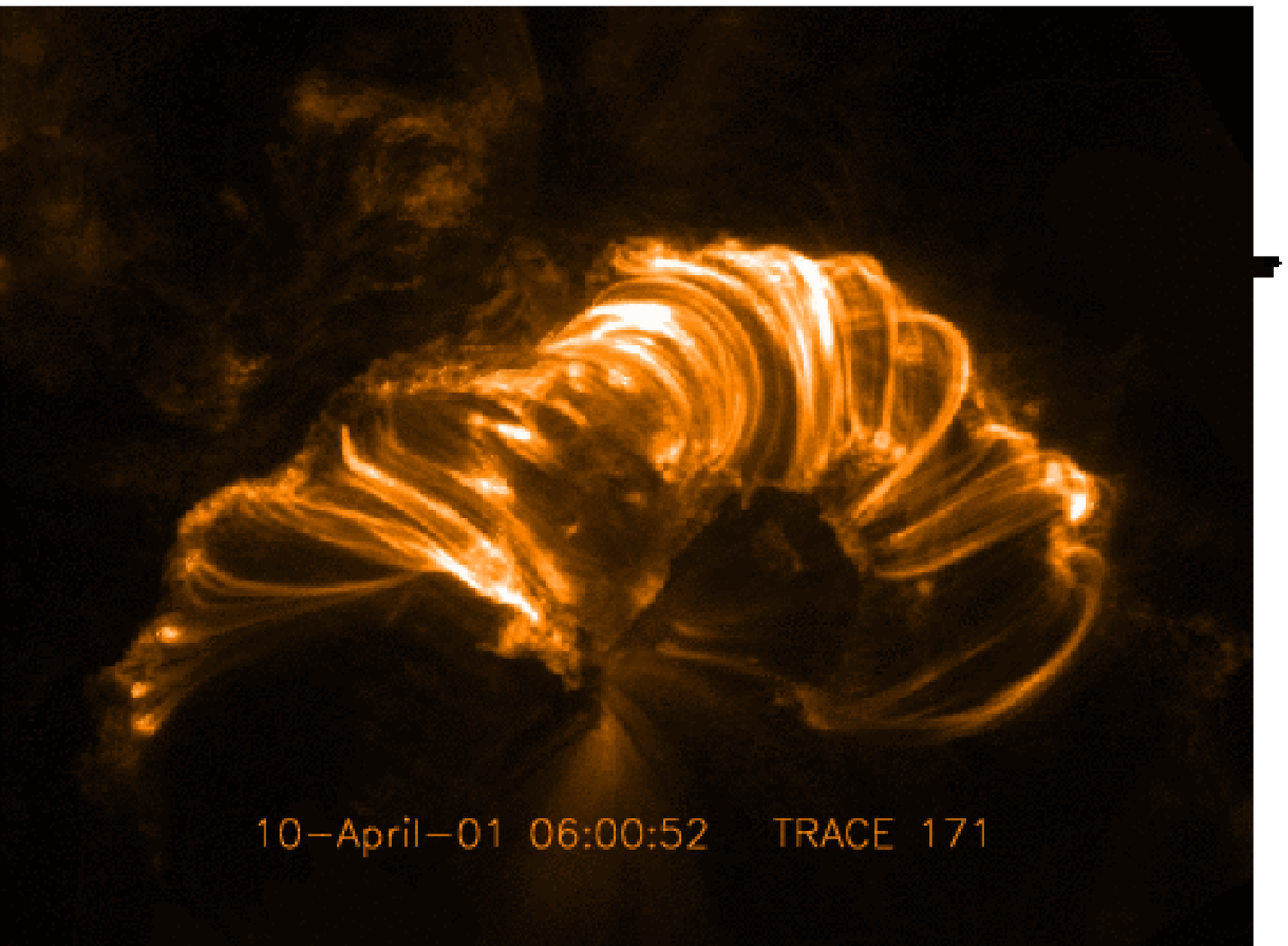}
\hspace{0.5cm}
\includegraphics[width=0.5\textwidth]{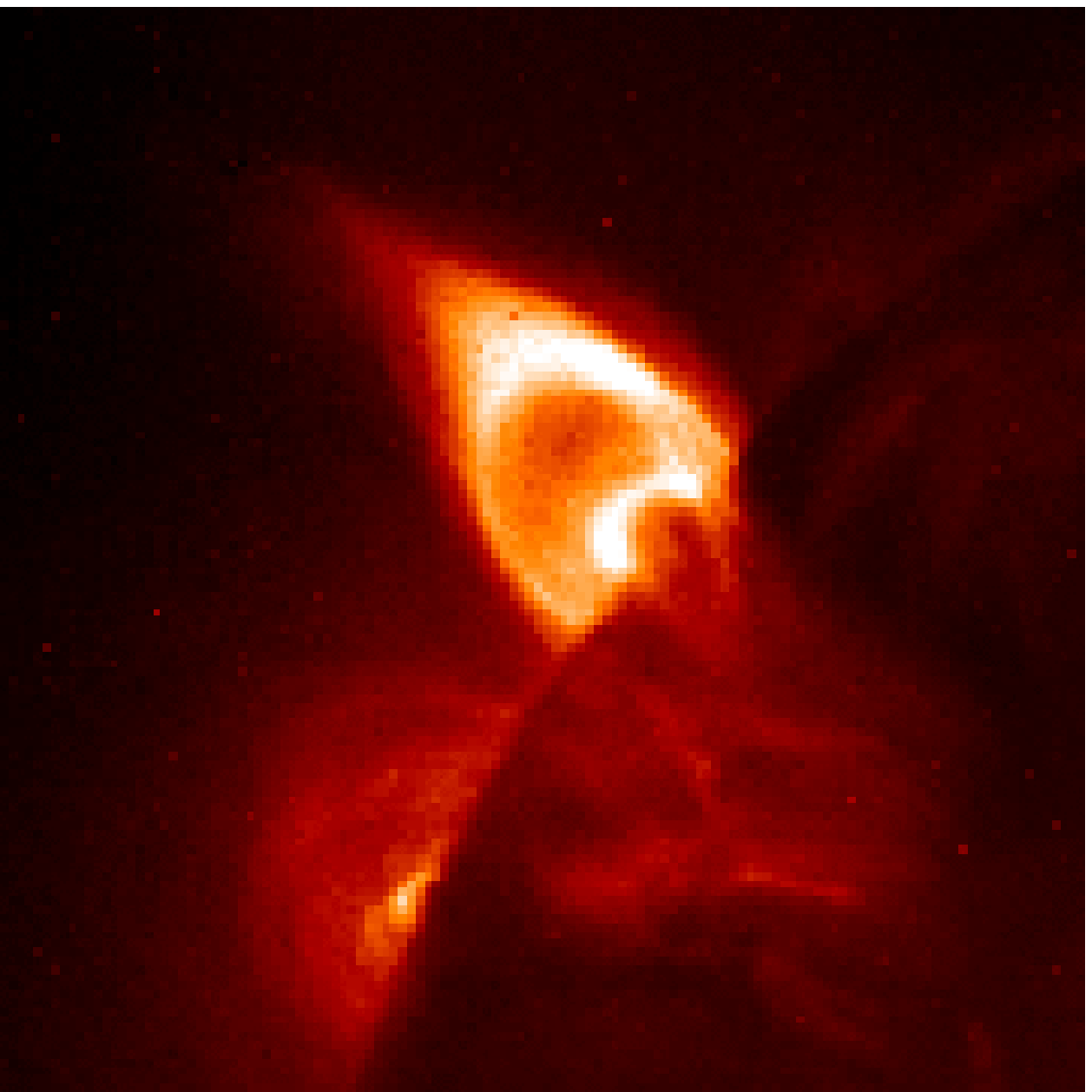}
}
\end{center}
\caption{\label{fig:trace_arcade} {\emph{Left}}: a beautiful
flare arcade (SOL2001-04-10T05:26, X2.3) seen here in the 171~\AA\ channel of \textit{TRACE}, revealing plasma at $\sim$1~MK emitting in lines of Fe~{\sc ix/x}.
\index{flare (individual)!SOL2001-04-10T05:26 (X2.3)!illustration}
{\emph{Right}}: a
large post-flare cusp structure observed several hours after the
impulsive peak of SOL1999-03-18T08:31 (M3.3)\index{flare (individual)!SOL1999-03-18T08:31 (M3.3)!illustration} 
by \textit{Yohkoh}/SXT
and reported by \cite{2001ApJ...546L..69Y}. The temperature in this
structure is 3-4~MK.} 
\index{arcade!illustration} 
\index{TRACE@\textit{TRACE}!flare arcade in EUV!illustration}
\index{magnetic structures!cusps!illustration}
\end{figure}

During the gradual phase, identified by its slowly decaying SXR and
microwave signatures, the effects of the flare on the corona become
apparent.  Loops and loop arcades emitting in SXRs and EUV form and
appear to grow (Figure~\ref{fig:trace_arcade}), filled (it is usually
assumed) by chromospheric plasma forced to expand into the corona
as the chromosphere is rapidly heated by particle energy deposition,
or by thermal conduction.  
\index{cooling!conductive}
\index{conduction!thermal}
\index{thermal conduction}
\index{chromospheric evaporation}
This expansion is known as \textit{chromospheric
evaporation} (see Section~\ref{sec:fletcher_fpevap}). 
The gas pressure of these flaring coronal loops can increase from
$\sim$0.1~dyne~cm$^{-2}$, typical of the quiet corona, to
$>$10$^3$~dyne~cm$^{-2}$, as shown by semi-empirical models based
on radiative transfer theory in chromospheric lines
\citep[e.g.,][]{1980ApJ...242..336M}. 
The pressure increase is
mainly due to the growth of density in the loops, but the new coronal
material is also at flare temperatures (10-20~MK), as opposed to
pre-flare coronal temperatures (1-3 MK). 
The loop arcades show a
gradient in temperature, with the outermost loops (i.e., those
corresponding to the outer edges of the ribbons) being the hottest
\citep{1996ApJ...459..330F}. The hottest outer loops sometimes
exhibit a ``cusp'' (Figure~\ref{fig:trace_arcade}) consistent with
the shape of the field that would be expected below a coronal current
sheet; this cusp is most pronounced later in the gradual phase of
the flare
\citep[][]{1992PASJ...44L..63T,2001ApJ...546L..69Y,2008PASJ...60..275H}.
\index{reconnection!circumstantial evidence for}
This is circumstantial evidence for coronal magnetic reconnection
during this phase. Later on, as the corona cools, the arcade becomes
visible at lower temperatures including EUV and
H$\alpha$~\citep[][]{1995SoPh..156..337S}. 
\index{cooling!parameter dependence}
\index{cooling!radiative}
\index{cooling!conductive}
Cooling occurs by both conduction
and radiation, depending on the flare loop length and plasma
parameters
\citep[][]{1970SoPh...15..394C,1995ApJ...439.1034C,2001SoPh..204...91A}.
Observational studies tend to find that early on the hottest plasma
cools for a few minutes by conduction \citep[e.g.,][]{1994SoPh..153..307C}
which may then be followed by dominant radiative cooling
\citep[e.g.,][]{2001SoPh..204...91A,2006SoPh..234..273V}. Models
have been formulated which also take into account the thermal energy
redistribution throughout the loop due to conduction, and the gentle
chromospheric evaporation that results \citep[][]{1995ApJ...439.1034C}.
As the loop plasma cools it begins to drain under gravity, and
H$\alpha$~downflows (``coronal rain'') become visible along the legs of
the arcade\index{downflows!coronal rain}. 
The plasma upflows\index{upflows} and
downflows have been detected spectroscopically in a number of events
\citep[e.g.,][]{1988ApJ...329..456Z,1999ApJ...521L..75C,2003ApJ...586.1417B}.

\begin{figure}
\begin{center}
\includegraphics[width=0.9\textwidth]{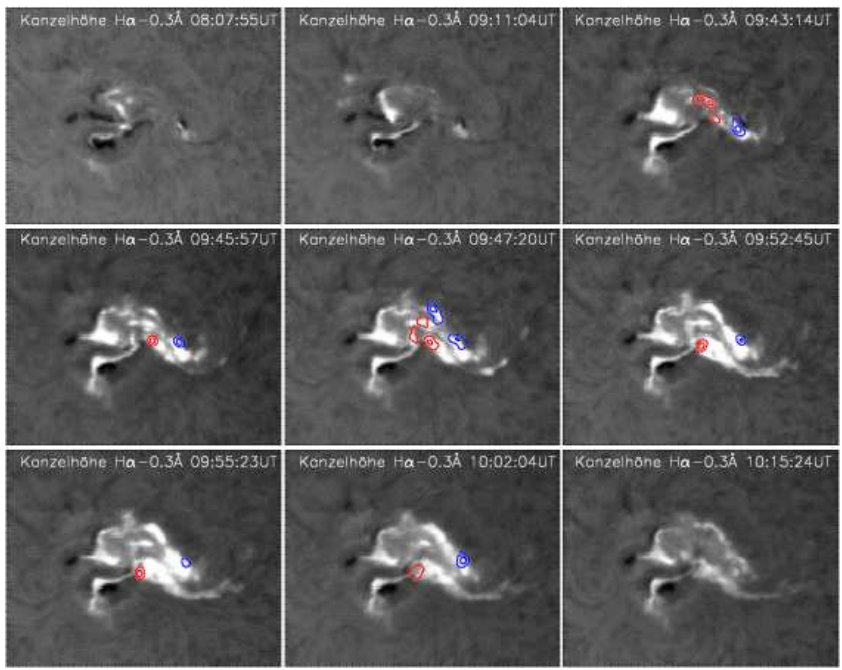}
\end{center}
\caption{\label{fig:v1} 
Sequence of H$\alpha$--$0.3$~{\AA} filtergrams taken at Kanzelh\"ohe
Observatory during SOL2005-01-17T09:52 (X3.8) in a
400$''$$\,\times\,$300$''$ field of view.  The contours are co-temporal
\textit{RHESSI} 30-100~keV images, and the color indicates the polarity of
the underlying photospheric magnetic field (red: positive, blue:
negative).  Before 09:35~UT, no \textit{RHESSI} data were available.  During
certain flare periods, up to five individual HXR footpoint sources were
observed simultaneously, located on different H$\alpha$ flare
ribbons,\index{ribbons!H$\alpha$} but two main footpoints prevailed during the overall flare impulsive
phase.  Note the general counterclockwise rotation of the line
connecting the opposite-polarity footpoints, an observation often
interpreted as a reduction in shear.  Adapted from
\cite{2007ApJ...654..665T}.  } 
\index{flare (individual)!SOL2005-01-17T09:52 (X3.8)!illustration} 
\end{figure}

Loop arcades observed at a particular wavelength appear to grow
upwards and outwards in time.  This can be seen in
H$\alpha$~\citep{1987SoPh..108..237S}, microwaves \citep{2005ApJ...629L.137L},
EUV, SXRs and even HXRs \citep[e.g.,][]{2002SoPh..210..341G}.
In the common interpretation, the point of magnetic reconnection moves
slowly upwards in the solar corona as the gradual phase proceeds.
\index{CSHKP}
\index{arcade} 
\index{flare models!CSHKP} 
\index{reconnection!standard flare model}
Successive shells of reconnected loops fill with hot plasma
expanding from the chromospheric footpoints, a pattern often described
as the ``CSHKP model'' after some of its major contributors
(H.~Carmichael, P.~Sturrock, T.~ Hirayama, R.~Kopp, and J.~Pneuman).
As new loops or their footpoints brighten in a particular wavelength,
their angle with respect to the magnetic polarity inversion line
may increase, as reported since \textit{Skylab} in many events
\citep[e.g.,][]{1992PASJ...44L.123S,1994PhDT.......335S,2001SoPh..204...55M,2001SoPh..204...91A,2006SoPh..236..325S,2006ApJ...643.1271S,2008ApJ...680..734J,2009ApJ...693..847L}.
This could be interpreted as due to reconnection starting in
highly-sheared field anchored close to the polarity inversion line,
and progressing to less-sheared field further from it.  
Figure~\ref{fig:v1} illustrates this process.

The gradual phase may last several hours, depending on the magnitude
of the flare.  In many events, the cooling timescales point towards
an additional energy source during the gradual phase, which could
come from ongoing slow reconnection and its associated heating
\citep[][]{1979SoPh...61...69M,1989SoPh..120..285F}, either above
the flare arcade or conceivably also between individual tangled
strands of the arcade as they shrink down.  The arcade appears to
grow in scale as newer and higher-altitude loops appear, while
individual loops physically shrink with time
\citep{1982SoPh...75..305S,1996ApJ...459..330F}.  
Multi-strand loop\index{magnetic structures!multi-strand loops}\index{flare models!multi-strand} modeling
\citep[][]{1998ApJ...500..492H,2002ApJ...579L..41W,2002ApJ...578..590R} may
explain cooling time profiles, and also the fact that the observed
velocity characteristics of evaporative upflows\index{upflows} tend to be smaller
than those predicted by radiation hydrodynamic simulations.\index{simulations!radiation hydrodynamic}

The gradual phase and the standard magnetic-reconnection model
nicely link the arcade formation with the occurrence of a filament
eruption and a CME; this would create the (approximately)
oppositely-directed field lines that then reconnect
(Section~\ref{sec:cme}).\index{flare models!CSHKP}\index{reconnection!standard flare model}\index{arcade}\index{CSHKP}
Further evidence for ongoing magnetic
coronal reconnection  comes from the supra-arcade
downflows\index{downflows!supra-arcade} \citep{1999ApJ...519L..93M,2004ApJ...616.1224S} observed mainly
during the gradual phase and apparently moving at a fraction of the
Alfv{\' e}n speed.  
\index{Alfv{\' e}n speed!and downflows}
\index{downflows} 
As we have seen already, such
downflows also occur in the impulsive phase of a flare.  The magnetic
geometry in which the reconnection and shrinkage occur may be
considerably more complex in the impulsive phase.

\subsection{The magnetic field}\label{sec:fletcher_magneticfield}
\index{magnetic field}

How well do we understand the structure of the magnetic field, which
underlies all aspects of solar activity?  There are routine
measurements of the Zeeman splitting\index{Zeeman effect} in the solar photosphere (see
Section~\ref{sec:opt}), especially for the line-of-sight component
of the field.  We can identify the magnetic environment of flare
occurrence morphologically -- a large, rapidly-forming sunspot group
containing a ``delta spot'' certainly has a higher probability of
flaring \citep[e.g.,][]{1987SoPh..113..267Z}.  
\index{sunspots!delta spots@$\delta$ spots}
\index{magnetic structures!delta spot@$\delta$ spot}
At the photospheric level, the polarity inversion line (``PIL'' or
``neutral line'') plays an important role morphologically.
\index{magnetic structures!neutral line} This line by definition must
separate the footpoints of coronal loops, such as flare loops.  The
field, again by definition, lies tangent to the plane of the
photosphere at the neutral line, and where it is concave up (the
``bald patch'' configuration) it does not lead to an overlying
arcade, but still may play an important role in flare development
\citep{1993A&A...276..564T}.  
\index{magnetic structures!bald patch}

Unfortunately the field in the photosphere, which can be observed
comprehensively, only maps into the corona in an elusive manner
\citep[e.g.,][]{2008ApJ...675.1637S}.  Many competing mathematical
and MHD-based approaches have been applied to the problem of
extrapolating the photospheric knowledge into the corona
\citep{2006SoPh..235..161S}, but several substantial problems prevent
these techniques from being quantitatively persuasive.  From the
perspective of flare research, one major objective (for example)
would be to characterize the ``non-potential energy''\index{non-potential energy} 
resulting
from current systems\index{magnetic structures!energy storage} 
linking the corona and the solar interior.
This energy exceeds the minimum level derivable from a ``potential
field'' representation, i.e., one without embedded currents.
\index{magnetic field!potential} This should decrease when a flare
happens, but the results of all analyses to date have been uncertain
on this point \citep[e.g.,][]{2002mwoc.conf..249M}.

The direct measurement of the field via line-splitting observations
in an active-region corona has met extreme difficulties, and
accordingly the results at the present time have great uncertainties.
In principle one needs to measure a full vector function {\bf B}({\bf
r}) at each point {\bf r} in the coronal volume. 
There is progress on this formidable task \citep[e.g.,][]{2008SoPh..247..411T}.
Radio techniques{\index{magnetic field!measurement!radio techniques} 
also have great promise to determine at least $|${\bf B}$|$, though,
since the Larmor frequency\index{frequency!Larmor} and its harmonics have clearly identifiable
and precise signatures \citep[e.g.,][]{1998ARA&A..36..131B}.  
In general the most exact knowledge of at least the plane-of-the-sky
projection of {\bf B}({\bf r}) comes from high-resolution imaging
in coronal emissions (visible, UV, EUV, X-ray), since they often
have fine striations that can be interpreted as due to the alignment
of the field. 
 A potential-field interpretation of the intensity in the corona above a large sunspot can be approximated by that of \index{sunspots!umbral field}
a simple current loop lying in the photosphere: $B(z)/B_0 = (z^2 + 1)^{-3/2}$, 
where z~is the height above the umbra in units of the umbral radius.  
For a radius of 10$^4$~km and an umbral field of
3500~G \citep[e.g.,][]{2000asqu.book.....C} this formula gives axial
fields of 300-1200~G at coronal altitudes of 1-2~$\times$~10$^4$~km.
For reference, the current in the loop has a magnitude of 5.5~$\times$~10$^{12}$~A.

When a solar flare happens, the observed photospheric field changes
in a stepwise manner \citep{1992SoPh..140...85W,2005ApJ...635..647S}.
\index{magnetic field!flare-related changes} This would be expected
from the extraction of energy by its dipolarization\index{dipolarization} following
large-scale reconnection, or generally just to reduce $\int{(B^2/8\pi)
dV}$ \citep{2000ApJ...531L..75H}; see Section~\ref{sec:fletcher_deep}.

\section{Status of observational techniques}\label{sec:techniques}
\index{observing facilities}

Flare emissions have been detected all the way from about $10^{-10}$~eV
(30~kHz, a typical plasma frequency\index{frequency!plasma} in the solar wind near the
Earth) out to some hundreds of MeV (the pion decay spectrum).\index{spectrum!pion decay@$\pi^0$-decay}
This
whole vast spectrum, in principle, could be broadly observed with
sensitive remote-sensing instruments, and in stereo.  We could thus
aspire to the observation of a many-dimensional data cube: x, y,
z, $\lambda$, polarization, and time for starters, and even the
directional components (limb-darkening function) of the emitted
radiation. Clearly the observations to date have only begun to
scratch the surface of this potential wealth of material. In each
parameter there is an implied sampling capability -- temporal
cadence, signal-to-noise ratio, contrast, and scattered light
considerations provide much further diversity.  The limits on the
available capabilities of course relate to somewhat intangible
matters such as community preferences, technical feasibility, and
cost. We note that solar observers tend (for several reasons) to
want to study the most  powerful events, whereas non-solar astronomers
may strive for sensitivity instead. Unfortunately, this leaves a
huge range of solar parameter space unobserved by the best
instrumentation. In the following we briefly review the capabilities
of the available multiwavelength observations existing at present,
broken down by wavelength.

\subsection{Radio}
\index{observing facilities!radio}

The radio observations broadly reveal both non-thermal and thermal
emissions via several mechanisms, and from a broad range of phenomena
occurring anywhere above the photosphere
\citep[e.g.,][]{1998ARA&A..36..131B}.  For flares the most important
of these are the cm-wave gyrosynchrotron radiation, for high-energy
electrons (the extension, to a few MeV, of the energetically important
electrons responsible for the impulsive-phase HXR emissions), and
the meter-wave plasma-frequency emissions that show many dynamical
processes in the corona.\index{frequency!plasma}

The past two decades have seen the improvement and deployment of
several instruments.  
\index{observatories!Nobeyama}
\index{observatories!Green Bank}
\index{observatories!Nan{\c c}ay}
\index{observatories!Owens Valley Solar Array (OVSA)}
The major dedicated cm/mm-wave observatory
is the radioheliograph at Nobeyama, Japan
\citep[e.g.,][]{1994PASJ...46L..33K}
which images at fixed frequencies
of 17~and 34~GHz with a beam size at 17~GHz of 10$''$. This is
accompanied by a non-imaging radio polarimeter operating at nine
fixed frequencies between 1~and 80~GHz. 
The Owens Valley Solar Array (OVSA) \citep{1984SoPh...94..413H,garyhurford99} has a smaller
number of dishes but superior frequency coverage, operating at 86
fixed frequencies from 1~to 18~GHz.  This allows the use of ``frequency
synthesis'' (i.e., the interpretation of the measured
(u,v)-plane\footnote{The radio astronomers'  ``(u,v)-plane'' is the
map of observed spatial frequencies; Fourier transformation of
measurements at the observed points in the (u,v)-plane yields the
source image projected on the plane of the sky.}  in terms of a
model source spectrum) to augment the coverage in the (u,v)-plane to a
certain extent.
\index{u,v@(u,v)-plane}\index{frequency synthesis}

\index{observatories!FASR}
\index{observatories!Phoenix-2}
\index{observatories!Tremsdorf}
\index{observatories!Hiraiso}
\index{observatories!LOFAR}
\index{observatories!El Leoncito}
At decimeter wavelengths the most productive facility is the
Nan\c{c}ay radioheliograph \citep[e.g.,][]{1989SoPh..120..193A},
typically operating at five fixed frequencies in the 150-450~MHz
range. Joint \textit{RHESSI}--radio observations have also been made with a
number of dynamic spectrographs spanning a wide range of frequencies,
such as Ondrejov at 0.1-4.2~GHz  \citep{1993SoPh..147..203J},
Huairou at 2.6-3.8~GHz \citep{2004SoPh..222..167F}, the Phoenix-2 instrument at 0.1-4~GHz
\citep{1999SoPh..187..335M}, Tremsdorf at 40-800~MHz \citep{Mannetal92},
the Greenbank Solar Radio Burst Spectrometer 
at 10-110~MHz \citep{2005ASPC..345..176W}
and Hiraiso at 25-2500~MHz \citep{1994CRLRv..40...85K}. 
There are plans for an ambitious new solar radio interferometer, the
Frequency Agile Solar Radio Telescope
\citep[FASR;][]{2003AdSpR..32.2705B} to cover the range 0.1 to
24~GHz, while low-frequency solar imaging and spectroscopy is planned
with the Low Frequency Array for Radio Astronomy
\citep[LOFAR,][]{2004P&SS...52.1381B}; see \cite{2003A&A...409..309C} and \cite{2005SoPh..226..121B}.
At submillimeter wavelengths a dedicated site now exists at El
Leoncito, Argentina, where the Solar Submillimeter Telescope observes
at frequencies of a few hundred GHz \citep{2003LNP...612..294K}.\index{submillimeter emission!observatories}

\subsection{Infrared}
\index{observing facilities!infrared}

The solar infrared spectrum extends from visible wavelengths out
to the 10~$\mu$m~band accessible to ground-based observations,
the mid-infrared accessible only from space, and the submm-THz range
again accessible from the ground.\index{spectrum!infrared}
This huge region contains thermal
free-free continuum and coronal line emission, plus other possible
contributors at the longest wavelengths.  
The IR has some advantages, for example, in terms of improved seeing and in magnetic sensitivity
\citep[e.g.,][]{1998BASI...26..253P}; on the other hand the diffraction
limit becomes severe for non-interferometric imagers.

No dedicated solar infrared observatories at wavelengths much
longward of 1$\mu$m exist at present although developments are
underway in imaging at 10~$\mu$m \citep[e.g.,][]{1975SoPh...45...69H},
with flares already having been detected \citep{2006PASP..118.1558M}.
Specific observations on the 1~m McMath Solar Telescope make use
of the mid-infrared to study photospheric magnetic fields
\citep{2007SoPh..241..213M}, and at the 0.76~m Dunn Solar Telescope
to probe flare activity at the ``opacity minimum" spectral region
\index{opacity minimum}
around 1.5~$\mu$m \citep{2004ApJ...607L.131X}.  
\index{magnetic field!coronal} 
Direct observations of the coronal magnetic field
are also becoming possible
 \citep{2004ApJ...613L.177L,2008SoPh..247..411T}.
\index{observatories!SOLIS}
\index{observatories!Purple Mountain}
\index{observatories!Purple Mountain!Multichannel Infrared Solar 
Spectrograph (MISS)}
Full disk observations are made daily in the Ca~{\sc ii} 8542~\AA~and  
He~{\sc i} 10830~\AA~lines by the SOLIS synoptic telescope (and by earlier
Kitt Peak telescopes), and flare spectroscopic observations at these
wavelengths are provided by the Multichannel Infrared Solar
Spectrograph \citep[MISS;][]{2002ChPhL..19..742L} at Purple Mountain
Solar Observatory (PMO), with a spectral resolution on the order of 1~\AA.

\subsection{Optical}\label{sec:opt}

\noindent{\bf{General:}}
\index{observing facilities!optical}
High-resolution observations have become much more commonplace from
ground-based observatories.
This results both from the availability of large apertures but especially
from active control of the optics to ameliorate the effects of
atmospheric seeing; in addition there are advanced post-processing techniques such
as speckle reconstruction  \citep{1993A&A...268..374V}.  
Specific telescopes include the Dunn
Solar Telescope, the Swedish Solar Observatory on La Palma, the
German Vacuum Tower Telescope on Tenerife, and the Dutch Open
Telescope.
\index{observatories!Dunn Solar Telescope}  
\index{observatories!Swedish Solar Observatory}  
\index{observatories!German Vacuum Tower Telescope}  
\index{observatories!Dutch Open Telescope}  
The best of these observations have substantially better
spatial resolution than contemporaneous telescopes in space, but
without the consistently perfect seeing conditions.

Several flexible optical/IR instruments operate in user mode at
ground-based observatories \citep[e.g.,][]{2008A&A...480..515C}.
With such instruments one can generate multidimensional data cubes
incorporating spatial, temporal, spectral, and polarimetric
measurements.  Being ground-based observatories, they may operate
with state-of-the-art adaptive optics and, having large apertures,
obtain high resolution while running at high cadence.  They also
have minimal constraints on data bandwidth and can flexibly study
a variety of spectral features.  
Future instrumentation is emphasizing
larger apertures, higher frame rates \citep{2007A&A...473..943J},
and the implementation of high-order adaptive optics
\citep[e.g.,][]{2003SPIE.4839..635R,2007PASP..119..170D}. Major new
facilities under study include the Advanced Technology Solar Telescope
(ATST) and the European Solar Telescope (EST).
\index{observatories!Advanced Technology Solar Telescope (ATST)}  
\index{observatories!European Solar Telescope (EST)}  

Optical observations from space have included the  white-light
telescope on \textit{Yohkoh}, 
\index{satellites!Yohkoh@\textit{Yohkoh}!white-light observations}
which despite being short-lived, observed
several solar flares in the Fraunhofer G~band\index{Fraunhofer lines!G band} \citep[][]{2003A&A...409.1107M}.
\index{satellites!SOHO@\textit{SOHO}!MDI}
In the \textit{RHESSI} era\index{eras!RHESSI@\textit{RHESSI}}, the Michelson Doppler Imager instrument
on the \textit{Solar and Heliospheric Observatory (SOHO)}\index{satellites!SOHO@!\textit{SOHO}!MDI} 
\citep[MDI;][]{1995SoPh..162..129S} provides white-light
full disk images, and the \textit{TRACE} satellite also has a ``white-light"
channel with a broad response ($\sim$1000-8000~\AA) which was
successfully used in the study of solar flares at high cadence
\citep[][]{2006SoPh..234...79H}.  
\index{satellites!Hinode@\textit{Hinode}}
\index{satellites!RHESSI@\textit{RHESSI}} 
\index{satellites!TRACE@\textit{TRACE}}
\index{satellites!Hinode@\textit{Hinode}!SOT}
However, the first full large-aperture
optical solar observatory in space, on board \textit{Hinode}
\citep{2002AdSpR..29.2009S}, was only launched in 2006.  This
telescope provides diffraction-limited observations with a 50~cm
primary and includes both narrow-band filter imaging and spectrographic
imaging, the latter permitting vector magnetograms
\citep[e.g.,][]{2001ASPC..236...33L}.

\index{observing facilities!H$\alpha$}
\index{observatories!Kanzelh\"ohe}
\index{observatories!Kharkov}
\index{observatories!Holloman}
\index{observatories!Learmonth}
\index{observatories!San Vito}
\index{observatories!Big Bear}
\index{observatories!Catania}
\index{observatories!Hida}
\index{observatories!Purple Mountain}
\index{observatories!Pic du Midi}
\index{observatories!Yunnan}
\index{observatories!Themis}
\bigskip
\noindent{\bf{H$\alpha$:}}
The 3$\rightarrow$2 Balmer transition line of neutral hydrogen,
centered at 6563~\AA, is historically important in flare observations
and continues to be observed and studied for its rich diagnostic
capability.  Many tens of observatories worldwide are engaged in
monitoring the Sun in H$\alpha$~for flares, making both full- and
partial-disk observations.  
The US National Oceanic and Atmospheric
Administration (NOAA) maintains a database of H$\alpha$~flares stretching
back decades.
Currently five stations provide H$\alpha$~data for this:
Kanzelh\"ohe (Austria), Learmonth (Australia), Holloman Air Force
Base (USA), San Vito (Italy) and Kharkov (Ukraine).  The major
H$\alpha$~instruments with facility for high cadence and high resolution
observations, which have been used in connection with \textit{RHESSI}
observations, are at Kanzelh\"ohe \citep{2003HvaOB..27..189O}, the
65~cm telescope at Big Bear Solar Observatory (removed from service
in 2006, and being replaced with a 1.6~m instrument), and the
H$\alpha$~Fine Structure Telescope at Purple Mountain Observatory.  The
Global High-Resolution H-alpha Network consists of the stations at
Big Bear, Kanzelh{\" o}he, Catania, Meudon, Pic du Midi, Huairou,
Yunnan, and Mauna Loa.  Spectroheliographs making H$\alpha$~observations
(other than daily images) include those on the THEMIS telescope
\citep{2002A&A...381..271M} and at Hida Observatory.

\bigskip
\noindent{\bf{Magnetographic:}}
\index{observing facilities!magnetographs}
It is difficult to measure the magnetic field in the solar corona
directly.  Photospheric glare makes observations against the disk
almost inconceivable except perhaps at radio wavelengths; above the
limb one has the high temperatures of the corona and line-of-sight
confusion to contend with.  The photospheric magnetic fields are
typically observed in a weak Fraunhofer line, Ni~{\sc i}~6758~\AA~for
MDI\index{Fraunhofer lines!Ni~{\sc i}~6758~\AA}.

Nevertheless, precise knowledge of the vector magnetic field in the
corona is fundamental to our understanding of flares and CMEs, and
much effort is expended in  extrapolating near-photospheric
measurements  into the corona.  To understand the coronal field,
the measurements would ideally be done above the solar chromosphere
where the magnetic field approaches the force-free condition and
can therefore be used as a reliable lower boundary for mathematical
extrapolations into the corona, if it remains static.  
Special observational
facilities exist for the observation of the solar magnetic field
in the photosphere via spectropolarimetric techniques, with perhaps
the earliest serious work done at the Crimean Observatory
\citep{1964SSRv....3..451S}.  These facilities have different
objectives, ranging from characterization of ``sun-as-a-star''
average line-of-sight field components
\citep[e.g.,][]{1983JGR....88.8095W}, to full vector measurements
at the highest possible resolution and at different heights in the
solar atmosphere.  The latter data provide the indispensable boundary
condition for the increasingly ambitious efforts to extrapolate the
structure of the coronal field in three dimensions
\citep[e.g.,][]{2006SoPh..235..161S}.  
\index{observatories!MSFC}
Ground-based magnetographs
include the early MSFC Vector Magnetograph \citep{1982SoPh...80...33H},
the Mees Imaging Vector Magnetograph \citep{1996SoPh..168..229M},
Huairou \citep{1986PBeiO...8....1A}, the 
\index{observatories!Huairou}
\index{observatories!Hida}
Hida Solar Magnetic Activity
Research Telescope \citep[SMART;][]{uenoetal04}, the BBSO Digital
Vector Magnetograph \citep{2001ASPC..236...65S},  the GONG Network,
including magnetographs located in 6~sites around the world
\citep{1996Sci...272.1284H}, and SOLIS \citep{2003SPIE.4853..194K}.
The \textit{Hinode} spectropolarimeter, which uses the  Fe~{\sc i}~lines
at 6302.1 and 6302.5~\AA, is the only high-resolution vector
magnetograph in space, while the HMI instrument on the \textit{Solar Dynamics Observatory} now provides high-cadence, full-disk photospheric vector magnetograms at a 1$''$
spatial resolution \citep{2003ASPC..307..131G}.
\index{observatories!SOLIS}
\index{observatories!GONG}
\index{satellites!Hinode@\textit{Hinode}}
\index{satellites!Hinode@\textit{Hinode}!SP}
\index{satellites!SDO@\textit{SDO}}

The coronal magnetic field, in spite of frequent small-scale activity,
normally is considered to be force-free.  Thus it contains currents,
so that a simple extrapolation of any photospheric magnetic field,
the (non-force-free) source of the stresses that drive these currents,
may incorporate systematic errors.  Accordingly several observatories
have begun systematic observations of the chromospheric magnetic
field in Na~{\sc i}~5896~\AA~\citep{1995ApJ...439..474M} or another
chromospheric line such as H$\beta$\index{Fraunhofer lines!H$\beta$}.  
The upper chromosphere should
be more force-free and these observations should enable more accurate
extrapolations.

\subsection{UV--EUV--X-ray wavelengths}
\label{sec:xeuv}

From the UV into the SXR domain (up to a few keV)\footnote{Roughly,
the UV and vacuum UV extend through the Lyman continuum; at shorter
wavelengths to about 44~\AA, the carbon K-edge, it would be EUV,
and shortward still soft X-rays.} one can use specialized astronomical
techniques involving normal-incidence or, at the shortest wavelengths,
grazing-incidence mirrors. Such observations need to be made in
space because of the large opacity of the Earth's atmosphere. 
Indeed, solar observations at these previously inaccessible wavelengths
were a part of the early history of space astronomy, leading to \textit{Skylab}.
\index{Skylab@\textit{Skylab}}
The following paragraphs summarize some of the observations concurrent with \textit{RHESSI}.

\bigskip
\noindent{\bf{Ultraviolet:}}
\index{observing facilities!ultraviolet}
For spectroscopy, SUMER\footnote{The Solar UV Measurements of Emitted Radiation instrument on \textit{SOHO}.}
\citep{1995SoPh..162..189W} has provided excellent 
data but with limited solar flare coverage for technical reasons.
\index{SUMER}
\textit{TRACE}\index{satellites!TRACE@\textit{TRACE}} has provided excellent imaging observations but with poor
spectral resolution \citep{1999SoPh..187..229H}. 
The \textit{TRACE} UV~bands
includes a narrow ($\sim$30~\AA) band centered on strong lines of
C~{\sc iv}, plus broader UV~channels covering Lyman-$\alpha$ and the
UV continuum to around 2500~\AA, and  a ``white light'' channel
covering the full range of spectral response of the Lumogen-coated
charge-coupled device 
(CCD)\footnote{Lumogen is a proprietary coating that extends the UV
response of the CCD.}.  
\index{Lumogen}
\index{CCD}
Nevertheless this rich spectral domain, the
most important for studying the chromosphere and transition region, still remains
underexploited; for example, there are no comprehensive observations
yet even of Lyman-$\alpha$, the strongest line.\index{transition region}

\bigskip
\noindent{\bf{EUV:}}
\index{observing facilities!EUV}
Again \textit{TRACE} has provided excellent EUV observations at the standard
normal-incidence narrow spectral bands centered at 171, 195, and
284~\AA, corresponding to transitions of Fe~{\sc ix/x},  Fe~{\sc xii}
(and Fe~{\sc xxiv} in flares), and Fe~{\sc xv}.  
Such observations, pioneered by the EIT\footnote{Extreme-EUV Imaging Telescope on \textit{SOHO}.} instrument \citep{1995SoPh..162..291D}, now continue stereoscopically \citep{2007SSRv..tmp..198K} 
on the \textit{STEREO} spacecraft \citep{2008SSRv..136...67H}.  
\index{satellites!STEREO@\textit{STEREO}}
\index{satellites!SOHO@\textit{SOHO}}
In addition to these broad-band imaging instruments, stigmatic slit spectrographs have flown on \textit{SOHO} \citep[310-380~\AA~and 520-630~\AA;][]{1995SoPh..162..233H},
and now on \textit{Hinode} \citep[170-210~\AA~and 250-290~\AA;][]{2007SoPh..243...19C}. 
These provide high resolution spectroscopy enabling detailed diagnostics for plasma density,
temperature, velocity and abundance.
\index{satellites!Hinode@\textit{Hinode}}
Currently the AIA\footnote{Atmospheric Imaging Assembly.} instrument \citep{2004SPIE.5171...53R} on \textit{SDO} provides comprehensive coverage of this type, with 10~s cadence, many passbands, and whole-Sun imaging with negligible gaps in time coverage.
\index{satellites!SDO@\textit{SDO}}
\index{satellites!SDO@\textit{SDO}!AIA}

\bigskip
\noindent{\bf{Soft X-rays:}}
\index{observing facilities!soft X-rays}
The SXR flux from a solar flare has become the definitive flare
observable via the \textit{GOES} flare classifications, and photometric SXR
observations have continued in an unbroken photometric sequence
since the 1970s. High-resolution imaging began with grazing-incidence
telescopes on rockets and then (with film readout) on \textit{Skylab}.
In 1991 \textit{Yohkoh} was launched with its SXT imager
\citep{1991SoPh..136...37T} which could image at spatial resolutions
down to 5$''$, and now \textit{Hinode} carries the improved XRT
instrument \citep{2007SoPh..243...63G}. 
\index{satellites!Yohkoh@\textit{Yohkoh}!SXT}
\index{satellites!Hinode@\textit{Hinode}!XRT}
The temperature coverage
of XRT is around 1-30~MK, and multiple filter combinations permit
a  temperature discrimination\index{soft X-rays!temperature coverage} 
as good as $\log{\rm T_e}$~=~0.2.

In terms of energy, the end of the soft X-ray range and the beginning
of the hard X-ray range is a somewhat arbitrary matter, which varies
from author to author in solar physics, and also varies from field
to field in astrophysics\index{soft X-rays!definition}. 
Roughly speaking, energies of 0.1-10~keV are counted as ``soft'' 
X-rays in solar physics, and energies above about 20~keV\index{hard X-rays!definition}  
are ``hard.''
Between these two, the description
``soft'' or ``hard''  can depend on whether the spectrum looks thermal
or non-thermal. 

\bigskip
\noindent{\bf{Spectroscopy:}}
\index{observing facilities!soft X-rays}
High-resolution spectroscopy (from infrared to the $\gamma$-rays)
provides the closest approach to learning the state of the flaring
plasma via remote sensing.  Generally these techniques provide
information about the electron energy distribution function, but
extracting information about the electron angular distribution is
extremely difficult, even adopting the common assumption of azimuthal
isotropy with respect to the magnetic field.  The SXR emission lines
have been particularly productive in flare research, with instruments
flown on many spacecraft -- most recently the \textit{Yohkoh} Bent
Crystal Spectrometer \citep{1992PASJ...44L..55L} and its follow-on
RESIK\footnote{\textit{REntgenovsky Spektrometr s Izognutymi Kristalami}.} 
\citep{2005SoPh..226...45S}.  
\index{satellites!Yohkoh@\textit{Yohkoh}!BCS}
\index{satellites!CORONAS-F@\textit{CORONAS-F}!RESIK}
\index{soft X-rays!emission lines}
The latter observes in narrow
bands around the principal X-ray emission lines of Si, S, Cl, Ar,
and K in the 3.1-6.6~\AA~range, with access to other features at
higher grating orders.

At longer wavelengths the definitive observations have come from
stigmatic slit spectrographs, which image in one dimension and thus
multiplex the second spatial dimension in time.  The definitive
instruments in this category are not particularly optimized for
flare observations but have produced many interesting results.
These are currently SUMER 
\citep{1995SoPh..162..189W}, which covers the VUV~range 500-1610~\AA, and EIS
\citep{2007SoPh..243...19C} which covers the XUV~range
170-290~\AA~(in two bands).  
\index{XUV spectral range}
This latter range conveniently overlaps
the standard spectral selections for normal-incidence imagers such
as those on \textit{SOHO} and \textit{TRACE} (see above)\index{satellites!TRACE@\textit{TRACE}}.

The \textit{Solar Dynamics Observatory} now also provides EUV spectral irradiance measurements for the Sun as a star via its EVE\footnote{EUV Variability Experiment.} instrument \citep{2010SoPh..tmp....3W}.
\index{satellites!SDO@\textit{SDO}!EVE}

\bigskip
\noindent{\bf{Hard X-rays and $\gamma$-rays:}} \index{observing
facilities!hard X-rays}\index{observing facilities!$\gamma$-rays}
At energies above a few keV, focusing optics have until recently
been prohibitively difficult.  
Accordingly simple counter spectrometers provided the main source of information.
Such HXR time series provide important information about the flare impulsive phase (and other
epochs of particle acceleration) even without imaging, and many instruments have contributed to this.
At present \textit{RHESSI} and \textit{Fermi}, through its GBM\footnote{Gamma-ray Burst Monitor.} instrument \citep{2010AAS...21640406S} provide this information routinely.
New spectroscopic data extending into the $\gamma$-ray range are available
from other instruments, including \textit{INTEGRAL}
\citep[e.g.,][]{2004ESASP.552..669G}.
\index{satellites!INTEGRAL@\textit{INTEGRAL}}

Over the years 1991-2001 \textit{Yohkoh} provided systematic HXR imaging for
the first time, using non-focusing optics. 
\index{satellites!Yohkoh@\textit{Yohkoh}!HXT}
\index{hard X-rays!non-focusing optics} 
The HXT instrument
\citep{1991SoPh..136...17K} thus anticipated some of \textit{RHESSI}'s results
in HXRs, providing four energy channels over $\sim$15--92~keV.  
True imaging spectroscopy for hard 
X-rays\index{hard X-rays!imaging spectroscopy}\index{imaging spectroscopy} 
and $\gamma$-rays began in 2002 with \textit{RHESSI} \citep{2002SoPh..210....3L}.  
The single \textit{RHESSI}\index{RHESSI@\textit{RHESSI}!configuration} 
instrument consists of nine high-purity germanium detectors, segmented
into front and rear segments for sensitive and simultaneous hard
X-ray and $\gamma$-ray measurements.  These detectors have resolutions
as good as about 1~keV (FWHM) over the range 3~keV to 17~MeV
\citep{2002SoPh..210...33S}.  Each of the detectors has a bigrid
modulation collimator\index{rotating modulation collimator} with different parameters (thickness and
angular resolution) \citep{2002SoPh..210...61H}, giving imaging (at
a basic 4-s time resolution governed by spacecraft rotation) down
to the arcsec range.

\subsection{Particles and fields}\index{observing facilities!solar energetic particles}
\index{observing facilities!particles and fields}

\bigskip
\noindent{\bf{Solar energetic particles:}} The Sun emits charged
particles, neutrons, and energetic neutral atoms as a part of
flare/CME physical processes, and these (as well as the solar wind)
provide independent samples of material or accelerated particles
with which to compare the remote-sensing observations. Ideally one
would be able to compare the separate populations of energetic
electrons and ions detected {\sl in situ} with  those inferred from
the \textit{RHESSI} HXRs and $\gamma$-ray observations.\index{in situ@\textit{in situ}!comparisons with remote sensing} 
The ``multiwavelength''
flare observations in a sense include the direct detection of
high-energy particles emitted both promptly and with delays by solar
flares and CMEs.  
These SEPs (Solar Energetic Particles) have a
rich history of observation from space, and have also been termed
``solar cosmic rays.'' 
\index{neutrons}\index{cosmic rays!solar}\index{acceleration!SEPs}\index{solar energetic particles (SEPs)} 
Many
spacecraft have observed SEPs in the past, and the current flotilla
includes \textit{ACE} (\textit{Advanced Composition Explorer}), \textit{SOHO}, the two \textit{STEREO} spacecraft, \textit{Ulysses}, \textit{WIND}, and other missions both in deep space and in near-Earth space\index{ground-level events}\index{neutron monitors}\index{neutron telescopes}.
The relativistic ``ground level event'' solar particles can also be
detected by neutron monitors at the Earth \citep[e.g.,][]{2006ApJ...639.1206C}, 
either as neutron secondaries or directly in a few cases\index{satellites!SOHO@\textit{SOHO}}\index{satellites!STEREO@\textit{STEREO}}\index{satellites!ACE@\textit{ACE}}\index{satellites!Ulysses@\textit{Ulysses}}\index{satellites!WIND@\textit{WIND}}.

\bigskip
\noindent{\bf{Neutrons and ENAs:}}
\index{observing facilities!neutrons}
Solar neutrons, produced mainly in (p, p), (p, $\alpha$), and
($\alpha$, p) reactions \citep{1987SoPh..107..351H}, can propagate
into interplanetary space and even arrive at the Earth's surface
if at high enough energies.  
\index{neutrons!half-life}
The natural decay time of a free neutron
is about 866~s.\footnote{A free neutron decays, with
a half-life of about 15 minutes, into a proton, and electron, and
an antineutrino.}  
An array of neutron monitors and neutron telescopes
\citep[e.g.,][]{2007APh....28..119M} can detect both direct neutrons
and secondary neutrons induced by charged particles creating cascade
showers.  The largest advances in our understanding of solar neutrons
will come from instruments flown within a few tenths of an AU of
the Sun, within the decay-time horizon for lower-energy neutrons that do not
survive to one AU.

Neutron-decay products (energetic protons, with a characteristic
spectrum) can also be detected in interplanetary space
\citep{1983ApJ...274..875E,1990ApJS...73..273E}, and in principle
fast electrons as well \citep{1996ApJ...464L..87D}.\index{spectrum!neutron-decay products}  
A further
channel for flare study has recently surfaced: one event (SOL2006-12-05T10:35, X9) has been found to have emitted detectable levels of
energetic neutral hydrogen atoms \citep{2009ApJ...693L..11M} in the
1.8-5~MeV range.  
\index{flare (individual)!SOL2006-12-05T10:35 (X9)!energetic neutral atoms}

\bigskip
\noindent{\bf{Interplanetary Coronal Mass Ejections:}} Finally, the
low-energy ejecta of an ICME (Interplanetary Coronal Mass Ejection)
may contain a large total energy and accelerate particles copiously
(see Section~\ref{sec:energetics}), as well as provide information
via their magnetic fields
\citep[][]{1982GeoRL...9.1317B,1998AnGeo..16....1B}.
\index{interplanetary coronal mass ejections (ICMEs)}

\subsection{Summary}

From the foregoing one can recognize that our observational power
has increased considerably during solar cycles 22 and 23, and in
fact provides some information in almost all wavelength bands.  We
list some of the dedicated whole-Sun instruments in
Table~\ref{tab:fletcher_instruments} for reference.  In spite of
all of the data available, it would be quite misleading to imagine
that any of the material is definitive, and we return to discuss
the missing things in Section~\ref{sec:missing}.  
The lack of imaging\index{imaging spectroscopy}
spectroscopy in the optical (other than a few prominent lines) and
the UV is especially painful, and the THz and $\gamma$-ray regimes
are very under-exploited. In general one can recognize that non-solar
observational facilities (e.g., NASA's Great Observatories such as
\textit{Chandra}) have highly desirable characteristics (resolution,
sensitivity, background levels) that are not at present achieved
by dedicated solar instruments.  In addition, the exciting era of
stereoscopic observations has now begun, but with only a limited
instrument complement.  
Clearly stereoscopic observations\index{stereoscopic observations}, 
including an out-of-the ecliptic capability,  would be of great value at all wavelengths;
for example, at HXR or radio bands such a capability would
make it possible to study the anisotropic radiation patterns of
flare emissions, as well as the three-dimensional structures of
their sources.

\begin{landscape}
\begin{table}
\caption{\bf Full-disk solar data sources$^a$}
\begin{tabular}{l l l l}
\hline
Observatory & Observation & Dates & URL \\
\hline
{\bf I. Radio} \\
\hline
Nan{\c c}ay & Decimetric imaging & 1997-present & \url{http://bass2000.bagn.obs-mip.fr/Tarbes/} \\
Nobeyama & 17 \& 35 GHz imaging & 1992-present & \url{http://solar.nro.nao.landac.jp/norh/} \\
SSRT (Irkutsk) & 5.73~GHz imaging & 2000-present & \url{http://ssrt.iszf.irk.ru/} \\
\hline
{\bf II. Infrared} \\
\hline
Kitt Peak & 8542~\AA, 10830~\AA~imaging& 2007-present& \url{http://solis.nso.edu/}\\
\hline
{\bf III. Optical} \\
\hline
GONG & Quasi-continuum images &1995-present & \url{http://gong.nso.edu/} \\
GONG & LOS magnetic field  &1995-present& \url{http://gong.nso.edu/} \\
Global H$\alpha$ & H$\alpha$~Imaging  & 2000-present & \url{http://swrl.njit.edu/ghn_web/} \\
\textit{SDO} & HMI & 2010-present & \url{http://jsoc.stanford.edu/} \\
\textit{SOHO}/MDI & Quasi-continuum images & 1995-2011 & \url{http://soi.stanford.edu/} \\
\textit{SOHO}/MDI & LOS magnetic field & 1995-2011 & \url{http://soi.stanford.edu/} \\
SOLIS & Vector magnetic field & 2008-present &  \url{http://solis.nso.edu/}\\
\hline
{\bf IV. UV/EUV} \\
\hline
\textit{SOHO}/EIT & 171, 195, 284, 304~\AA~imaging& 1995-2011 & \url{http://umbra.nascom.nasa.gov/eit/} \\
\textit{STEREO}/EUVI &  171, 195, 284, 304~\AA~imaging& 2006-present &\url{http://secchi.lmsal.com/EUVI/} \\
\textit{SDO} & AIA & 2010-present & \url{http://jsoc.stanford.edu/} \\
\hline
{\bf V. Soft X-ray} \\ 
\hline
\textit{Yohkoh}/SXT &$\sim$ 1~keV imaging & 1991-2001 & \url{http://solar.physics.montana.edu/ylegacy/} \\
\textit{GOES}-12/SXI & $\sim$ 1~keV imaging& 2001-2007 &  \url{http://www.swpc.noaa.gov/sxi/index.html} \\
\textit{GOES}-13/SXI & $\sim$ 1~keV imaging& 2006 &  \url{http://www.swpc.noaa.gov/sxi/index.html}\\
\textit{Hinode}/XRT & $\sim$ 1~keV imaging& 2006-present & \url{http://sdc.uio.no/sdc/welcome} \\
\textit{GOES}-14/SXI & $\sim$ 1~keV imaging& 2009-present & \url{http://www.swpc.noaa.gov/sxi/index.html} \\
\hline
{\bf VI. Hard X-ray/$\gamma$-ray} \\ 
\hline
\textit{Yohkoh}/HXT &$\sim$ 13-91~keV maging & 1991-2001 & \url{http://solar.physics.montana.edu/ylegacy/} \\
\textit{RHESSI} & 3~keV - 17~MeV imaging spectroscopy & 2002-present & \url{http://hesperia.gsfc.nasa.gov/hessi/} \\
\hline
\end{tabular}
\label{tab:fletcher_instruments}

\smallskip
$^a$ This is only a selection, emphasizing stable full-disk observations; please note that there
are many more resources with limited fields of view and/or coverage.
\end{table}
\end{landscape}

\section{Footpoints and ribbons}\label{sec:fletcher_footpoints}
\index{footpoints}
\index{ribbons!UV}
\index{footpoints!white light}
\index{ribbons!EUV}

\subsection{Overview}\label{sec:fpoverview}

Traditionally we call the HXR brightenings, often observed as paired
compact sources, \textit{footpoints}, and the elongated structures
originally found in H$\alpha$~images \textit{ribbons}.\index{ribbons!H$\alpha$}  
In fact both features can be observed at many wavelengths; the ribbons are
prominent in the UV and EUV (observed, for example, by 
\textit{TRACE}\index{satellites!TRACE@\textit{TRACE}}
\citep{2001ApJ...560L..87W,2006ApJ...641.1197S}).
The footpoint sources tend to be  distinguished by their strong HXR and/or white light
emission.  
\index{flare types!white light}
Either way, these features occur pairwise separated by
a magnetic polarity inversion line in the photosphere, which separates
the two polarities of the vertical magnetic field.  
Regions connected
by coronal magnetic fields are called \textit{conjugate points}.
Figures~\ref{fig:timeprofile_overview},~\ref{fig:v1}, and~\ref{fig:RH_fps} show the
context.  
\index{conjugacy} \index{magnetic structures!neutral line}
\index{hard X-rays!conjugacy}
\index{hard X-rays!footpoints}

The principal physical distinction between the footpoint and ribbon\index{ribbons!physics}
morphologies may reflect the idea that the footpoints are the direct
result of non-thermal processes (traditionally interpreted in terms
of the thick-target model, in which electrons lose their energy
collisionally in the chromosphere), whereas the ribbons also show
excitation which could be due to thermal conduction from the overlying
coronal arcade,  or from weak particle precipitation.\index{arcade}\index{thick-target model}
\index{chromospheric evaporation!conduction-driven}
Careful observation and modeling of individual flare events
demonstrates that the location and evolution of the ribbons is
clearly related to the magnetic topology in which the event occurs.\index{ribbons!topology}
In particular, flare ribbons reflect the projection of the separatrix
or quasi-separatrix structure where flux transfer between magnetic
domains is occurring
\citep[e.g.,][]{1991A&A...250..541M,1997A&A...325..305D} (see
Section~\ref{sec:fletcher_fpmagnetic} for illustrations and further
discussion).\index{magnetic structures!separatrices}\index{magnetic structures!quasi-separatrix layers}\index{quasi-separatrix structure}\index{separatrix} 
In H$\alpha$~and the
UV/EUV, the flare ribbons have a tendency to spread systematically
outwards from the magnetic polarity inversion line, with their
appearance becoming more ordered as the flare progresses.\index{ribbons!motions}

\begin{figure} \begin{center}
\includegraphics[width=\textwidth]{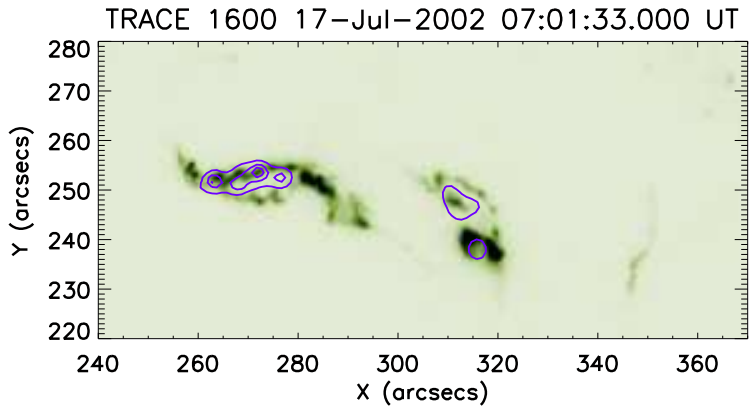} 
\end{center} 
\caption{ \textit{RHESSI} and \textit{TRACE} flare
observation of SOL2002-07-17T07:13 (M8.5), showing UV ribbons made
up of multiple small sources, and simultaneous \textit{RHESSI} 25-50~keV
footpoints superposed (blue contours, at 25\%, 50\% and 75\% of
maximum).  The \textit{RHESSI} contours were made using the Pixon algorithm
\citep{1996ApJ...466..585M}, including Grid~1 which is capable of
showing spatial detail at 2.3$''$ scales.  
The \textit{RHESSI} sources correspond well to the ribbons overall, but 
the ribbons are significantly more extended.  
\index{flare (individual)!SOL2002-07-17T07:13 (M8.5)!illustration}} 
\index{flare (individual)!SOL2002-07-17T07:13 (M8.5)!\textit{TRACE}}
\index{hard X-rays!and UV ribbons!illustration}
\index{ribbons!illustration}
\label{fig:RH_fps}
\end{figure}

On the other hand, the HXR footpoint sources frequently move along
the H$\alpha$~or UV/EUV flare ribbons, rather than away from the polarity
inversion line, and are found at locations distinguished by their
high magnetic field strengths 
\citep[][]{2007ApJ...654..665T} or
their high magnetic flux transfer rates
\index{flux transfer}\index{ribbons!and shear}
\citep{2008ApJ...672L..69L,2009A&A...493..241F}.  
\index{reconnection!flux transfer} 
\index{hard X-rays!footpoint sources}
HXR footpoints, generated by electron-ion bremsstrahlung
of non-thermal electrons in the chromospheric plasma, are well-correlated
in space and time with white light sources, confirming them as
locations of intense energy deposition \citep[e.g.,][]{2006SoPh..234...79H}.
\index{hard X-rays!and white light}
\index{flares!white light}
\index{flare types!white light}
Though it was for a long time believed that white-light emission
\index{big-flare syndrome}
\index{syndromes!big-flare}
is a ``big flare'' phenomenon, targeted observations at high cadence
and high spectral resolution seem to suggest that even the smallest
flares have a white-light counterpart. The generation mechanism for
the white-light emission is not known, one issue being that if the
white-light emission is a photospheric or lower chromospheric
phenomenon this places perhaps unreasonably strong demands on beam
excitation models.  Observations in the infrared at 1.56 $\mu$m,
corresponding to the wavelength at which the solar opacity is a
minimum, also suggest strongly that the deep atmosphere is involved
\citep{2004ApJ...607L.131X}. 
\index{footpoints!and \textit{TRACE} diffraction patterns}
\index{satellites!TRACE@\textit{TRACE}}
\index{hard X-rays!and EUV}
In EUV, the counterparts to the HXR
footpoints are often observed as particularly intense brightenings,
and in \textit{TRACE} observations this leads to characteristic diffraction
patterns caused by the metal grids supporting the EUV filters, such
that even if the CCD is saturated at the footpoint pixel some limits
to its intensity can be calculated
\citep{2001SoPh..198..385L,2006SoPh..239..531G,2007A&A...472..945M}.

One of the significant observational discoveries by \textit{RHESSI} has been
$\gamma$-ray footpoints in a small number of solar flares, imaged
in the 2.223~MeV neutron capture line \index{footpoints!$\gamma$-rays}
\citep[][]{2003ApJ...595L..77H,2006ApJ...644L..93H}.
\index{RHESSI@\textit{RHESSI}!gamma-ray imaging@$\gamma$-ray imaging} The 2.223~MeV line is produced
by the radiative capture on hydrogen of a thermalized neutron.  The
neutron itself results from an earlier interaction by a primary
energetic ion undergoing inelastic scattering resulting in a
neutron-emitting isotope \citep{1987SoPh..107..351H}.\index{scattering!inelastic}
It takes a substantial column depth to slow the initially fast neutrons, so
the 2.223~MeV radiation presumably forms in the dense lower atmosphere.

The \textit{RHESSI} 2.223~MeV $\gamma$-ray observations should determine the
centroid location of the final neutron captures to within about an
arcsecond \citep[][]{2003ApJ...595L..77H}.  These observations are
difficult to make, involving long integrations because of the low
intensity of the radiation, and the high energy photons are modulated
only by the thickest \textit{RHESSI} grids, providing limited spatial
information. The four flares that have been imaged in this way so
far show single or double (in one case) $\gamma$-ray image components at
roughly, but not exactly, the same location as the HXR footpoints.
In three out of the four events, there is a statistically significant
displacement between the 2.223~MeV sources and the 200-300~keV
sources. This is dealt with in \cite{Chapter4}.

Strong evidence for the impact of solar flares on the deep atmosphere
- the lower chromosphere or photosphere --  is also present in two
other signatures.
Flare-induced seismic waves\index{waves!seismic} 
ripple out across the photosphere as detected by 
helioseismic techniques \citep{1998Natur.393..317K}, and
strong changes occur in the photospheric magnetic field, most readily
visible in the line-of-sight component.
The changes are essentially simultaneous with the flare 
\citep[e.g.,][]{2005ApJ...635..647S}
and with emission at the 1.56~$\mu$m ``opacity minimum'' height
\citep{2004ApJ...607L.131X}\index{opacity minimum}\index{magnetic field!flare-related changes}.
There is no question that the effects of the flare reach
deep into the solar atmosphere\index{global waves!seismic}.

The behavior of the chromospheric footpoint  plasma during the
impulsive phase remains ill-understood theoretically.  At the same
time it is in one sense the most important flare problem, since as
we will describe in Section~\ref{sec:energetics} the white-light
and UV continuum in this phase may dominate the radiated energy of
a flare.  
The problem has been repeatedly treated in the 1-D~``radiation
hydrodynamics'' approach \citep{1975SvA....18..590K} in which gas
dynamics is coupled with some treatment of radiative transfer.
Recent developments in this area \citep{2005ApJ...630..573A} include
an elementary formulation of beam heating, along with ionization
calculations, line and continuum radiative transfer, and 1-D
hydrodynamics. However the limitations of even this advanced treatment
(one dimension; no self-consistent treatment of particle or wave
energy transport) strongly suggest a need for additional development\index{waves!energy transport}\index{flare models!radiation hydrodynamic}.
The coupling of radiation and matter is decisive for
the radiation signatures, of course, and also is necessary to
understand the upwards mass flow and the seismic signatures.
Self-consistent modeling should also include wave energy transport
\citep{1982SoPh...80...99E,2008ApJ...675.1645F}.

\subsection{Timing}

The HXR sources can fluctuate rapidly\index{hard X-rays!short time scales}, 
on time scales below 1~s
\citep[e.g.,][]{1995ApJ...453..973K}. 
Such short timescale variations could reflect transport or acceleration processes. 
For reference, a 30~keV electron at $v = 0.3 c$ would move $\sim$10$^4$~km in of
order 0.1~sec.  The timing measurements are especially interesting
in comparison with radio measurements, which show myriad fine
structures in their spectrograms\index{radio emission!spectrograms}.  
Figure~\ref{fig:aschwanden} shows
a representative example of a decimeter-wavelength dynamic spectrum
(unflitered and frequency-filtered) compared with HXRs
\citep[][]{1995ApJ...455..347A}, in which detailed correlation
between spikes is seen.  It is more difficult to interpret the
decimeter bursts because of their non-linear generation mechanisms;
the timescale for the development of the kinetic instability related
to the beam transport time  (the ``bump-on-tail'' instability) is
\index{plasma instabilities!bump-on-tail}
important. 
In some flares the
decimeter wavelengths and HXR correlations are very good, but not
in all -- possibly due to the complexities of the plasma radiation
generation and absorption/Razin suppression \citep{1998ARA&A..36..131B}.
\index{radio emission!Razin suppression}\index{Razin effect}
At centimeter wavelengths there can be a much more precise relationship.

\begin{figure}
  \begin{center}
        \includegraphics[width=0.8\textwidth]{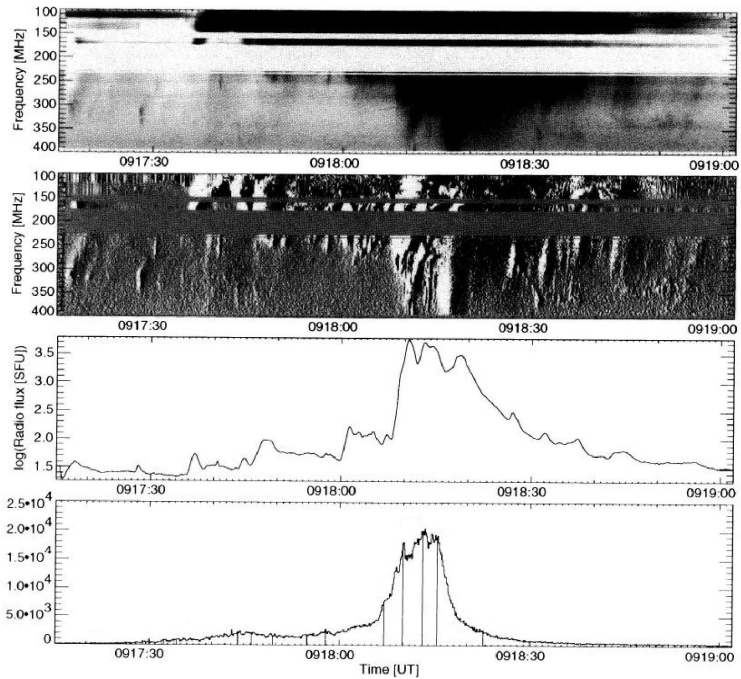}
  \end{center}
  \caption{ \label{fig:aschwanden} Comparison of decimeter-wave and HXR time variations \citep{1995ApJ...455..347A}; data from the Ikarus spectrograph \citep{1982SoPh...81..197P} and the \textit{SMM}/HXRBS instrument \citep{1980SoPh...65...25O}, respectively.
  The upper two panels show the radio spectrogram for SOL1980-03-29T09:18, with the lower
  one filtered to show temporal gradients; the bottom two panels show the radio flux integrated
  over 100-400~MHz and the HXR flux.
  }
\end{figure}
\index{flare (individual)!SOL1980-03-29T09:18 (M1)}

\cite{1996AdSpR..17...67S} have used HXR timing measurements to
establish conjugacy (see Section~\ref{sec:fpoverview}) statistically,
drawing the inference that a coronal particle acceleration region
sends bremsstrahlung-producing fast electrons simultaneously (to a
few tenths of a second) into the two footpoint regions.  At present
there is no analysis of \textit{RHESSI}  data along these lines.
\cite{1995ApJ...447..923A} have extended HXR timing analysis in an
effort to establish time-of-flight delays between HXRs at different
energies.  
\index{hard X-rays!time-of-flight analysis} 
\index{time-of-flight analysis} 
Their decomposition of
\index{Masuda flare}
\index{flare (individual)!SOL1992-01-13T17:25 (M2.0)}
the time series of the Masuda flare, to name only one example, is
consistent with signatures of loop-top injection superposed on a
slowly varying envelope; note though that the interpretation of
such a decomposition is not unique \citep{1998ApJ...509..911B}.

\subsection{Morphology of flare footpoints and ribbons}
\index{ribbons}
\index{footpoints}

There are distinctions between the impulsive (early) and gradual
(late) phase behavior of flare ribbons and footpoints.\index{ribbons!early vs. gradual phase}
In the
gradual phase of a flare, the emission tends to be in two roughly
parallel ribbons, visible particularly well in H$\alpha$~and~UV (see, for
example, Figure~\ref{fig:asai}).  These tend to separate slowly from
one another and from the magnetic polarity inversion line as the
flare proceeds, although the HXR footpoints can have more complicated
apparent motions as discussed below.  
The roughly parallel expansion
of the ribbons led to the standard magnetic reconnection model.
\index{flare models!CSHKP}\index{CSHKP}
\index{reconnection!standard flare model} 
In the gradual
phase of the flare, footpoint emission is thought to be powered
primarily by electron beams, but may instead result from thermal
conduction from the overlying hot loops \citep[see the analysis
of][]{2001ApJ...552..849C}.\index{beams!vs. conduction fronts}
The flare ribbons in H$\alpha$~can be some
tens of arcseconds wide, particularly in the gradual phase, and
show internal structure (also visible in some cases in UV).  
The gradual phase ribbon morphology in EUV appears similar to a
patterning associated with hot, high pressure coronal loops
(in this case flare loops).  
In such structures conduction from the loops causes plasmas at transition-region
or coronal temperature to appear at chromospheric heights.  
\index{moss}
\index{soft X-rays!moss}
\index{magnetic structures!moss}
The resulting pattern of spatially-intermittent and dynamic hot
plasma and spicules is known as ``moss'' \citep{1999ApJ...519L..97B}.
The leading edges of the ribbons, illuminated in EUV, are considerably
narrower \citep[e.g.,][]{2003ApJ...586..624A}, and are also where
the brightest H$\alpha$~emission \citep{1982SoPh...75..305S} and the HXR
footpoint sources are found in the impulsive phase.  Hard X-ray
bremsstrahlung is not normally a feature of the ribbon-associated
gradual phase, athough late-phase non-thermal emission with different
morphologies is now known to be common \citep{2008A&ARv..16..155K}.

\begin{figure}
  \begin{center}
	\includegraphics[width=0.4\textwidth]{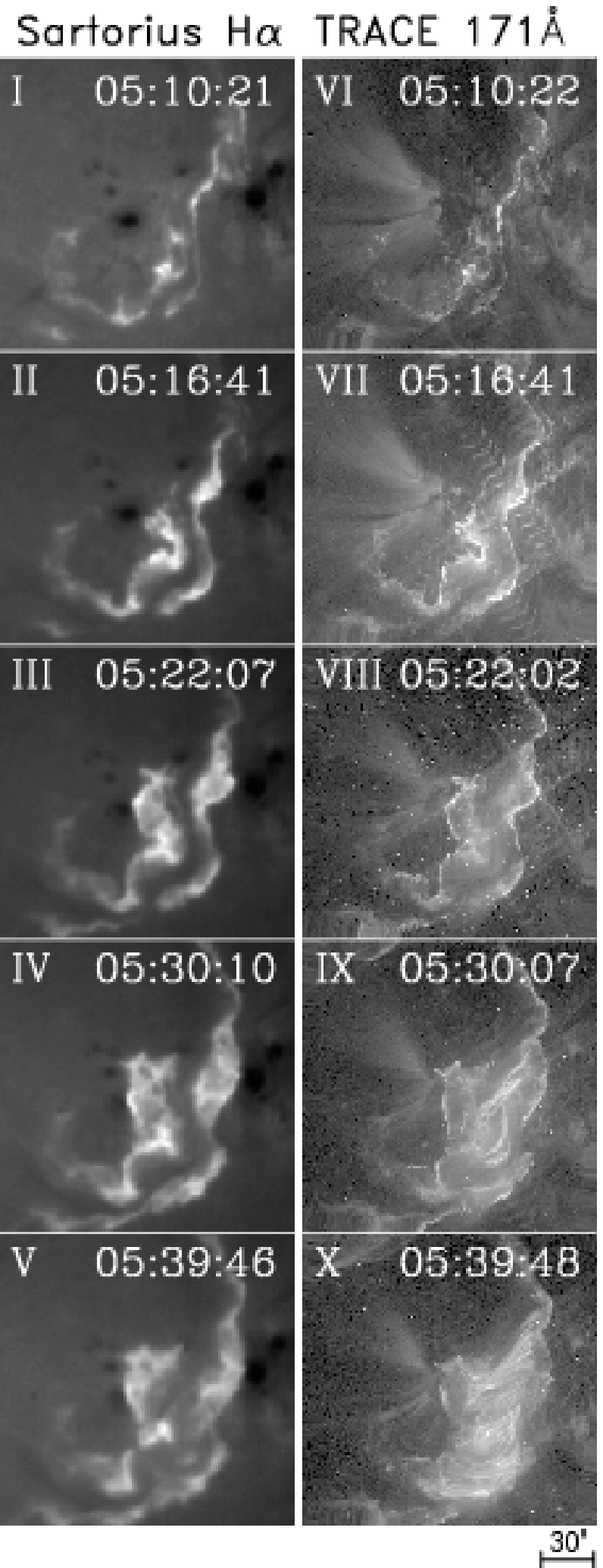}
  \end{center} \caption{ \label{fig:asai}  
  The flare SOL2001-04-10T05:26 (X2.3)\index{flare (individual)!SOL2001-04-10T05:26 (X2.3)!illustration}
\index{arcade!illustration} \citep{2003ApJ...586..624A} in H$\alpha$,
showing chromospheric emission, and \textit{TRACE} 171~\AA~ showing a mixture
of transition-region ribbons and loops.  This sequence shows clearly
the spreading of the flare ribbons as the flare progresses. This
event had an extended impulsive phase consisting of several HXR
spikes; the cross diffraction patterns at the footpoints are
characteristic of \textit{TRACE} impulsive phase EUV observations.  The
thickening of the H$\alpha$~ribbons is clear, and where these thickened
inner portions map to the \textit{TRACE} emission in panel VII, there is a
``moss''-like appearance. }
 \end{figure}
 \index{ribbons!TRACE@\textit{TRACE} EUV observations!illustration}
 \index{ribbons!H$\alpha$!illustration}

In the flare impulsive phase the picture is more complicated. For
example, more than two UV or H$\alpha$~ribbons may be seen in magnetically
quadrupolar flares\index{magnetic structures!quadrupolar configuration},
\citep[e.g.,][]{1985SoPh..102..131T,2007ApJ...655..606S},  as well
as a small number of bright footpoints visible in white-light, UV,
HXR or H$\alpha$~kernels\index{flare kernels}. 
In general the HXR sources\index{hard X-rays!identified with flare kernels} 
are confined to
localized areas situated on the outer edges of the elongated flare
ribbons observed in UV and H$\alpha$ (Figure~\ref{fig:v1}) and are
predominantly associated with bright H$\alpha$/UV kernels
\citep{2002ApJ...578L..91A,2004ApJ...611..557A,2005AdSpR..35.1707K}. Often
there are two main HXR footpoints, but sometimes there are more
\citep[e.g.,][]{2002SoPh..210..307F,2003ApJ...595L..69L,2007ApJ...654..665T}.
The HXR emission is direct confirmation of the presence of non-thermal
electrons in the lower chromosphere, which also heat and (further)
ionize the chromospheric plasma as they stop collisionally.

A basic property of flare footpoints is that they are compact. In
wavelengths from the infrared through the extreme ultraviolet,
the dimension of the brightest flare footpoints matches the miniumum
scale resolvable by the instrumentation used.  
\index{satellites!Hinode@\textit{Hinode}}
Only with the launch
of \textit{Hinode} do we have hints that we are approaching the
basic scale for the optical flare kernels.  
Sub-arcsecond structure\index{white-light flares!sub-arcsecond structure}
has been detected in optical flare sources which are seen to consist
of a bright emission core with a FWHM of around 500~km (corresponding
to an area of around $10^{16}\ \rm{cm}^2$), surrounded by a diffuse
halo of emission having greater extent \citep{2007PASJ...59S.807I}.
This diffuse halo is interpreted as radiation from the core
backscattered by deeper atmospheric layers. Arcsecond-scale widths
(diffraction limited) for flare ribbons were also observed in the
infrared at 1.56~$\mu$m \citep{2004ApJ...607L.131X}.  
Spatial resolution at UV/EUV wavelengths is not so high, 
but \textit{TRACE}~UV observations\index{TRACE@\textit{TRACE}!flare UV kernels}
of flare kernels\index{flare kernels!UV} in the 1600~\AA~band, where not saturated, are
consistent with them being on the scale of the telescope point
spread function \citep[see images in
e.g.,][]{2001ApJ...560L..87W,2006ApJ...640..505A}.  The size scale
of HXR images is harder to quantify, but in some flares \textit{RHESSI}
imaging reveals HXR footpoint sizes comparable with the resolution
capability of the finest grids (2.3 arcseconds FWHM)
\citep[e.g.,][]{2007ApJ...656.1187F,2007SoPh..240..241S,2008A&A...489L..57K,2009ApJ...698.2131D},
corresponding to an area on the order of $10^{17}\ \rm{cm}^2$.  The
smaller optical sizes may imply that the optical excitation is
taking place deeper down, in a converging magnetic field, or it may
mean that the HXR footpoints simply have not been resolved yet at
the best \textit{RHESSI} resolution.

During the {\it Yohkoh} era, the {\it Yohkoh}/SXT observed impulsive
SXR footpoint sources indicating heating of upper chromosphere or
transition region plasmas to around 10~MK
\citep{1993ApJ...416L..91M,1994ApJ...422L..25H,2004A&A...415..377M}.\index{eras!Yohkoh@\textit{Yohkoh}}\index{transition region}
The SXR emission is not consistent with an extrapolation of the
bremsstrahlung power law spectrum to low energies, and originates
in the chromosphere. Thus it appears to correspond to strong heating
of the chromosphere \citep[][]{1994ApJ...422L..25H}.\index{models!radiation hydrodynamic}\index{radiation hydrodynamics!and beam energy}
In beam-driven radiative hydrodynamic simulations, to achieve such temperatures
at the appropriate heights in the atmosphere requires beam energy
fluxes on the order of $10^{11}\ \rm{erg~cm}^{-2}~s^{-1},$ or
alternatively a lower electron flux for hundreds of seconds
\citep[][]{2005ApJ...630..573A}.

Although the impulsive-phase HXRs and optical emissions are
well-correlated in space and in time, the  relationship between HXR
and UV/EUV emission is not so clear. 
\index{hard X-rays!sources uncorrelated with UV/EUV}
There are often pre-flare
brightenings in UV which remain bright during the flare, and the
HXR footpoints occur only at locations which were not bright in~UV
before the flare \citep{2001ApJ...560L..87W}. There are good temporal
UV/HXR correlations during the flare, but as noted the UV ribbons\index{ribbons!UV}
are  more extended than the HXR footpoints \citep{2006ApJ...640..505A}.
So there are evidently only a few locations in the flare magnetic
field which are involved in the acceleration of a large number of
non-thermal particles. 
Looking next at the UV/EUV and HXR sources
which are at the same locations during the flare itself, there is
a relatively  good correlation\index{TRACE@\textit{TRACE}!UV correlation with HXRs} 
between the \textit{TRACE} 1600~\AA~channel
flux and the \emph{Yohkoh}/HXT 33-53 keV flux, and a weak
anticorrelation between the \textit{TRACE} UV (1600~\AA) and \textit{Yohkoh} HXR spectral
index \citep{2007A&A...472..945M}. 
An anticorrelation would be
expected due to the plasma heating in the upper chromospheric levels
produced by an electron beam in which low energies (with short
collisional stopping depths) dominate. 
The \textit{TRACE} 1600~\AA~channel
is rather broad in wavelength, so it is not clear which lines or
continua may be dominating the UV flux. This may explain in part
the relative weakness of this correlation. However, the weak
correlation may well indicate the importance of heating mechanisms
other than beam heating (e.g., thermal conduction) in producing EUV
footpoints. The study has not yet been repeated using the better
spectroscopic capabilities offered by \textit{RHESSI}.

A small number of isolated footpoints is the dominant impulsive-phase
HXR morphology, with elongated HXR flare ribbons rarely observed.
\index{ribbons!hard X-ray}
\index{hard X-rays!ribbons}
The first report of HXR ribbons was in \textit{Yohkoh} HXT observations of
SOL2000-07-14T10:24 (X5.7)\index{flare (individual)!SOL2000-07-14T10:24 (X5.7)}
\citep{2001SoPh..204...55M}.  In the well-observed SOL2005-05-13T16:57 (M8.0)
flare, 
\index{flare (individual)!SOL2005-05-13T16:57 (M8.0)} the \textit{RHESSI} HXR
sources evolve from footpoints concentrated in strong magnetic field
areas along the \textit{TRACE} UV ribbons in the HXR rising phase to ribbon-like
HXR source structures\index{hard X-rays!ribbon-like structures} closely matching the UV flare ribbon morphology
after the HXR peak (see Figure~\ref{veronig_liu_ribbon}).  The
simpler structure of the HXR ribbons presumably reflects more closely
the pattern of electron acceleration along the flare loop arcade
during its formation \citep[][]{liu07,2008ApJ...672L..69L}.
\index{arcade} Figure~\ref{veronig_liu_ribbon} suggests that \textit{RHESSI}
has sufficient resolution to resolve discrete footpoint features
that would be interpreted as multiple simultaneous footpoint
brightenings.  We should note that the appearance of elongated HXR
ribbon-like features might also result from source motions during
the relatively long HXR integration times ($>$4~s for \textit{RHESSI}
imaging).

\begin{figure}
  \begin{center}
        \includegraphics[width=0.7\textwidth]{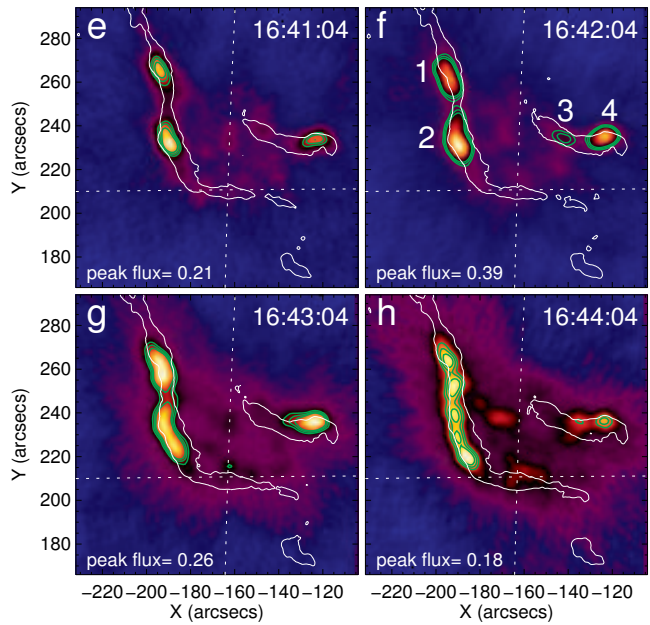}
  \end{center}
  \caption{ \label{veronig_liu_ribbon}Time sequence of \textit{RHESSI} 25--50 keV HXR images together with
    \textit{TRACE} 1600~{\AA} image contours (white lines) showing the
    evolution of the HXR source emission from localized footpoints to
    ribbon-like emission. 
    Flare SOL2005-05-13T16:57 (M8.0), adapted from Liu et al.\ (2007).}
\index{flare (individual)!SOL2005-05-13T16:57 (M8.0)!illustration}
\index{ribbons!hard X-ray!illustration}
\end{figure}

\subsection{Motions of ribbons and footpoints}\label{sec:fletcher_fpmotions}
\index{footpoints!motions}
\index{reconnection!footpoint motions}

As flare reconnection proceeds in the standard model, different
elements of the magnetic field move into the reconnection region.\index{standard model}
This leads to the expectation that the H$\alpha$/UV flare ribbons or
footpoint sources move.\index{ribbons!motions}  
In many flares the footpoints appear to
move away from one another and from the magnetic polarity inversion\index{magnetic structures!polarity inversion line}
line as the flare loop system grows \citep[for recent studies see,
e.g.,][]{2002SoPh..210..307F,2002ApJ...565.1335Q,2003ApJ...595L.103K,2004ApJ...611..557A,2006A&A...446..675V,miklenic07,2007ApJ...654..665T}.
The flare ribbons are understood to somehow map to the energy release
site in solar flares, and the movement of the ribbons and kernels
across the photosphere, and their relationship to magnetic fields,
is an important means by which the magnetic reconnection process
can be explored. For example, under the assumption of magnetic flux
conservation the progress of the flare sources across the magnetic
flux of the photosphere can be used to measure the magnetic flux
transfer rate.

However, much more complex HXR footpoint motions are observed than
the straightforward separating ribbon motions mentioned above. In
{\it Yohkoh}/HXT data only about 13\% of HXR flares exhibit HXR
footpoint motions corresponding strictly to separation with respect
to the polarity inversion line \citep{bogachev05}, and are more
likely to have a component of motion along the ribbon direction.
They can also approach one another\index{hard X-rays!and polarity inversion line}.  
Both converging and separating
footpoints travel at  tens of kilometers per second. A recent
statistical study of footpoint motions in 27~\textit{RHESSI} X- and M-class
flares by \cite{2009ApJ...693..132Y} has found that parallel/antiparallel
motions are more likely during the SXR rise phase than during the
flare peak, where separating motions become more prominent.  Many
individual examples confirm that the HXR footpoint motions do not
always agree with the standard-model predictions of separating
footpoints
\citep[e.g.,][]{2002SoPh..210..307F,2003ApJ...595L.103K,2005ApJ...625L.143G,2006ApJ...636L.173J}.
\index{flare models!CSHKP}\index{standard model}\index{reconnection!standard flare model}\index{CSHKP}
As in the
2-D flare model, the interpretation of the footpoint motion is still
in terms of magnetic reconnection, but in a complex magnetic field.
The complicated footpoint motions are likely to be linked to the
projection(s) of the locus of reconnection.  Particular examples
of  this can be found in sheared arcade models \citep{2002ApJ...579..863S},
and the ``slip-running'' reconnection model
\citep[][]{2006SoPh..238..347A}\index{reconnection!slip-running}\index{arcade}\index{flare models!slip-running reconnection}.

\bigskip
\noindent{\bf{Footpoint motion parallel to the magnetic polarity  inversion  line:}}
An example of footpoint motion parallel to the magnetic inversion
line is shown in Figure~\ref{grigis_fpim1}\index{footpoints!parallel motions}. 
The \textit{SOHO}/EIT\index{satellites!SOHO@\textit{SOHO}!EIT} image of
SOL2002-11-09T13:23 (M4.9)  
\index{flare (individual)!SOL2002-11-09T13:23 (M4.9)!footpoint motion} 
shows a postflare arcade,  with the centroid position of the
\textit{RHESSI} HXR footpoints superimposed. As the event progresses, in
several emission spikes, the HXR footpoint pairs move along the
arcade. Source motion specific to the emission spikes shows up as
deviations from the overall trend of the footpoint motions, and are
decomposed into parallel and perpendicular components relative to
this trend.  Footpoint motion is directed parallel to the ribbons
and is smooth at these scales, in contrast with the bursty evolution
of the HXR flux. There is no evidence for a systematic trend in
outward perpendicular displacement, or of discontinuities during
the transitions from one spike to the next. The emission spikes
originate at different sources along the arcade. The overall picture
is that the HXR footpoint motion may be a consequence of a moving
trigger, possibly caused by an asymmetric eruption of a filament
\citep{2006A&A...453.1111T}, or a ``domino effect'' where energy
release in part of the field triggers activity in its neighbors.

\begin{figure}
\begin{center}
\includegraphics[width=0.45\textwidth]{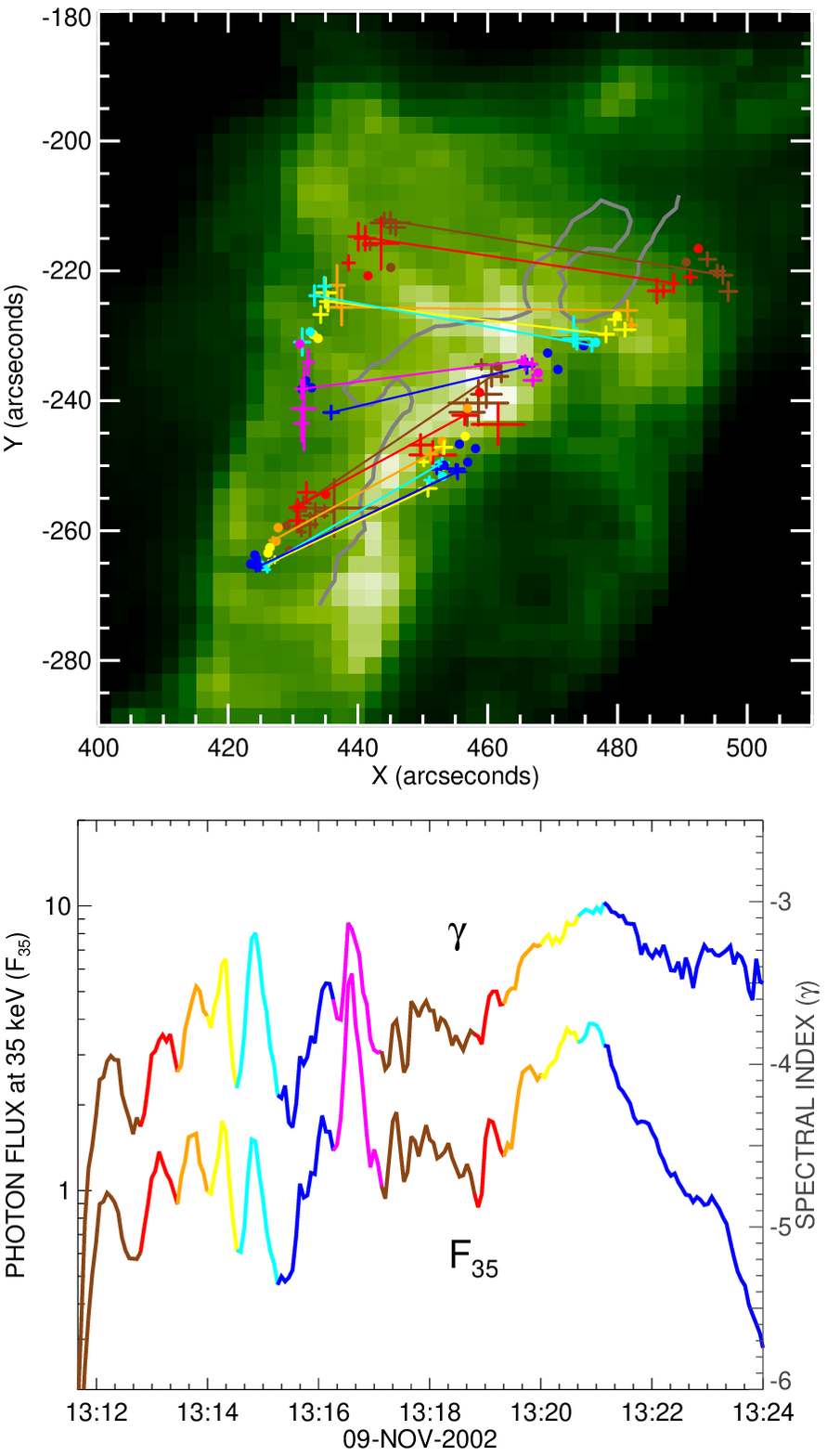}
\includegraphics[width=0.45\textwidth]{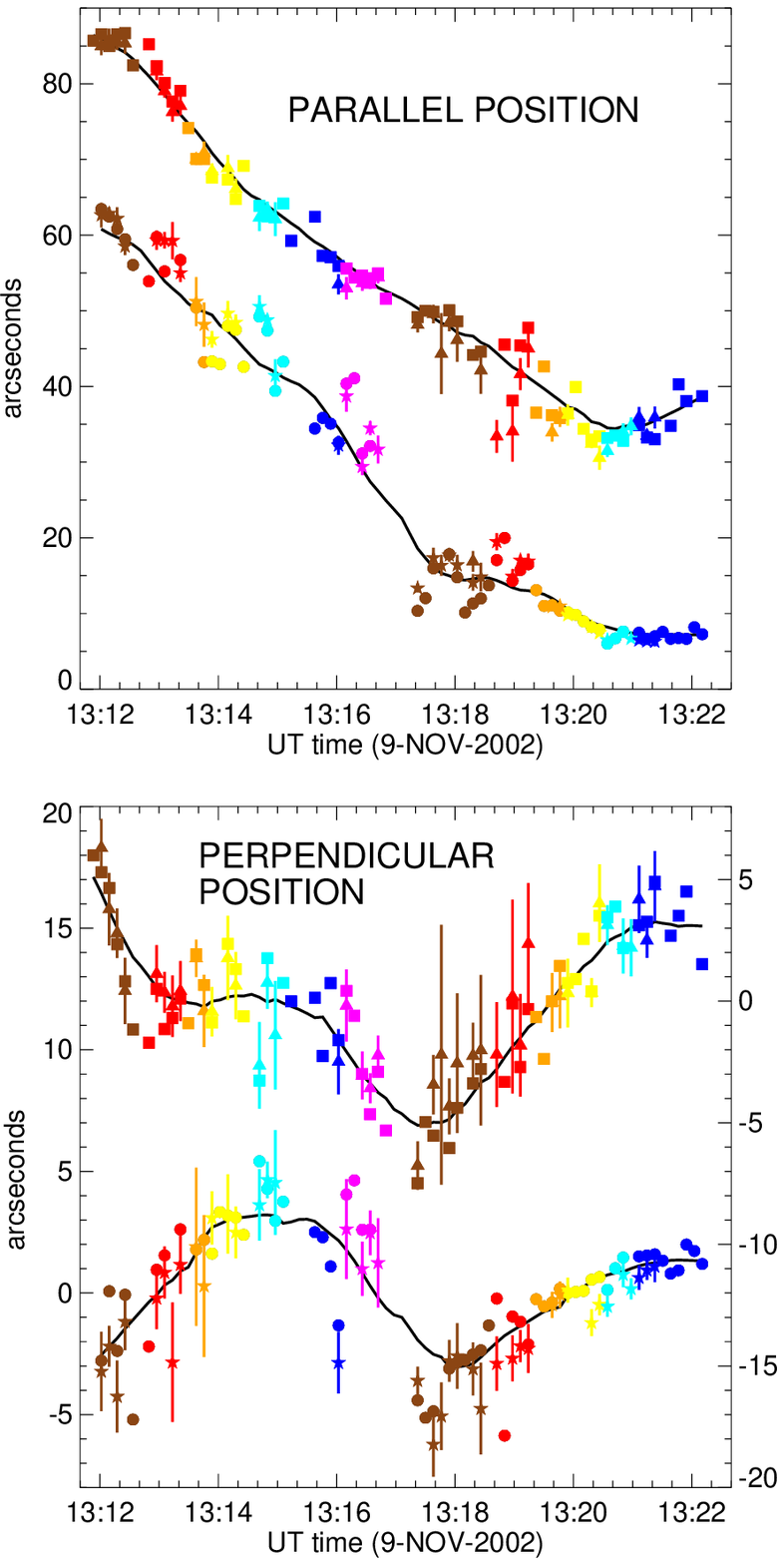}
\end{center}
\caption{\label{grigis_fpim1}{\it Top left}: \textit{SOHO}/EIT 195~\AA~image of post-flare
loops with the
  \textit{RHESSI} HXR source positions superimposed for SOL2002-11-09T13:23 (M4.9). 
  The positions of the
  20-50~keV sources from the CLEAN images are represented by crosses with
  arm lengths equal to the errors, positions from the PIXON images are
  given by circles. Simultaneous footpoints are connected and color
  coded according to the time intervals defined in the bottom
  part. The neutral line is shown in gray. {\it Bottom left}: Time evolution
  of the flux and spectral index. {\it Right}: Time evolution of the source
  positions relative to the trend lines. Triangles and stars with
  error bars refer to values derived using CLEAN, squares and circles
  using PIXON, for the western and eastern footpoints, respectively.}
\end{figure}
\index{flare (individual)!SOL2002-11-09T13:23 (M4.9)!illustration}
\index{footpoints!apparent motions!illustration}

\bigskip
\noindent{\bf{Converging Footpoints:}} 
In its first couple of minutes, SOL2002-11-09T13:23 (M4.9)
also exhibits a convergence of the flare 
\index{flare (individual)!SOL2002-11-09T13:23 (M4.9)!footpoint convergence} 
footpoints\index{footpoints!converging motions}, explained in terms of the
sequential activation of a flare arcade which varies in width along
its length \cite[][]{2005ApJ...625L.143G}.  
A different kind of converging footpoint motion, corresponding to footpoints travelling
anti-parallel to one another and along the ribbons, was seen earlier
with {\it Yohkoh}/HXT \citep{bogachev05}, and now frequently in
\textit{RHESSI} \citep[e.g.,][]{2002SoPh..210..307F}. However, a new feature
of such events became apparent in \textit{RHESSI} observations showing the
accompanying HXR coronal source. A number of events observed in
HXRs, H$\alpha$~and UV/EUV showed approaching footpoints in the early
impulsive phase, accompanied by a projected downward motion of the
coronal HXR source, and followed by separation of the footpoints
and a projected rise in the coronal source ~(Figure~\ref{ji_fig2}).
These include SOL2002-03-14T01:44 (M5.7)
\index{flare (individual)!SOL2002-03-14T01:44 (M5.7)!footpoint motion}
\index{footpoints}, SOL2003-10-29T20:49 (X10.0)
\citep{2008AdSpR..41.1195Z,2008ApJ...680..734J,2009ApJ...693..847L},
\index{flare (individual)!SOL2003-10-29T20:49 (X10.0)!footpoint motion} 
SOL2002-09-09T17:52 (M2.1), and SOL2004-11-01T03:22 (M1.1)
\citep{2004ApJ...607L..55J,2006ApJ...636L.173J}.  
\index{flare (individual)!SOL2002-09-09T17:52 (M2.1)!footpoint motions} 
\index{flare (individual)!SOL2004-11-01T03:22 (M1.1)!footpoint motions} 
The initial converging motion has also been noted in UV/EUV
ribbons \citep{2008AdSpR..41.1195Z}. The footpoint convergence phase
lasts for a few minutes, and both footpoints and coronal sources
move with a projected speed of some tens of km~s$^{-1}$.
\cite{2009ApJ...693..847L} found an example of simultaneous height
decrease and footpoint convergence.  In the cases of antiparallel
footpoint convergence, the empirical shear (determined from the
angle relative to the neutral line made by the line joining footpoints)
may decrease no matter whether the footpoints move inward or outward
with respect to the polarity inversion line.

The downward motion of the coronal sources is consistent with the
extraction of magnetic energy from the field \citep{2000ApJ...531L..75H}.
It has been interpreted in two related ways -- as an initial shrinkage
of the field in a 2-D ``collapsing trap'' 
\index{collapsing magnetic trap}
immediately following reconnection
\citep[e.g.,][]{2004A&A...419.1159K,2006A&A...446..675V} and as the
consequence of the relaxation of shear in a 3-D arcade model \citep[see
also][]{2007ApJ...660..893J,2002ApJ...579..863S}. In the latter
model, the less-sheared field reaches a lower altitude in the
corona than more-sheared field, and its relaxation can in
principle explain both the decrease in coronal source altitude and
the converging footpoint motion.  
We discuss the relationship between
ribbon motions and coronal dynamics extensively in
Section~\ref{sec:ltmotion} below.

\begin{figure}
  \begin{center}
  \includegraphics[width=0.7\textwidth]{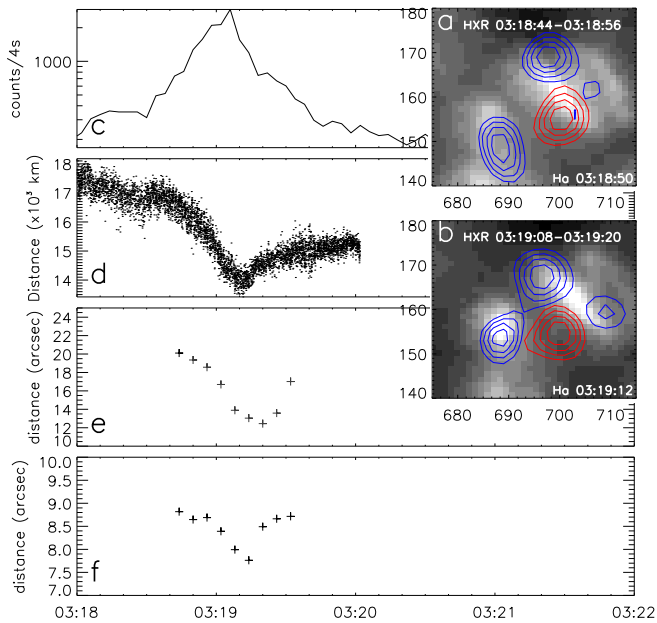}
   \end{center}
   \caption{\label{ji_fig2}Converging footpoints in SOL2004-11-01T03:22 (M1.1):
   (a)-(b) \textit{RHESSI} HXR contours overlaid on H$_\alpha-$0.5~{\AA} filtergrams
   taken at PMO, with the blue contours at 25-60~keV and the red ones at 3-6~keV,
   (c) \textit{RHESSI} 50--100~keV time profiles, (d) distance between the two
conjugate H$\alpha$~kernels or HXR conjugate
footpoint sources, and e) projected height of the \textit{RHESSI} loop-top
source. From \cite{2006ApJ...636L.173J}.}
\end{figure}
\index{flare (individual)!SOL2004-11-01T03:22 (M1.1)!illustration}
\index{flare kernels!H$\alpha$!illustration}

\subsection{Flare footpoints and the magnetic environment}\label{sec:fletcher_fpmagnetic}

It has long been known that the magnetic structure of the solar
corona is reflected in the distribution and evolution of flare
footpoint sources\index{impulsive phase}.
The simplest example of this is the straightforward
mapping between the pre- and post-reconnected field in the 2-D
standard model, and the spreading H$\alpha$~ribbons.\index{standard model!2-D}
Even in more
magnetically complex configurations, in principle each X-ray
footpoint, or white-light/UV kernel, maps via the coronal field to a
conjugate counterpart.  In practice this has been difficult to
demonstrate quantitatively.  
The flare impulsive phase is characterized
by complex magnetic geometries, and recent years have also seen
great advances in breaking down the active region coronal field
into its topological elements -- separatrix and quasi-separatrix
layers, separator field lines and null-points.\index{magnetic structures!separatrices}\index{magnetic structures!null}\index{separatrix}
This remains an area of intense theoretical
activity, as part of an overall effort to understand how magnetic
reconnection takes place in three dimensions. The hope is that the
observed evolution of footpoints can aid in this overall goal.
During the flare impulsive phase, where the HXR and WL footpoints
are typically observed, the magnetic geometry is not readily
interpreted from EUV loop observations (as is the case in the gradual
phase). 
The magnetic field is presumably stressed, and
therefore the relatively straightforward potential field extrapolations
\index{magnetic field!extrapolation!worries about}\index{caveats!photospheric field extrapolation}
may provide a misleading picture of the the overall coronal
structure (though as we see below they have been used to explore
certain aspects of the flare geometry). 
Since the \textit{Skylab} era\index{eras!Skylab@\textit{Skylab}}
\citep{1973SoPh...32..173Z,1975SoPh...45..411P} it has been clear
that later loops in the gradual phase of a flare look more
potential-like, and make a larger angle to the neutral line, and
this is often glibly taken as evidence for the reduction of shear
expected to reduce the stored magnetic energy\index{flares!energy content!magnetic}. 
Clearly it is not that simple; the observed
changing pattern of shear is determined by the amount of shear in
the pre-flare field as a function of distance from the neutral line
before the flare, as well as its reduction as a function of time
during the flare.  
The distribution of magnetic shear\index{magnetic field!shear distribution} within the
flaring volume reflects the paths taken by coronal current systems,
about which we have little knowledge.

\begin{figure}
\begin{center}
\includegraphics[width=0.7\textwidth]{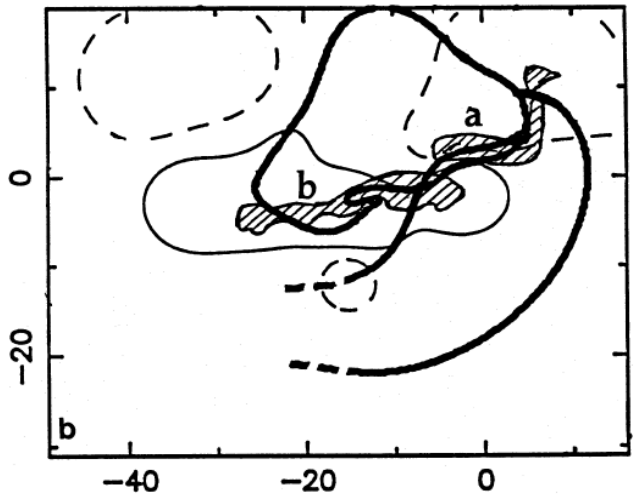}
\end{center}
\caption{\label{fig:mandrini_seps} Portions of the photospheric
projection of separatrix structures from a linear force-free magnetic
extrapolation are shown by \cite{1995A&A...303..927M} to correspond
to the locations of flare H$\alpha$\ ribbons. Contours show $\pm$400~G
levels of magnetic field strength. Note, the correspondence between
separatrix and ribbon positions is not found for a potential model.
The H$\alpha$\ ribbons are found by \cite{1995A&A...303..927M} to correspond
to the regions of highest current density in their model.}
\index{magnetic structures!separatrices!illustration}
\index{separatrix!illustration}
\index{ribbons!H$\alpha$!illustration}

\end{figure}

\bigskip
\noindent{\bf{Footpoints and magnetic topology:}}
The earliest observational studies demonstrated that the locations
of H$\alpha$~ribbons could be explained (with a suitable arrangement of
magnetic charges\footnote{This is ``magnetic charge topology,'' in
which an array of fictitious magnetic monopoles is used as a best
fit to a photospheric magnetogram, as a basis for 3-D potential-field
extrapolations
\index{magnetic structures!extrapolation!magnetic charge topology}
\citep{1980SoPh...67..245B,1993A&A...271..292D,1996SoPh..169...91L}.},
extrapolated in a potential approximation) as the intersection of
coronal separatrix surfaces with the photospheric boundary
\citep[][]{1989SvA....33...57G,1991A&A...250..541M,1996SoPh..169...91L}\index{magnetic charge topology}\index{magnetic structures!separatrices and flare ribbons}.
An example of this is shown in
Figure~\ref{fig:mandrini_seps} \citep{1995A&A...303..927M}\index{ribbons!high current density}.
Many flares have been modeled using similar approaches with increasing
degrees of complexity -- for example, incorporating linear force-free
fields \citep{1994SoPh..150..221D} and non-linear force-free fields
\citep{2002A&A...392.1119R,2006A&A...451..319R,2008ApJ...675.1637S},
and using increasingly precise representations of the photospheric
field \citep{2005ApJ...629..561B} and chromospheric magnetic fields
\citep{2005ApJ...623L..53M} as input.  Overall the correlations
between the photospheric mappings of separatrices, quasi-separatrices
and separators, and the observed location of flare footpoints are
convincing, but certain aspects still evade a clear explanation.
\index{magnetic structures!separatrices} 
These mappings have not yet allowed us to fully understand the reason why the HXR or WL footpoints are few and compact, while the H$\alpha$ ribbons are extended, nor the
properties of the reconnection that determine the motions of the footpoints. 
But they are providing us with fascinating clues.

\begin{figure}
  \begin{center} \hbox{
  \includegraphics[width=0.5\textwidth]{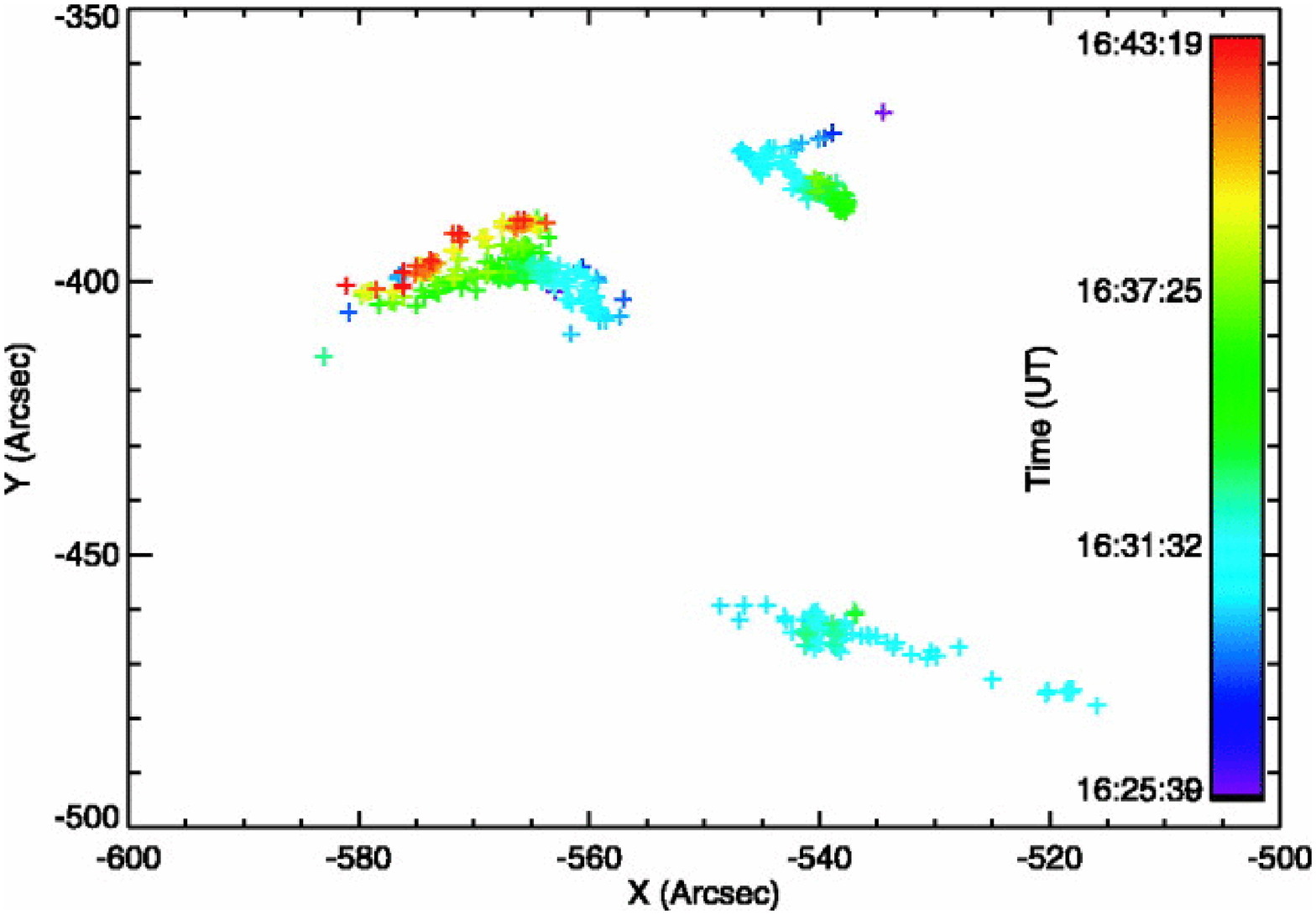}
 \includegraphics[width=0.5\textwidth]{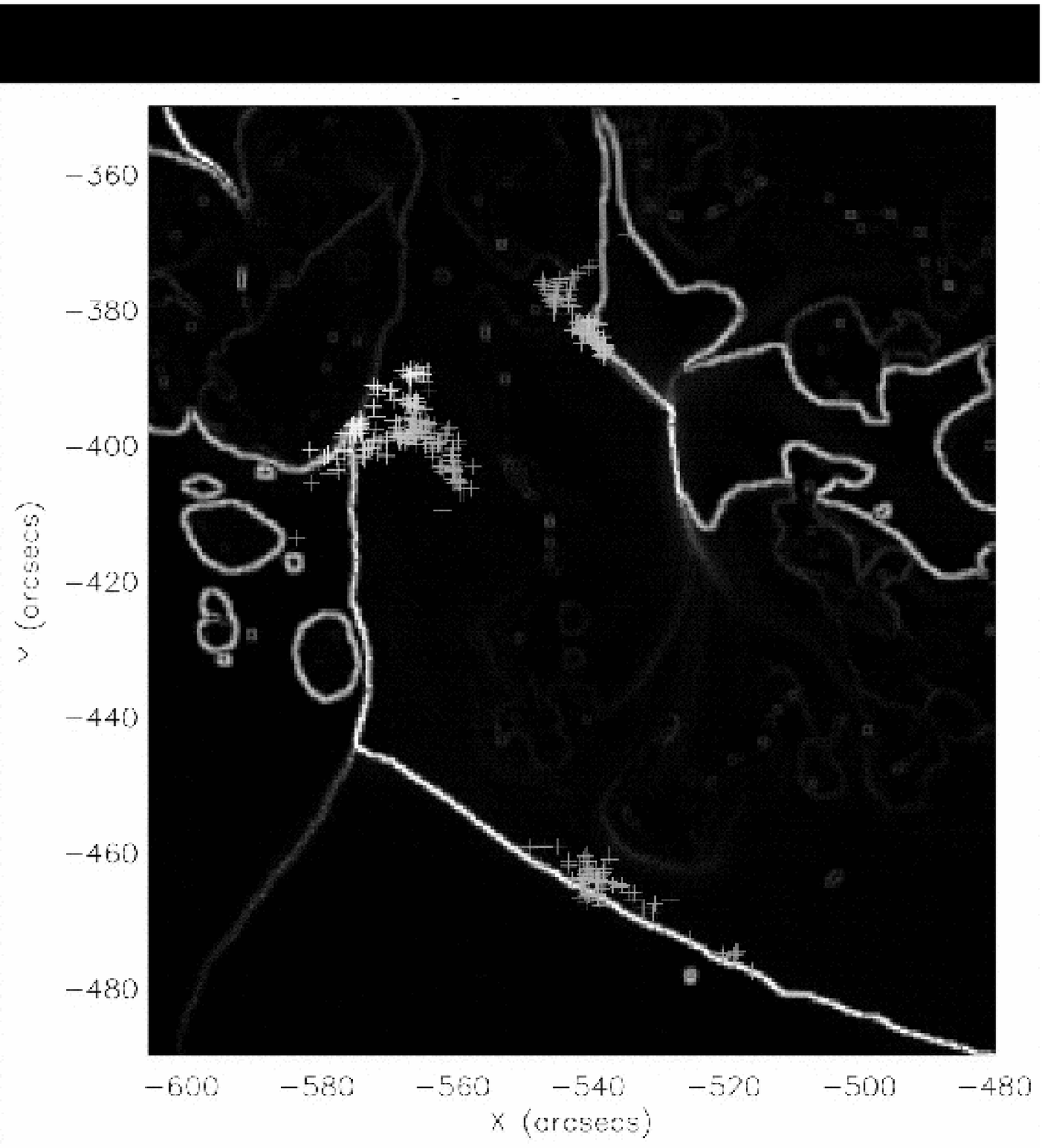} }
\end{center}
   \caption{\label{fig:metcalf}
 \textit{Left:} positions of the HXR footpoints in SOL2001-08-25T16:45 (X5.3) \citep{2003ApJ...595..483M},
  with time color-coded on the right; data from the \textit{Yohkoh}/HXT
  instrument \citep{1994PhDT.......335S}.
\textit{Right:} locations of the separatrix surfaces projected onto
the photospheric plane, with the shading representing the magnitude
of the separatrix discontinuity.  The upper footpoints
illustrate the complex motions along the flare ribbons, while the
lower points show rapid motion.  
The authors showed that this motion coincided with the photospheric intersection of
a magnetic separatrix structure.} 
\end{figure} 
\index{flare (individual)!SOL2001-08-25T16:45 (X5.3)!illustration} 
\index{magnetic structures!separatrices!illustration}
\index{separatrix!illustration}

Prior to \textit{RHESSI}, a small number of studies had been carried out in
which the relationship between magnetic field and HXR footpoint
location was investigated. As well as the investigations into
footpoint asymmetry described below, there was some early work on
HXR sources in the context of magnetic topology. For example,
Figure~\ref{fig:metcalf} shows the {\it Yohkoh} HXR and 
\textit{TRACE} WL\index{TRACE@\textit{TRACE}!white light}
source motions observed in SOL2001-08-25T16:45 (X5.3),
\index{flare (individual)!SOL2001-08-25T16:45 (X5.3)}
compared to the projections of
coronal separatrix structures onto the photosphere
\citep{2003ApJ...595..483M}. 
This Figure reveals strong resemblances
between observed and calculated features; in particular the lower
source moves extremely rapidly almost in coincidence with one of
the separatrix intersections, in a manner suggesting the ``slip-running''
reconnection model of \cite{2006SoPh..238..347A}.\index{flare models!slip-running reconnection}\index{separatrix}
Evidence was also
found for a coronal null and reconnection of external field through
a separatrix ``dome'' \citep{2001ApJ...554..451F}.  \textit{Yohkoh}-era
observations also showed that HXR sources tended to avoid sites of
high vertical current density, preferentially occurring adjacent
to them \citep{1997ApJ...482..490L}.

\textit{RHESSI} data have led to more studies relating HXR footpoint behavior
and magnetic fields. 
For example, in SOL2005-01-17T09:52 (X3.8), 
\index{flare (individual)!SOL2005-01-17T09:52 (X3.8)}
which exhibits
four H$\alpha$~ribbons and corresponding HXR footpoint sources, the two
strongest and long-lived H$\alpha$~kernels and HXR footpoints are observed
to tend to avoid the strongest fields, and move approximately along
\index{sunspots!and flare ribbons}
iso-Gauss contours\index{flare kernels!on iso-Gauss contours}, 
along the border between the sunspot's umbra and
penumbra \citep{2007ApJ...654..665T}. 
\index{footpoints!and H$\alpha$ ribbons}
Also in this event, the
magnetic reconnection rates derived for flare ribbon locations
\index{reconnection!and HXR footpoints}
showing HXR footpoints are higher (by two orders of magnitude) than
those in flare ribbon locations not showing HXRs.  The strongest
HXR sources were preferentially located in those regions of the
ribbons with the strongest magnetic field, although this cannot
readily be seen from Figure~\ref{temmer_hard X-ray_motion}. Similar
results were obtained from \textit{Yohkoh}/HXT studies
\citep{2002ApJ...578L..91A,2004ApJ...611..557A}.

For three major \textit{RHESSI} flares, it has been demonstrated
\citep[][]{2009ApJ...693.1628J} that the path of HXR footpoints
corresponds to a particular type of topological structure, a subset
of the photospheric spine lines identified in magnetic charge
topology models (see Longcope, 2005, and the discussion
in Section~\ref{sec:fletcher_fpmagnetic}).
\index{magnetic charge topology}
\nocite{2005LRSP....2....7L}
These are lines that join two magnetic sources of the
same sign (``charge'') via a magnetic null\index{magnetic structures!null}. 
The implication is that
the footpoint movement reflects the changing length of the separator
joining the nulls on the two spine lines, as the coronal reconnection
proceeds and the reconnection region moves.

The huge differences derived in the local energy release rates for
flare ribbon locations with/without HXR footpoints in combination
with the limited dynamic range of present HXR instruments (of
order~10:1) can explain the different flare morphologies typically
observed in HXRs (compact footpoints) and H$\alpha$/UV (extended ribbons).
However, it is still implied that a large fraction of the electrons
is accelerated into spatially confined subsystems of magnetic loops
as outlined by the HXR footpoints, and only a minor fraction goes
into the large flare arcade outlined by the H$\alpha$/UV ribbons and EUV
postflare loops \citep{2007ApJ...654..665T}\index{arcade}.\index{ribbons!energy release rate}

Although white light is an important indicator of the locations of
strongest energy input \citep[e.g.,][]{1989SoPh..121..261N}, and
although it is substantially simpler to image than HXRs, we do not
have adequate systematic observations.  In particular the white-light
footpoint motion has only rarely been studied.  
\index{white-light flares!energy input}
SOL2001-08-25T16:45 (X5.3) \citep{2003ApJ...595..483M} 
\index{flare (individual)!SOL2001-08-25T16:45 (X5.3)!white light} 
is one such example.  
The SOL2002-09-30T01:50 (M2.1)
\index{flare (individual)!SOL2002-09-30T01:50 (M2.1)!white light}
white-light flare \index{white-light flares} studied by
\citet{2006ApJ...641.1217C} is another. 
It showed systematic footpoint
motion in the white-light continuum, following roughly that of the
corresponding HXR source.  Footpoints at both wavelengths zigzag
back and forth,  primarily parallel to the magnetic neutral line
(see Figure~\ref{fig:chen_qr_fig2})  in a manner which may be
explained by the particular magnetic configuration in the flaring
region.
\index{flare (individual)!SOL2002-09-30T01:50 (M2.1)!white light}

\begin{figure}
  \begin{center}
      \includegraphics[width=\textwidth]{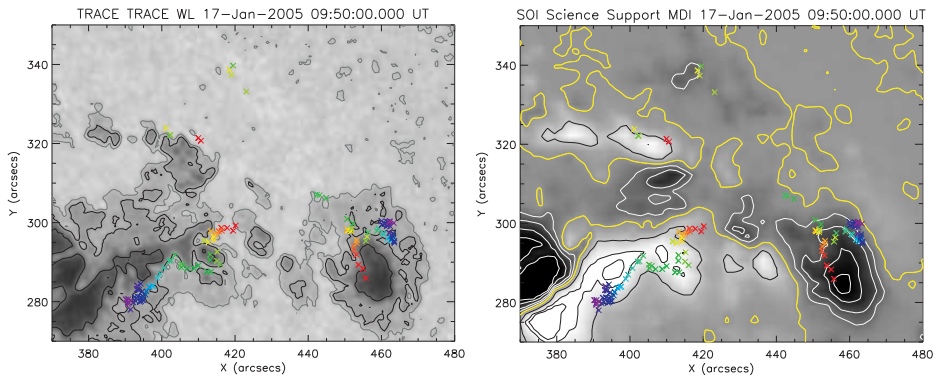}
   \end{center}
   \caption{\label{temmer_hard X-ray_motion}\textit{RHESSI} hard X-ray source centroids overlaid on a \textit{TRACE} white-light 
   image (left) and MDI magnetogram (right)
   for the X4 flare SOL2005-01-17T09:52. The subsequent occurrence of the HXR sources from
   09:43:20 to 10:04:10~UT is color-coded from red to blue. 
   \textit{Left:} \textit{TRACE} white-light contours roughly
   outline the umbrae as well as inner and
    outer penumbrae of the sunspots. 
    \textit{Right:} Isogauss lines at $-2000,  -1600,  -1300$, and $-600$~G
    (white contours) and +600, +1300, and +1500~G (black contours).
    The yellow line marks the magnetic inversion line. 
    Adapted from  \cite{2007ApJ...654..665T}.}
\end{figure}
\index{flare (individual)!SOL2005-01-17T09:52 (X3.8)!illustration} 
\index{sunspots!and flare ribbons!illustration}
\index{flare types!white light!illustration}

\begin{figure}
\begin{center}
\includegraphics[width=0.6\textwidth]{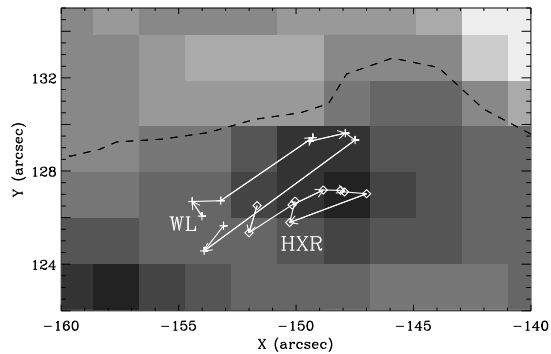}
\end{center}
\caption[]{\label{fig:chen_qr_fig2}Footpoint motion history in the white-light 
continuum ({\it pluses}) and
12-25~keV HXR emission ({\it diamonds}) superposed on the MDI magnetogram
in the white-light flare SOL2002-09-30T01:50 (M2.1).
The magnetic neutral line is plotted as the dashed line
 \citep{2006ApJ...641.1217C}.
 The points generally describe a clockwise motion, and cover a time interval of
 about five minutes.
}
\index{flare (individual)!SOL2002-09-30T01:50 (M2.1)!illustration}
\index{white-light flares!illustration}
\end{figure}

\bigskip
\noindent{\bf{Footpoint Asymmetry:}}
\index{footpoints!asymmetry}
A basic prediction of flare models invoking electron acceleration
in the corona and precipitation to the chromosphere is that regions
with stronger magnetic field convergence (i.e., a stronger chromospheric
or photospheric magnetic field) should be locations of weaker HXR
footpoint sources, because the higher mirror ratio leads to a larger
fraction of accelerated electrons mirroring before they reach the
thick-target footpoints. The ratio of brightness in footpoint pairs
should thus be inversely correlated with the ratio of magnetic field
strengths at the location of those footpoints. This tendency was
demonstrated systematically in early analyses of a small number of
double-footpoint flares observed with {\it{Yohkoh}}/HXT, using both
line-of-sight \citep[e.g.,][]{1994kofu.symp..169S} and vector fields
\citep{1997ApJ...482..490L}, but later work revealed counter-examples.
Stronger HXR footpoints were found in stronger magnetic field regions
in at least one-third of 32 flares examined
\cite[e.g.,][]{2004A&A...423..363G}\index{footpoints!asymmetry!statistics}. 
A detailed study by
\cite{2009ApJ...693..847L} of \textit{RHESSI} footpoint pairs in SOL2003-10-29T20:49 (X10.0) \index{flare (individual)!SOL2003-10-29T20:49 (X10.0)}
showed the expected relationship in the first few minutes of the
event, but thereafter it disappeared.  
Furthermore, although the
sign of the correlation was as expected early in the event, the
magnitude was not consistent with simple predictions of the magnetic
mirroring model.\index{flare models!magnetic mirror}
They also found that collisional losses\index{electrons!collisional losses} due to
asymmetric column densities from the looptop\index{looptop sources} (assumed to be the
acceleration region) to the footpoints alone cannot explain the
totality of the observed HXR fluxes and spectra.  This is consistent
with the result of a statistical study of \textit{RHESSI} footpoint asymmetry
carried out by \cite{2008SoPh..250...53S}, though these authors did
not examine the footpoint magnetic fields.  
\index{footpoints!model requirements}
As \cite{2009arXiv0906.2449L}
suggest, more detailed modeling including mirroring, collisional
losses, and other particle transport effects (such as nonuniform
target ionization, relativistic beaming, photospheric albedo, and
return currents) may provide a resolution to the above discrepancies
(see Figure~\ref{fig:LiuW2009_asym_fig9}).  
An alternative investigation
of the footpoint asymmetry intrinsic to the acceleration process
has been pursued \citep{2005ApJ...619.1153M} to explain the
observations of \cite{2002SoPh..210..323A}.

 \begin{figure}     
 \begin{center}
 \includegraphics[width=0.6\textwidth]{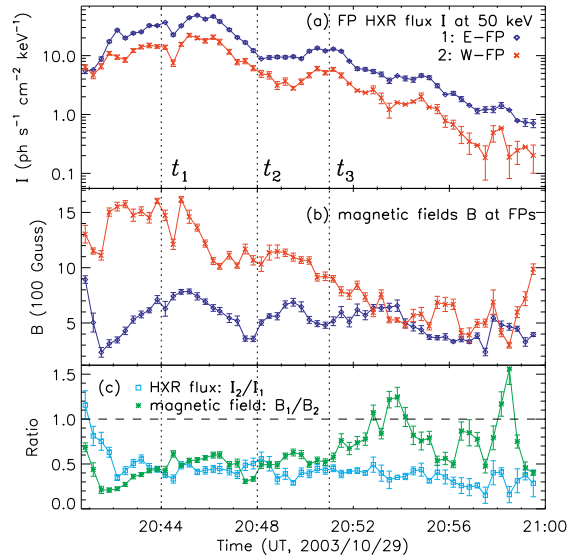}
 \end{center}
 \caption[History of the HXR fluxes of the two footpoints and associated magnetic fields]
 {Time profiles of X-ray and magnetic field parameters of the conjugate footpoints in SOL2003-10-29T20:49 (X10.0).
 \index{flare (individual)!SOL2003-10-29T20:49 (X10.0)!illustration}
   ({\it a}) HXR fluxes at 50~keV of the eastern footpoint (E-FP, {\it blue diamonds}) and western footpoint 
 (W-FP, {\it red crosses}) obtained from power-law fits in the 50--150~keV range.
    ({\it b}) \textit{SOHO}/MDI magnetic field strengths registered at the two footpoints.
   ({\it c}) Ratios of the 50~keV fluxes (W-to-E) and magnetic fields (E-to-W) of the two footpoints. The expected
 correlation between these two ratios only holds for the first half of the flare duration
 \citep[from][]{2009ApJ...693..847L}.
 }	\label{fig:LiuW2009_asym_fig9}
 \end{figure}

Most of the microwave emission in flares is gyrosynchrotron from
non-thermal electrons, the intensity of which depends on the electron
energy and on the ambient magnetic field.
Thus the stronger-field footpoint\index{footpoints!asymmetry!and microwaves} 
should correspond to stronger microwave sources (and
weaker HXR footpoints), giving a complementary view of footpoint
asymmetry. This relationship has been found to hold in some recent
investigations \citep{1995ApJ...454..522K,1995ApJ...453..505W}, but
the modeling presents the complicated problem of understanding the
microwave absorption. Without understanding absorption it is not
possible to get at information about the microwave footpoints, and
so any conclusions drawn from HXR and microwave comparisons are
premature.

\bigskip
\noindent\textbf{Inference of the properties of magnetic reconnection:}
The product of the footpoint apparent speed and the line-of-sight
magnetic field, expressed as a flux transfer rate $\dot{\Phi}$, has
been used as a measure of the coronal reconnection rate by various
authors
\citep[e.g.,][]{2002ApJ...565.1335Q,2004SoPh..222..279F,2005ApJ...620.1085J,2007ApJ...654..665T}\index{reconnection!flux transfer in ribbons}\index{ribbons!flux transfer}.
The quantity $\dot{\Phi}$ can be estimated from the observations
as \citep[e.g.,][]{forbes00}: 
\begin{equation} \dot{\Phi} =
\frac{\partial}{\partial t} \int B_{\rm ph} da, 
\end{equation} 
where $da$ denotes the ribbon area and $B_{\rm ph}$ the normal component
of the photospheric magnetic field\index{footpoints!and reconnection rate}.

In the gradual phase of large two-ribbon flares where the magnetic
configuration is well-approximated by a 2- or 2.5-D field (i.e., no
significant shear or twist component of the field), the coronal
reconnection  rate is also equal to an equivalent electric field\index{electric fields!and reconnection rate}
$E_c$ \citep{1986lasf.symp..453P,forbes00}, i.e.,
\index{magnetic structures!2.5-D approximation}

\begin{equation}
E_c = v_{\rm fp} B_{\rm ph} .
\end{equation}
where $E_c$ is the convective electric field~at the magnetic
reconnection site, ~$v_{\rm fp}$ the observed speed of the apparent
HXR footpoint or H$\alpha$/UV flare ribbon and $B_{\rm ph}$ the vertical
component of the photospheric magnetic field.  
Moreover the energy release rate equals the Poynting flux into the current sheet
\citep[e.g.,][]{2006ApJ...647..638L}.\index{footpoints!and UV ribbons}\index{electric fields!convective}\index{Poynting flux!estimate from footpoint motions}
Estimates of the flux can be
obtained from the motions of footpoints and the line-of-sight
magnetic field strength at the flare footpoints
\citep[e.g.,][]{isobe02,2004ApJ...611..557A,2007ApJ...654..665T}.

In the impulsive phase, which observationally is far from
two-dimensional, the relationship between footpoint motion, magnetic
field and flux transfer rate should be preserved (this follows from
magnetic flux conservation) but the coronal reconnection electric
field is not so readily obtained.  Nor, in an environment of strong
twist and shear, will there be a straightforward relationship between
$\dot{\Phi}$, the Poynting flux and the energy release rate.  We
note that a study with a prescribed 3-D coronal field
\citep{2005ApJ...631.1227H} showed a relationship between the
instantaneous reconnected magnetic flux at a field line and the
``field-line-integrated'' parallel electric field\index{electric fields!parallel} along that field
line, suggesting that the endpoints of field lines with high values
of this electric field correspond to locations of chromospheric
excitation.  Although the relationship between $\dot\Phi$ and the
parallel field is determined by the coronal magnetic configuration,
which is not generally known, this analytic work can give us some
confidence that  $\dot\Phi$ calculated from footpoint motions is a
meaningful quantity, also in the impulsive phase.

Observationally there are interesting correlations between $\dot\Phi$
and properties of the footpoint radiation. 
Several studies reveal
correlations between the HXR flux evolution and the derived
reconnection quantities $E_c$ and $\dot{\Phi}$, and also with the speed
of the footpoint separation or flare loop growth
\citep{2002ApJ...565.1335Q,
2002SoPh..210..307F,2003ApJ...595L.103K,2004ApJ...611..557A,2004ApJ...611L..53L,2005AdSpR..35.1707K,2006A&A...446..675V,2006ApJ...641.1197S,miklenic07,2007ApJ...654..665T}.
Figure~\ref{fig:veronig_krucker_fp} shows the result for SOL2003-10-29T20:49 (X10.0), with
derived $v_{\rm fp}$, $v_{\rm fp} B_{\rm ph}$ and $v_{\rm
fp}B_{\rm ph}^2$ curves correlated with the \textit{RHESSI} HXR flux\index{reconnection!footpoint motions} \citep[][]{2005AdSpR..35.1707K}.\index{flare (individual)!SOL2003-10-29T20:49 (X10.0)}
In this flare the HXR flux exponentially correlates with
the magnetic field strength at the footpoints, which may scale with
the field strength in the coronal reconnection region
\citep{2009ApJ...693..847L}.
 Figure~\ref{veronig_miklenic_rec} shows the relationship in time
 of $\dot{\Phi}(t)$ derived from SOL2003-11-18T08:31 (M3.9).
 \index{flare (individual)!SOL2003-11-18T08:31 (M3.9)!flux transfer}
Each of the strongest three \textit{RHESSI} HXR peaks is well reflected in
the derived $\dot{\Phi}(t)$ time profiles but shifted in time by
1--2~min \citep[for a discussion of this effect see][] {miklenic07}.
The correlation in time and space between locations of high $v_{\rm
fp} B_{\rm ph}$ and footpoint intensity has also been demonstrated
in detail using \textit{TRACE} UV footpoints\index{TRACE@\textit{TRACE}!UV footpoints}  
\citep{2009A&A...493..241F}.

 \begin{figure}
  \begin{center}
    \includegraphics[width=0.7\textwidth]{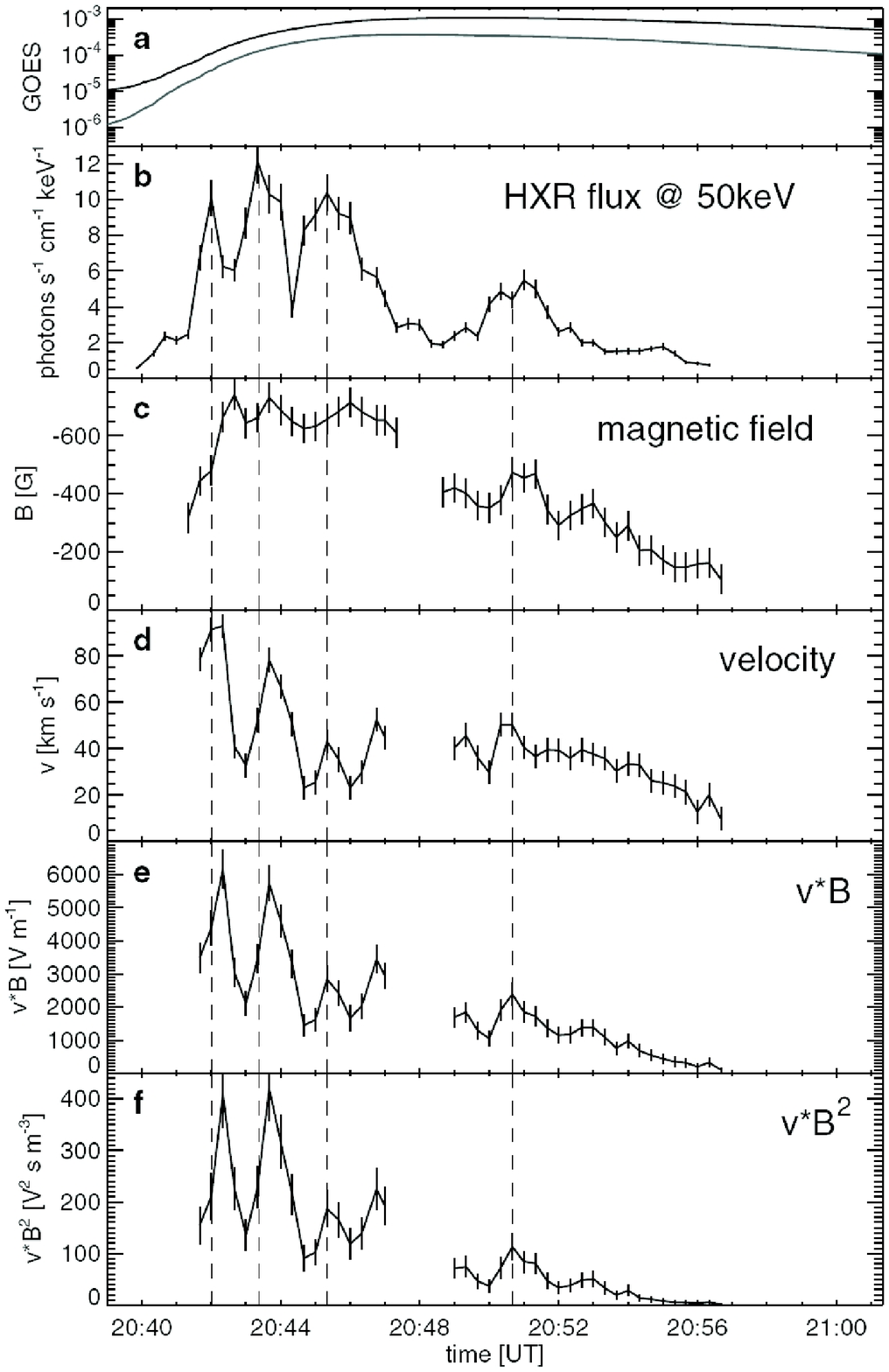}
   \end{center} \caption{ \label{fig:veronig_krucker_fp}Magnetic
   reconnection analysis of SOL2003-10-29T20:49 (X10.0).
     Time series of: (a)~\textit{GOES} SXR flux, (b)~\textit{RHESSI} HXR flux at 50
     keV of the Eastern footpoint source, (c)~photospheric magnetic
     field~$B$ at the instantaneous footpoint location, (d)~velocity~$v$
     of the HXR footpoint source, (e)~product $vB$ (magnitude of convective
     electric field~$E_c$), (f) product $vB^2$ (measure of the
     energy release rate~$\dot{W}$) \cite[from][]{2005AdSpR..35.1707K}.
     }
\end{figure} 
\index{reconnection!illustration}
\index{ribbons!flux transfer!illustration}
\index{flare (individual)!SOL2003-10-29T20:49 (X10.0)!illustration}

Interestingly, in the rare example of the HXR ribbon flare
\index{hard X-rays!ribbons}
SOL2005-05-13T16:57 (M8.0), which has at first glance quasi-2-D properties,
there is a better correlation between the HXR intensity and the
derived local magnetic reconnection rate and energy release rate
when there are only a few isolated HXR footpoints, than when the
ribbon-like HXR emission appears
\citep[][]{2007ApJ...664L.127J,2008ApJ...672L..69L}.
\index{flare (individual)!SOL2005-05-13T16:57 (M8.0)!hard X-ray ribbons}

\begin{figure}
  \begin{center}
    \includegraphics[width=0.7\textwidth]{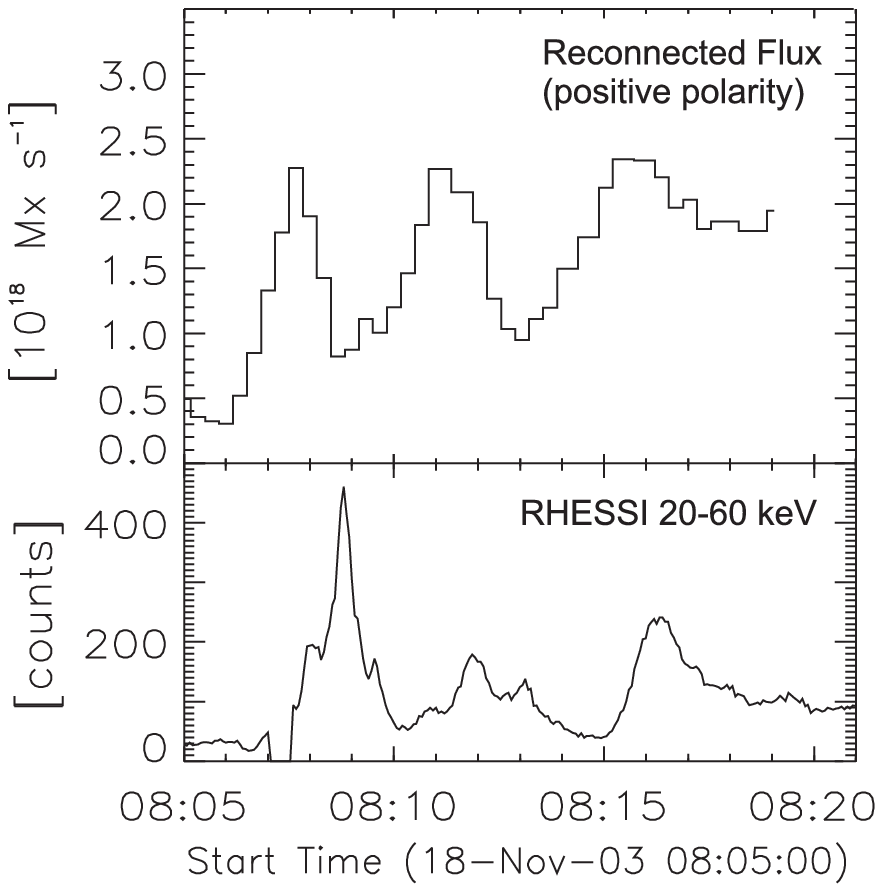}
  \end{center}
  \caption{\label{veronig_miklenic_rec}Magnetic flux change rate $\dot{\varphi}(t)$ together with the \textit{RHESSI} 20--60~keV HXR flux
    for SOL2003-11-18T08:31 (M3.9). Adapted from \cite{miklenic07}. }
\index{flare (individual)!SOL2003-11-18T08:31 (M3.9)!illustration}
\end{figure}

Further substantial progress in determining the reconnected flux
and Poynting flux will be extremely difficult. 
\index{Poynting flux!and reconnection}
It will require not
only the measurement of the chromospheric vector magnetic field (to
assess time-dependent field perturbations at the boundaries) but
also microwave observations of gyroresonance emission leading to
improved knowledge of the coronal magnetic field. Using multi-frequency
microwave observations, such as are planned with FASR 
\index{observatories!FASR}
\citep[][]{2003AdSpR..32.2705B}, isosurfaces
of magnetic field strength can be computed. Coupled with plasma
flow information from EUV spectrscopic diagnostics, ``before and
after'' changes of the field around a flare would give an independent
view of the energy extracted from the field, as well as some
information about how it moves through the configuration. 
Direct
\index{Hanle effect}
\index{magnetic structures!need for coronal magnetography}
observations of the coronal magnetic field at the limb using infrared
Zeeman splitting\index{Zeeman effect!infrared} \citep{2000ApJ...541L..83L} are proceeding now with the CoMP instrument \citep{2008SoPh..247..411T}.
\index{CoMP}
 \index{magnetic field!coronal}
The Hanle effect\index{Hanle effect} in the UV \citep{2009ASPC..405..429R} is also utilized.

\subsection{Excitation of the deep atmosphere}\label{sec:fletcher_deep}
As is well known, the first recorded observation of a solar flare
\citep{1859MNRAs..20...13C} was in the optical or ``white light''
wavelength range\index{flare types!white light}\index{flares!Carrington}\index{white-light flares}\index{flare (individual)!SOL1859-09-01T11:18 (pre-\textit{GOES})}.
Since such a flare is visible over and above the bright photospheric
radiation (6.27~$\times$~10$^{10}$ erg~cm$^{-2}$ s$^{-1}$ in the
quiet Sun),  roughly doubling it, this emission is a significant
component in the flare energy budget.  
A small number of direct\index{flares!total irradiance}
measurements of the flare total irradiance now exist for large
flares \citep{2006JGRA..11110S14W,2008cosp...37.1617K}, in which
the total radiant energy of the flare is measured to be a few times
$10^{31}$ to $10^{32}$ erg.  
White-light emission\index{flares!white light} can also be
present in relatively weak flares, down to low \textit{GOES} C class
\citep[][]{2003A&A...409.1107M,2006SoPh..234...79H,2008ApJ...688L.119J}.

The generation of flare optical radiation (IR/visual/UV continuum)
is not yet well explained, and it may be that there are different
processes operating in different flares. 
Where spectroscopic
observations are available, observed white-light flares have been
split into two types \citep{1986lasf.conf..483M,1989SoPh..121..261N}.
Type I flares\footnote{Type~I and Type~II as used here should not
be confused with the meter-wave radio bursts
\citep[e.g.,][]{1963ARA&A...1..291W}, nor with the spicules
\citep{2007PASJ...59S.655D}.} show intense and broad Balmer lines
and Balmer\index{continuum!Balmer} and Paschen\index{continuum!Paschen} edges (resulting from recombination), and are thought to occur in a heated chromosphere.\index{white-light flares!type II}\index{white-light flares!type I}\index{recombination radiation}
Type~II flares, much
less frequent, do not show these features and may arise from enhanced
$H^-$ continuum\index{continuum!h@$H^-$}. 
The location at which the Type~II flare radiation
is produced is not known.  
Generally it is hard to see how the deep
layers of the photosphere could be directly excited by electron
beams without requiring rather unreasonable electron energy budgets
\citep{1986A&A...156...73A}.\index{protons!beams}\index{beams!at photosphere}
Excitation by proton beams\index{beams!proton} with energies of
10-20~MeV has also been proposed
\citep[e.g.,][]{1970SoPh...13..471S,1978SoPh...58..363M}, since
protons have a greater penetration depth than electrons of the same
velocity.  
A small velocity dispersion is required to focus the
energy deposition over a narrow range of depths, as well as to find
agreement with proton fluxes at $\sim30$~MeV implied by $\gamma$-ray
observations\index{gamma-rays!and white-light flares}.
Energetic electrons would need to be present in
addition, to account for the HXR flux.

A popular model for the Type~I events, which does not require
electron beams to reach the deepest layers of the chromosphere, is
the ``radiative backwarming'' model. 
In this model energy is deposited
in the upper chromosphere generating a strong Balmer-Paschen continuum
by recombination, which warms lower levels \citep{1990ApJ...350..463M}.
\index{backwarming}\index{continuum!Balmer}\index{recombination radiation}
This idea is close to the original suggestion of \cite{1972SoPh...24..414H}
to bypass the complicated problems of radiative transfer with the
``specific ionization'' approximation\index{specific ionization approximation}, which implies secondary
ionizations.  \cite{1972SoPh...24..414H} and \cite{1986A&A...156...73A}
also note that the non-thermal ionization of hydrogen would also
strongly enhance the continua\index{models!non-LTE}.
Non-LTE simulations suggest that a
purely chromospheric temperature rise may be insufficient to produce
the continuum intensity enhancements seen \citep{1996A&A...314..643D}
and an enhancement near the temperature minimum region\index{temperature minimum} 
may still be necessary.\index{simulations!non-LTE radiative transfer}  
Evidence for the effect of energy deposition was
found in the white-light flare SOL2002-09-29T06:39 (M2.6)
\index{flare (individual)!SOL2002-09-29T06:39 (M2.6)}
\index{white-light flares}
\citep[][]{2003ApJ...598..683D,2005ApJ...618..537C}. 
This event had two HXR footpoints, one with weaker HXR emission but stronger
white-light continuum emission, and a relatively weak, centrally-reversed
H$\alpha$~profile \citep[see e.g.,][for a discussion of flares at optical
wavelengths]{1966SSRv....5..388S}.  
This profile indicates that at the weaker HXR footpoint the atmosphere
had not been fully heated, and under such conditions it is possible
that an electron beam could effectively penetrate the chromosphere
and produce the observed continuum emission via radiative backwarming.
By contrast, the local atmosphere at the other footpoint had been
appreciably heated, producing a high coronal pressure
\citep{1984ApJ...282..296C}. 
Electrons would thus be prevented from penetrating into the deeper atmosphere.


Direct evidence for excitation of the deep atmosphere during a flare comes from the flare seismic waves (``sunquakes'')  first observed by  \cite{1998Natur.393..317K}, and thereafter in several other flares of M~and X~class\index{global waves!seismic}\index{tsunami}\index{sunquakes}.

\begin{figure}[h]
  \begin{center}
          \includegraphics[width=1\textwidth]{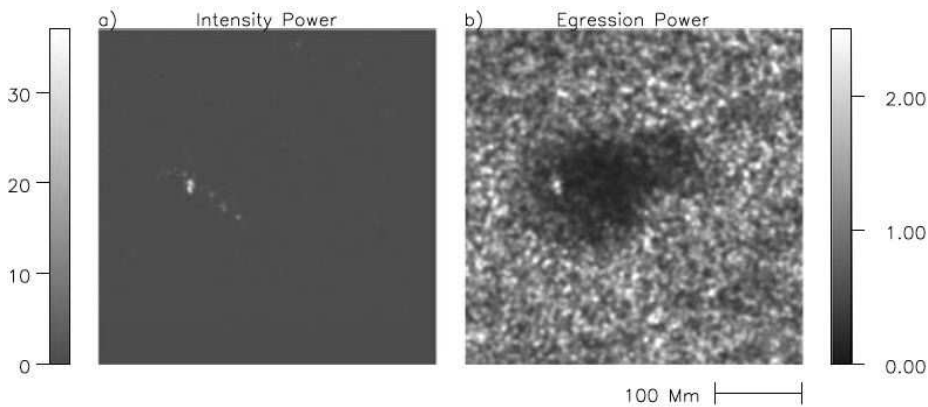}
 \end{center}
  \caption{Signatures of SOL2003-10-29T20:49 (X10.0) filtered in the 5-7~mHz band, 
  ie at frequencies above most of the p-mode power.
  \textit{Left:} intensity; \textit{right:} ``egression power,'' showing the source of the seismic
  waves observed from this flare.
  The main seismic source is within the area of the large sunspot.
  From \cite{2008SoPh..251..627L}.
   }\label{fig:seismic}
\end{figure}
\index{flare (individual)!SOL2003-10-29T20:49 (X10.0)!illustration} 
\index{flare (individual)!SOL2003-10-29T20:49 (X10.0)!egression power} 
\index{sunspots!and seismic sources!illustration}

These waves refract through layers deep in the convection zone and
appear as surface ripples, traveling at apparent speeds of only
some tens of km~s$^{-1}$, far from the flare site. 
Seismic waves
have been observed from several flares, but they are still comparatively
rare -- most large flares do not produce detectable wave amplitudes
\citep[][]{2005ApJ...630.1168D}. 
However, the energy required to
produce the disturbances is found to be small, on the order of
$10^{-4}$ of the total flare energy.
The sources of this seismic emission can be located holographically
(Figure~\ref{fig:seismic}), and a large proportion of the seismic sources
are located within the penumbrae of sunspots.  The sources appear
\index{sunspots!and seismic sources}
to coincide with the HXR footpoints and white-light flare kernels\index{flare kernels!white light}
\citep{2008MNRAS.389.1905M}, and move with them
\citep[][]{2006SoPh..238....1K}.  
They do not appear to be so
strongly associated with the $\gamma$-ray sources -- i.e., the
accelerated ions \citep[][]{2007ApJ...670L..65K} -- but the
$\gamma$-ray imaging is much inferior to the HXR imaging\index{waves!seismic!and $\gamma$-rays}\index{gamma-rays!and seismic waves}.  
Flare seismic waves may be associated with downwards-moving material
in the MDI Doppler data\index{shocks!sunquake excitation}\index{shocks!radiative damping}.
Theory suggests that a shock could be produced by intense heating of the chromosphere by an electron beam,
and initially it was proposed that the waves resulted from this
shock impacting on the photosphere.  
However, the momentum required
to produce the seismic disturbance is substantially higher than
that observed directly in the plasma downflows, and the shock
propagation time to the photosphere is inconsistent with observations
of the seismic pulse onset versus the HXR peak \citep[][and references
therein]{2008SoPh..251..641Z}.  
\index{downflows} It is also likely
that such shocks would be radiatively damped before reaching the
chromosphere \citep[][]{2008SoPh..251..627L}.  
Thus another method for delivering momentum into the deep photosphere may be required.  
Proton beams have been proposed \citep[][]{2007ApJ...664..573Z}.
\index{sunquakes!and proton beams}\index{beams!proton}

\begin{figure}      
 \begin{center}
 \includegraphics[width=0.7\textwidth]{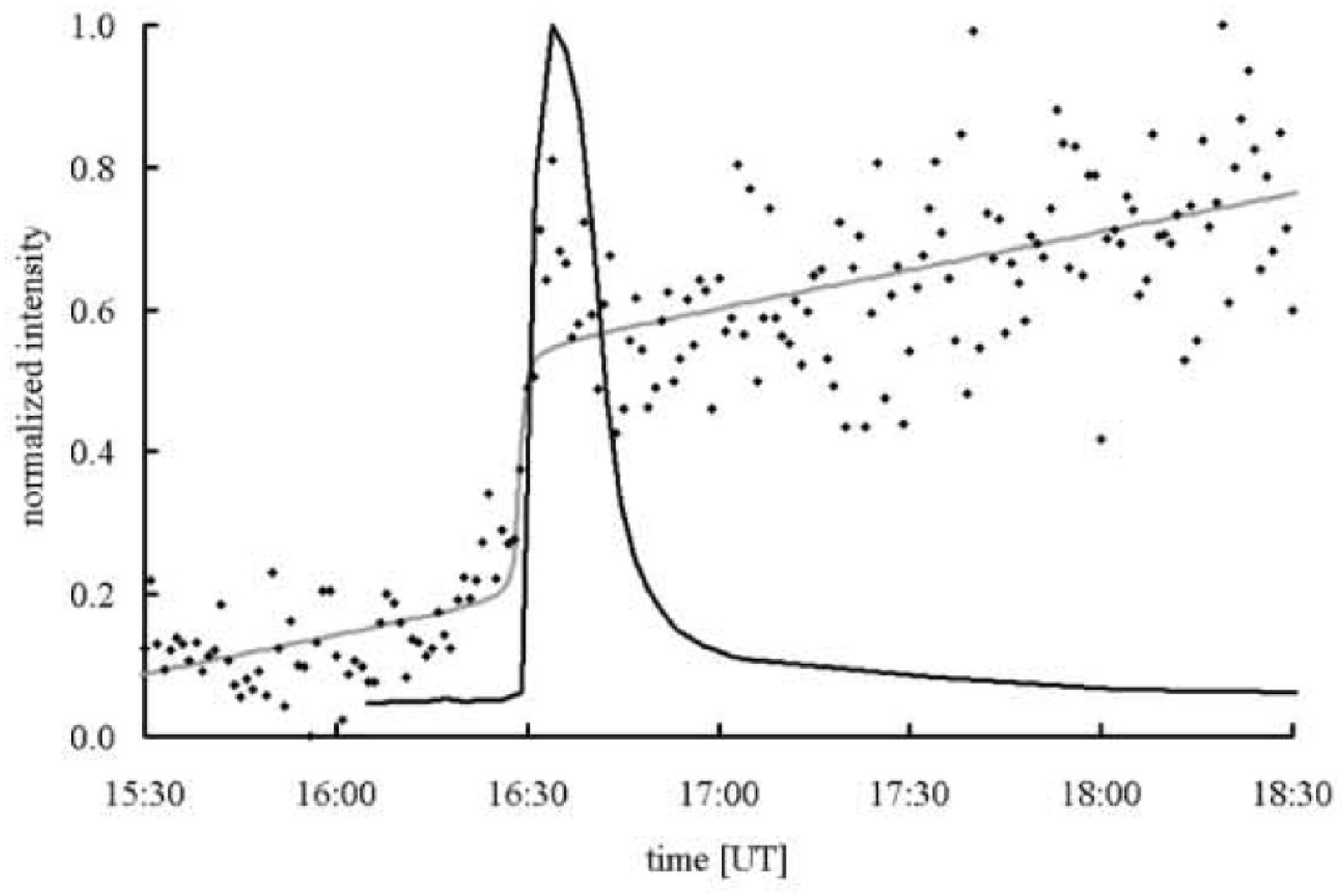}
 \end{center}
 \caption {\label{fig:sudolharvey} Points and gray line show the
 normalized time variations of the line-of-sight magnetic field
 measured by GONG at a point, during SOL2001-08-25T16:45 (X5.3)\index{flare (individual)!SOL2001-08-25T16:45 (X5.3)!illustration}.
 The line is the \textit{TRACE} 1600~\AA~intensity measured co-spatially with
 the changing magnetic field, to within around 20$''$. The line-of-sight
 magnetic field changes abruptly at the time of the flare impulsive
 phase \citep[from][]{2005ApJ...635..647S}.  } \end{figure}
 \index{magnetic field!flare-related changes!illustration}

An alternative view is that the seismic waves are launched by a
``jerk'' of the magnetic field,  caused by field re-organization in
the corona, imparting momentum to the ions and collisionally-coupled
neutrals at the photosphere \citep[][]{2008ASPC..383..221H}\index{jerks}\index{global waves!seismic!and Lorentz force}\index{global waves!seismic!and shock waves}\index{global waves!seismic!and backwarming}.
The jerk is the Lorentz force imparted at the photosphere and is thus
capable of launching an interior seismic wave.  
\index{waves!energy transport}
In the standard
reconnection flare model, the perturbation producing the jerk would
originate in the corona and propagate as a wave into the photosphere
\citep[e.g.,][]{2008ApJ...675.1645F}.  
Initial analyses cast doubt
on the viability of this mechanism \citep[][]{2009MNRAS.395L..39M},
but it is difficult to disentangle this mechanism and the others
proposed: the original idea of a hydrodynamic shock
\citep{1975SvA....18..590K,1998Natur.393..317K}, and the more recent
discussion of direct photospheric heating, e.g., through radiative
backwarming \cite[e.g.,][]{2008SoPh..251..627L}.
\index{backwarming}\index{shocks!hydrodynamic}

The Lorentz-force jerk \index{magnetic field!jerk} is
consistent with the non-reversible magnetic field changes in the
line-of-sight magnetic field observed in many flares.  
\index{magnetic field!flare-related changes}
Early observations showed permanent changes in the vector field around
the magnetic neutral line \citep{1994ApJ...424..436W}, in one case
the observations being separated by only a few minutes before and
after the flare. Similar observations for a flare on the limb found
variations in the line-of-sight magnetic field (i.e., the component
tangential to the photosphere for a limb flare), the importance of
this being that the line-of-sight component does not suffer from
the 180$^\circ$ directional ambiguity of the vector field
\citep{1999ApJ...525L..61C}. 
\index{magnetic field!flare-related changes} 
Further work on irreversible changes
to the line-of sight magnetic field followed using 
\textit{SOHO}/MDI\index{satellites!SOHO@\textit{SOHO}!MDI}
\citep{2006SoPh..238....1K} and they have now been confirmed to
occur in all large flares, usually close in time to the flare
impulsive phase \citep[][]{2005ApJ...635..647S}, and close spatially
to the HXR footpoint sources. Figure~\ref{fig:sudolharvey} shows
an example of such field changes, the typical magnitude of of which
is 100-200~G (or on the order of 10\% of the photospheric field in
the region).  In many cases, the field changes are also associated
with visible evolution in the sunspot, particularly to a disappearance
of a part of the penumbra
\index{sunspots!penumbral disappearance}
\citep{1993SoPh..147..287A,2004ApJ...601L.195W,2005ApJ...622..722L}.

Changes of this magnitude at the photosphere imply that the overlying
field undergoing rearrangement -- presumably in the low corona --
must be strong, and energy transported by magnetic disturbances
propagating through the chromosphere to the photosphere may be an
important component in the flare chromosphere energization
\citep[][]{1982SoPh...80...99E,2008ApJ...675.1645F}.
\index{waves!energy transport}
An interesting
aspect of such observations is that many flares show an \textit{increase}
in the observational shear along the magnetic polarity inversion
line \citep{1994ApJ...424..436W}, counter to what would be expected
in a scenario in which the active region free energy should decrease
to power the flare.  \index{magnetic field!free energy} 
However,
\emph{Hinode}/SOT\index{satellites!Hinode@\textit{Hinode}!SOT} 
observations of a flare show that below about
8000~km above the photosphere the shear increases after
a flare, whereas above this altitude it decreases
\citep[][]{2008ApJ...676L..81J}. It is possible that the shear
increase close to the polarity inversion line is associated with
flux emergence, as part of the ongoing build-up of magnetic free
energy in a repeatedly flaring region.

\subsection{Chromospheric Evaporation}\label{sec:fletcher_fpevap}
\index{chromospheric evaporation}

The arcades of loops characteristic of the gradual phase of solar
flares are filled with hot, dense plasma, usually interpreted as
chromospheric plasma which expands to a new equilibrium following
chromospheric heating in the impulsive phase.  This process is
termed chromospheric evaporation.  It has been studied for almost
40~years, since it was first proposed to explain the delay between
the peaks of SXR (gradual) and microwave (impulsive) emissions
\citep{1968ApJ...153L..59N}.  Latterly, the Neupert effect usually
refers to the often-observed relationship between time-integrated
HXR flux and the SXR flux.  \index{Neupert effect} The Neupert
effect gives indirect evidence for chromospheric evaporation; more
direct support comes from  observations of blueshifted emission of
high-temperature plasma, often correlated with impulsive HXR bursts
as discussed further below. 
Early on,  spatially-unresolved
300-400~km~s$^{-1}$ upflows\index{upflows} in resonance lines of Ca~{\sc{xix}} and
Fe~{\sc{xxv}} were observed using instruments on board the 
\textit{Solar Maximum Mission}
\citep{1980ApJ...239..725D,1983SoPh...86...67A,1988ApJ...324..582Z}, and
since then confirmed in spatially-resolved observations with \textit{SOHO}/CDS\footnote{Coronal Diagnostic Spectrometer.}
\citep[e.g.,][]{1999ApJ...521L..75C,2003ApJ...588..596T,2004ApJ...613..580B,
2006SoPh..234...95Z} and more recently 
\textit{Hinode}/EIS\index{satellites!Hinode@\textit{Hinode}!EIS} \citep{2008ApJ...680L.157M}.
\index{satellites!SOHO@\textit{SOHO}!CDS}
It has  also been suggested that the heating of quiescent active-region
loops  is not actually ``coronal heating''\index{coronal heating!chromospheric} at all, 
but happens in the
chromosphere and stems from the principles of flare-induced
chromospheric evaporation \citep{2007ApJ...659.1673A}.

SXR spectra of flares often show a dominant stationary component
as well as the upflow.  This is a puzzle for evaporation theory in
a single loop excitation, but it could be  explained by a filamentary
structure, in which many sub-resolution magnetic loops are activated
in succession, each having such a small cross-section that it
produces undetectably small amounts of emission. Emission would
then be detected only after some time, when a number of these loops
are emitting together, and the evaporated plasma in each has come
to rest at the looptop\index{looptop sources} \citep{2005ApJ...629.1150D}.  It has also
been argued that hot dense plasma exists in the flare corona in
advance of the flare impulsive phase \citep{1990ApJ...364..322F,2010ApJ...725L.161C}.
It is usually assumed that the hot emission comes from the flare
corona, but recently, stationary Fe~{\sc xxiv} emission has been
detected at loop footpoints 
\citep{2009ApJ...699..968M}\index{footpoints!stationary hot component}. 
There are also observations which suggest that the upflows\index{upflows!height limit} do not reach high into the corona, and that the coronal density increase occurs as a
result of compression \citep{2004ApJ...609..439F,2010ApJ...725L.161C}.  
An adiabatic compression\index{adiabatic compression} 
would in fact create a \textit{negative} microwave flare\index{flares!negative}
in the free-free continuum because the free-free emissivity scales
as $T^{-0.5}$.  On the contrary, the observations show a good
correlation between the radio and X-ray continuum emission 
measures\index{emission measure!radio and SXR},
so a negative flare is contrary to the observations
\citep[e.g.,][]{1965sra..book.....K,1972SoPh...23..155H}; see White
et al. (in this volume) for more detail\index{radio emission!negative microwave bursts}.
Note that negative microwave
bursts do occur, but they can be explained by intervening absorptions
\citep{1973SoPh...33..439C}.

Early in the flare the chromosphere is heated rapidly and impulsively,
primarily by energetic electrons which lose energy collisionally
in the chromosphere.  
Thermal conduction from the corona may also
play a role in heating the chromospheric\index{chromospheric heating} plasma, particularly in
pre-impulsive \citep{2009A&A...498..891B} or gradual
\citep[e.g.,][]{1988ApJ...329..456Z} phases.  \cite{2009arXiv0906.2449L}
also suggest that conduction may play an important role in the
impulsive phase for a flare with a substantial low-energy component.
The heated atmosphere can radiate or conduct away the energy, and can
also expand upwards and downwards\index{chromospheric evaporation!gentle}\index{chromospheric evaporation!explosive}\index{flare models!radiation hydrodynamic}\index{radiation hydrodynamics}.
Whether this evaporation is
gentle or explosive depends on the energy deposition rate by
accelerated electrons as treated in the 1-D~radiation hydrodynamics
calculations
\citep{1985ApJ...289..434F,1985ApJ...289..425F,1985ApJ...289..414F,1999ApJ...521..906A}.
For energy input rates of less than $\sim$3$\times\ 10^{10}$~erg~cm$^{-2}$~s$^{-1}$, 
theory suggests that the evaporation
is gentle,  with upward plasma flows at several tens of kilometers
per second. 
Gentle evaporation can also be conductively driven. At
high non-thermal electron rates ($>3 \times 10^{10}$
erg~cm$^{-2}$~s$^{-1}$), the chromosphere is unable to radiate at
a sufficient rate and consequently expands rapidly. 
This condition
is met when the heating time-scale\index{chromospheric heating!in flares} 
is less than the hydrodynamic
expansion time-scale: \begin{equation} \frac{3kT}{Q} > \frac{L_0}{c_s}
\end{equation} where $Q$ is the flare heating rate per particle,
$T$ is the final temperature of the heated plasma, $c_s$ is the
corresponding sound speed, and $L_0$ is the length-scale of the
flaring region.  If this condition holds, the heated chromospheric
plasma expands upward at hundreds of km~s$^{-1}$ in a process known
as ``explosive'' evaporation. The overpressure of the flare plasma
relative to the underlying chromosphere causes cooler, more dense
material to recoil downward at tens of km~s$^{-1}$ (known as
``chromospheric condensation'')\index{chromospheric condensation}.
In addition to the magnitude of
the energy deposition rate, \cite{1985ApJ...289..434F} also stipulate
that the direction of flows in the transition region/upper chromosphere
determines whether the evaporation is gentle or explosive.\index{transition region!and evaporation}

\begin{figure}      
 \begin{center}
 \includegraphics[angle=90,width=0.8\textwidth]{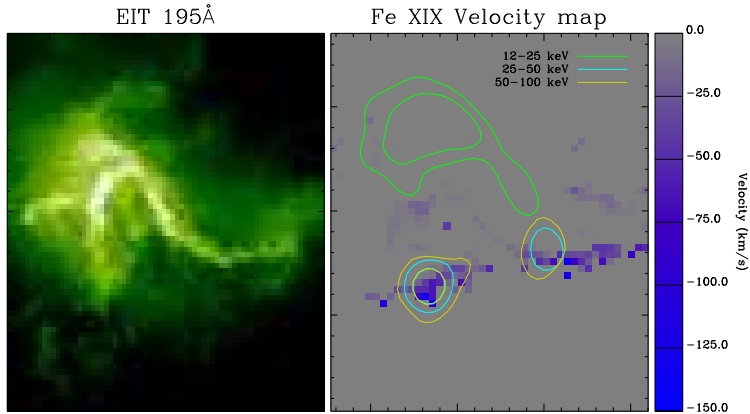}
 \end{center}
 \caption {\label{fig:explos_evap} {\it Left:} EIT image of a flare
 observed by \textit{SOHO}/EIT showing a loop-like structure and possibly
 an extended EUV ribbon to the west.  {\it Right:} the event was
 also observed by \textit{RHESSI} and the \textit{SOHO}\index{SOHO@\textit{SOHO}!CDS}
 Coronal Diagnostic Spectrometer.
 The \textit{RHESSI} high energy footpoints (yellow contours) coincide with
 regions of upflow in Fe~{\sc xix} of formation temperature  ($\log{\rm
 T_e} =  6.9$) , with speeds of up to 150~km$^{-1}$~s$^{-1}$. This provides
 evidence for explosive evaporation driven by electron heating.  }
 \end{figure}
 
\bigskip\noindent{\bf Explosive Evaporation:}\index{chromospheric evaporation!explosive}
 \index{chromospheric evaporation!explosive!illustration}
Spatially-resolved spectroscopic EUV observations of explosive
upflows\index{upflows!explosive} during the impulsive phase of a flare are relatively rare,
since they require the spectrometer slit to be located at the flare
footpoints exactly at the time of strong energy deposition.  
However, this was managed in SOL2003-06-10T14:36 (M2.2).  
\index{flare (individual)!SOL2003-06-10T14:36 (M2.2)} 
\textit{SOHO}/CDS was used to detect simultaneous
strongly blueshifted ($\sim$250~km~s$^{-1}$) Fe~{\sc xix}
emission \citep[peak formation temperature $\log{\rm T_e} =  6.9$;][]{1998A&AS..133..403M} from
footpoints, supportive of explosive evaporation, along with weakly
redshifted He~{\sc i} ($\log{\rm T_e} =  4.5$) and O~{\sc v} ($\log{\rm T_e} =  5.4$) emission \citep{2006ApJ...638L.117M}.
Coordinated \textit{RHESSI} imaging confirmed that the origin of these
flows was at the flare HXR footpoints (see, for example,
Fig~\ref{fig:explos_evap}). The combination of images and spectra
from \textit{RHESSI} also allows estimates to be made of the energy flux
contained in the non-thermal electrons (in erg~cm$^{-2}$~s$^{-1}$)
in order to make a direct comparison with the predictions of theory,
under the assumption of the collisional thick-target model.\index{thick-target model!collisional}
Subject
to uncertainties in the area of the unresolved HXR footpoints, this
event revealed an electron energy flux greater than
$4\times10^{10}$~erg~cm$^{-2}$~s$^{-1}$ in agreement with the
predictions of \cite{1985ApJ...289..434F}.

Other observations without HXRs have instead used the
time derivative of the \textit{GOES} flux (assuming the Neupert effect) to
identify the impulsive phase
\citep{2006A&A...455.1123T,2006SoPh..234...95Z}\index{Neupert effect!use in place of HXRs}.
Spectroscopy from CDS
at high time cadence (10~s) of a ``flare-like transient'' has also
been obtained \citep{2007ApJ...659L..73B}\index{Neupert effect}.
All authors report line profiles in hot lines (e.g., Si~{\sc{xii}}
with peak formation temperature $\log{\rm T_e} =  6.3$,  Fe~{\sc {xvi}}
at  $\log{\rm T_e} =  6.4$ and  Fe~{\sc{xix}} at $\log{\rm T_e} =
6.9$) consistent with upflows\index{upflows} on the order of 100-200 $\rm{km~s}^{-1}$.
In the transition lines of O~{\sc{v}} at $\log{\rm T_e} =  5.4$
and He~{\sc{i}} at $\log{\rm T_e} =  4.5$ downflows of a few tens
of $\rm{km~s}^{-1}$ were most commonly observed, though upflows are
also reported \citep{2006SoPh..234...95Z}.  \index{downflows!evaporative}
Recent observations also confirm the cospatial downflows in H$\alpha$~spectra
and upflows in Ca~{\sc xix} \citep[first observed
by][]{1994ApJ...424..459W} of a few $\rm{km~s}^{-1}$
\citep{2006A&A...455.1123T}\index{momentum balance}.
These can be used to determine the
expected momentum balance between the evaporated and condensing
material \citep{1987Natur.326..165C}.

\bigskip\noindent{\bf Gentle Evaporation}:\index{chromospheric evaporation!gentle}
Plasma flows attributed to gentle evaporation driven by thermal
conduction from the overlying hot corona have frequently been
observed during the gradual/decay phase of flares, after the
non-thermal beam heating has ceased and upflows\index{upflows!and beam heating} 
are sustained by the thermal conduction fronts set up by the steep temperature
gradients \citep[e.g.,][]{1987ApJ...317..956S,1988ApJ...329..456Z,
1999ApJ...521L..75C,2001ApJ...552..849C}\index{conduction fronts}.
However, gentle evaporation due
to a weak non-thermal electron flux has only recently been observed
by \textit{SOHO}/CDS in conjunction with \textit{RHESSI} in SOL2002-07-15T11:55 (C9.1) \index{flare (individual)!SOL2002-07-15T11:55 (C9.1)}
\citep{2006ApJ...642L.169M}. 
Doppler shifts of lines formed at a
range of temperatures showed upflows\index{upflows!temperature dependence} 
of~\lapprox100 km~s$^{-1}$ at all temperatures, consistent with gentle evaporation. 
The upflow
velocity of the Fe~{\sc xix} material was a factor of two higher
when the electron energy flux was an order of magnitude greater
\citep{2006ApJ...638L.117M} (see Figure~\ref{milligan_evap_fig}).
The absence of any redshifted lines supports the hypothesis that
only a large flux of electrons is capable of driving the downflows
in the transition region and upper chromosphere associated with
explosive evaporation.

\begin{figure}[!t]
  \begin{center}
          \includegraphics[angle=90,width=0.8\textwidth]{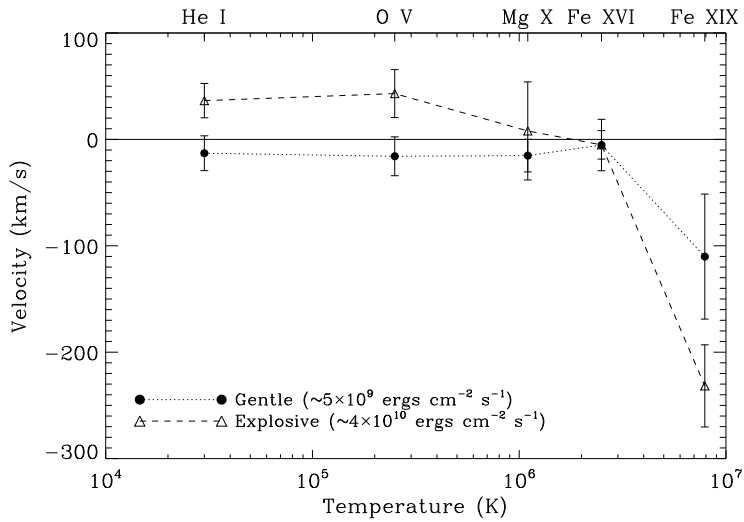}
  \end{center}
  \caption{ \label{milligan_evap_fig} Plasma velocity as a function
  of temperature for each of the five
emission lines observed using CDS during the impulsive phases of
two flares, plotted against their characteristic temperatures.
Positive velocities indicate downflows (redshifts), while negative
values indicate upflows.  The data points plotted with filled circles
denote the case of gentle evaporation, while the open triangles
illustrate the case of explosive evaporation. 
From \cite{2006ApJ...642L.169M}. } 
\index{chromospheric evaporation!illustration}
\end{figure}

The improved spatial, spectral and temporal capabilities of the
Extreme Ultraviolet Imaging Spectrometer (EIS)\index{satellites!Hinode@\textit{Hinode}!EIS} onboard \textit{Hinode} has
also been used to study chromospheric evaporation using emission
lines formed over a broad range of temperatures
\citep{2009ApJ...699..968M}.  
During the impulsive phase of SOL2007-12-14T14:16 (C1.1), 
\index{flare (individual)!SOL2007-12-14T14:16 (C1.1)!blueshifts} 
\index{chromospheric evaporation!temperature-dependent flows}
blueshifted emission (coincident with \textit{RHESSI} HXR emission) 
was observed in six emission lines (Fe~{\sc xiv--xxiv})
formed over the temperature range $\log{\rm T_e}$ = 6.3-7.2. 
These upflows were found to scale with temperature 
over the range 8-18~MK, reaching speeds of $>$250~km~s$^{-1}$ in the Fe~{\sc
xxiv} line.  
This dependence on temperature exists as chromospheric
material, heated to a range of different temperatures by a distribution
of electron energies, will be subject to different pressure gradients
relative to the overlying corona and therefore rise at different
rates. A new finding was that cospatial material formed at temperatures
from $\log{\rm T_e} =  4.7-6.2$ was redshifted by several tens
of km~s$^{-1}$.

At this time the origin of these higher-temperature downflows is
unclear and presents a challenge for current evaporation models.
\index{downflows!high temperature}\index{spectrum!electrons!and efficiency of evaporation}
They may be related to the
spectrum of the electron beam, with a soft spectrum resulting in
energy deposition higher in the atmosphere (and downflowing
transition-region plasma) and a hard spectrum resulting in energy
deposition low in the atmosphere (leading to upflowing transition
region plasma).  
However, a recent model by \cite{2009arXiv0906.2449L}
suggests that this may be a result of sustained chromospheric
heating\index{chromospheric heating}, rather than a single heating burst as is used in most
modeling.

\bigskip\noindent{\bf Imaging of Evaporation}:\index{chromospheric
evaporation!imaging} 
Direct imaging observations of the expected
fronts of multi-million K upflowing plasma in the act of filling
the loops are rather hard to come by \citep{1996ApJ...459..823D}.
One reported observation with {\it{Yohkoh}}/SXT \citep{1997ApJ...481..978S}
has features suggestive of upwards plasma flows at the rather slow
speed of around 60~km~s$^{-1}$  in the impulsive phase.
Evidence for evaporative upflows has been claimed in \textit{RHESSI}
observations of SOL2003-11-13T05:01 
\index{flare (individual)!SOL2003-11-13T05:01 (M1.6)!evaporation} 
\citep{2006ApJ...649.1124L}.
This event has a pair of footpoint/loop leg sources in the range
12 to 30~keV  (see Figure~\ref{fig:LiuW2006_evapor_fig2}), which
converge into a single source near the center of the loop at a speed
of some hundreds of $\rm{km\ s}^{-1}$, perhaps as much as $10^3\
\rm{km\ s}^{-1}$. The emission centroids in this event shift
systematically toward the footpoints with increasing energies up
to $\sim$70 keV, and the upward source motion occurs first at low
energies and progresses to higher energies\index{hard X-rays!coronal thick target}\index{coronal thick target}.
The source motion is
thus also consistent with the behavior expected in a coronal thick
target (see Section~\ref{sec:thick}).  
This is accompanied by an
increase in density at the looptop\index{looptop sources} (see also \cite{2008PASJ...60..835J}
who observe the appearance of a coronal loop at high energies).

 \begin{figure}      
 \begin{center}
 \includegraphics[width=0.7\textwidth]{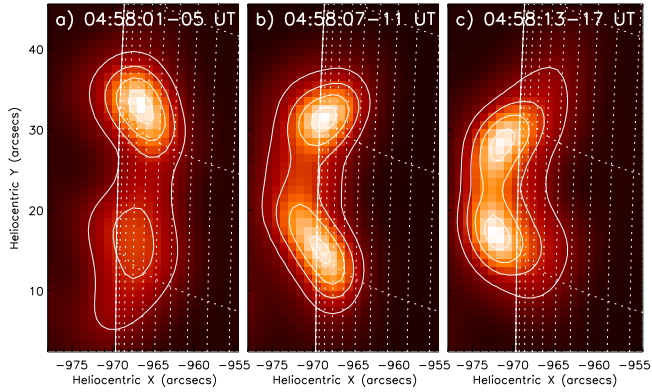}
 \end{center}
 \caption[]{\label{fig:LiuW2006_evapor_fig2} \textit{RHESSI} hard X-ray
 images of SOL2003-11-13T05:01 (M1.6) at 12-15 keV integrated
 over three 4~s time intervals in sequence. The contours are at
 30\%, 50\%, 70\% and 90\% of the image maximum.  The two sources
 rapidly move from the footpoints toward the looptop. This is
 consistent with the flows of hot plasma in  chromospheric evaporation,
 and the Neupert effect is present in this flare as expected
 \citep[adapted from][]{2006ApJ...649.1124L}. 
\index{flare (individual)!SOL2003-11-13T05:01 (M1.6)!illustration} 
} 
\end{figure}
 \index{Neupert effect!illustration}

Finally, a recent observation of a weak B1.7-class flare using
\textit{RHESSI} and \textit{Hinode}/EIS found hot ($\sim$2~MK) upflowing
plasma at one footpoint, but hot downflowing plasma at the other
\citep{2008ApJ...680L.157M}, whereas both theory and previous
observations at lower spatial resolution have only shown upflows\index{upflows}
at this temperature.  An interpretation in terms of a siphon-like
flow is inconsistent with the apparent filling of the flare loop
observed with {\it{Hinode}}/XRT\index{satellites!Hinode@\textit{Hinode}!XRT}.

\section{Coronal Sources}\index{hard X-rays!coronal sources}
\index{coronal sources}

The presence of coronal hard X-ray sources was first inferred in
disk-occulted events (flares with HXR footpoints behind the solar
limb) using data from HXR spectrometers on the OSO-5 and OSO-7
satellites.\index{satellites!OSO-5@\textit{OSO-5}}\index{satellites!OSO-7@\textit{OSO-7}} \citep{1971ApJ...165..655F,1978ApJ...224..235H}
The radial height above the likely active regions in these cases could
be estimated at  more than 2~$\times$~10$^4$~km.  This placed them
above the mean heights of flare SXR sources
\citep{1973ApJ...185..335C,1981ApJ...244L.157V}.  Similar HXR source
altitudes were inferred using  observations from multiple spacecraft
\citep{1979ApJ...233L.151K}.  The earliest HXR imaging (from
\textit{Hinotori}) directly showed coronal source energies up to
at least 25~keV \citep{1983ApJ...270L..83T}.

The launch of the \emph{Yohkoh} satellite led to several further
reports of coronal emission. 
\index{satellites!Yohkoh@\textit{Yohkoh}}
Observations of an impulsive, and by
implication, non-thermal coronal component, with energies up to the
33-53~keV band, were first made by using \emph{Yohkoh}/HXT
\citep{1994Natur.371..495M}. 
More {\textit{Yohkoh}}/HXT coronal
sources were subsequently found, though few with quite the same
remarkable properties as the ``Masuda flare.''\index{Masuda flare}
HXR coronal sources
were demonstrated to exhibit both gradual and impulsive HXR characters,
with the impulsive spikes being more energetic and having harder
spectra \citep{2001A&A...366..294T}.

The \textit{RHESSI} observations have substantially added to the literature
on coronal HXR sources, and it is one area in which \textit{RHESSI} has made
a tremendous impact. 
This body of new observational work has recently
been summarized, along with an overview of the theoretical ideas,
by \cite{2008A&ARv..16..155K}. 
This paper should be consulted for
more detail on this rich and relatively new field than can be
provided below.

\subsection{Thick-target looptop sources}\label{sec:thick}
\index{looptop sources!thick target}

\begin{figure}    
\begin{center}
          \includegraphics[width=0.7\textwidth]{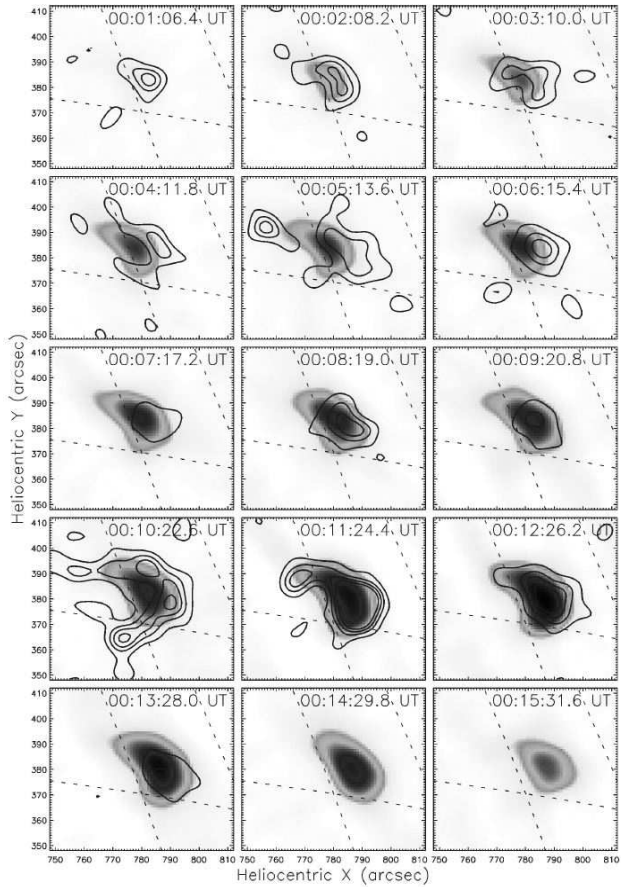}
\caption{\label{fig:bone_veronigbrown} \textit{RHESSI} images of a thick-target
looptop source in SOL2002-04-15T00:15 (M3.7) \citep{2004ApJ...603L.117V}:
images 6-12~keV, contours 25-50~keV with minimum level 0.17 times
the maximum in each frame.  The HXR contours show little evidence
for footpoint brightening.}
\end{center}  
\end{figure}
\index{flare (individual)!SOL2002-04-15T00:15 (M3.7)!illustration} 
\index{coronal sources!thick-target!illustration}

Coronal densities are generally too low to produce HXR
emission efficiently, which is why footpoint sources normally dominate the
images\index{Masuda flare}.
The observation of the Masuda source inspired the suggestion
that it could be explained by a partially thick coronal target
\citep{1995SoPh..158..283W}.  Subsequent observations with \textit{RHESSI}
\citep{2004ApJ...603L.117V} uncovered a new type of coronal HXR
emission embedded in coronal loops, rather than ``above the loop
top'' as in the Masuda source.  Several events could be interpreted
as ``coronal thick target'' sources (Figure~\ref{fig:bone_veronigbrown}),
identifiable by their lack of compact footpoint HXR sources.  In
the first case, SOL2002-04-15T00:15 (M3.7)\index{flare (individual)!SOL2002-04-15T00:15 (M3.7)!coronal thick target}, the 
spectrum observed with \textit{RHESSI} was rather steep ($\gamma \geq\ 6 $) and the
column density N$_{loop}$, estimated using \textit{GOES}, was high ($N_{loop}\
\geq\  10^{20}\ $cm$^{-2}$), leading to the interpretation that
plasma in the corona was dense enough to act as a thick target.
This can happen when the stopping energy $E_{loop}={\sqrt{3KN}}\approx 8.8\sqrt{N_{19}}$~keV
is greater than the energy of an electron, where N$_{19}$ is the
column density in 10$^{19}$~cm$^2$.  A 25~keV electron has a range
(stopping column density)  of  about $10^{20}$cm$^{-2}$
\citep{1973SoPh...31..143B}\index{coronal thick target}\index{hard X-rays!coronal thick target}.
In the Veronig-Brown flare such a
column density appears already to have been present within the loop
at the onset of the flare, and during the flare it increased to
several times $10^{20}$~cm$^{-2}$ which allowed electrons up to
50~keV to be fully stopped.

A high pre-flare coronal density\index{flares!pre-flare coronal density} is puzzling, because if the material is at coronal temperatures and static, the resulting high equilibrium
pressure will make it a bright X-ray source.  This consideration
would not apply to a prominence or a loop structure at intermediate
temperatures.  Having a high density at the start of a flare is
also a difficult  problem for the standard flare model, which
envisions the opening of the field prior to the reconnection.\index{flare models!CSHKP}\index{CSHKP}
\index{reconnection!standard flare model}
An  earlier
flare in the same region could  create an enhanced coronal column
density by evaporation in a loop that subsequently flares again
\citep{2005ApJ...621..482V}, or, similarly, slow pre-flare heating
could lead to evaporation into a system which then becomes unstable.
However, at least in the standard 2-D scenario, electrons would
propagate down a different set of field lines from the closed,
post-reconnection loops onto which the plasma has already evaporated,
precluding the scenario in which energetic electrons are accelerated
onto dense loops as a result of subsequent reconnection.  A more
complicated magnetic geometry or a less direct link between
reconnection and acceleration is needed.

The SOL2002-04-15T00:15 (M3.7) ``coronal thick target'' event could also be
observed via Nobeyama microwave imaging at 17~GHz.\index{observatories!Nobeyama}
\index{hard X-rays!coronal thick target!microwaves}\index{coronal thick target!microwave imaging}
These data include a circularly polarized component\index{radio emission!microwaves!circular polarization}\index{polarization!circular}
\citep{2005ApJ...621..482V,BoneEA07}, establishing the presence of
non-thermal electrons, visible in the whole loop through unpolarized
thermal emission and consistent with the high density needed for
the coronal thick target \citep{BoneEA07}. 
A rather non-standard
explanation is that the previously flaring dense loops could become
\index{plasma instabilities!high beta}
unstable due to a high-beta instability resulting in a second flare
\citep{2001ApJ...557..326S}.

The idea of a coronal thick target might extend to a seemingly
separate class of events, namely the ``soft-hard-harder'' coronal
HXR sources discussed below in Section~\ref{sec:coronal_gradual}.
This possibility might require the stable trapping of high-energy
\index{plasma instabilities!loss-cone}
\index{plasma instabilities!loss-cone!and soft-hard-harder pattern}
electrons in a theoretically unstable loss-cone distribution
\citep{1976ApJ...208..595W}.  
Other than high density and magnetic
trapping, turbulence or plasma waves \citep[e.g.,][]
{1977ApJ...211..270B,1996ApJ...461..445M,2004ApJ...610..550P},
generated as a consequence of magnetic reconnection, 
provide an alternative mechanism that could accelerate electrons and, at the
same time, confine them to the region near the looptop\index{acceleration!stochastic}\index{looptop sources}\index{magnetic structures!turbulent particle confinement}.

\subsection{X-ray observations suggesting coronal current sheets}
\index{current sheets!X-ray observations}

\textit{RHESSI} observations revealed the existence of double coronal X-ray sources, 
interpreted in terms of the current sheet expected in the large-scale reconnection model\index{magnetic structures!current sheet!and double coronal sources}\index{soft X-rays!double coronal sources}.

\begin{figure}
\begin{center}
	  \includegraphics[width=0.8\textwidth]{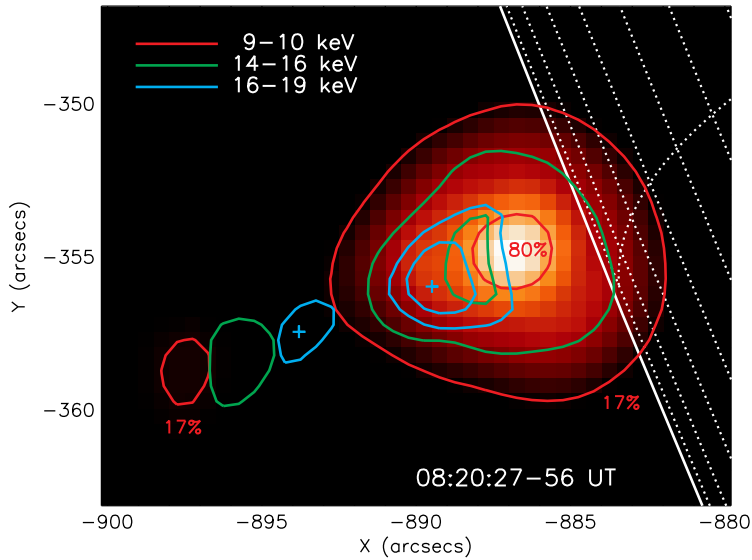}
\caption{\textit{RHESSI} soft X-ray observations \citep{2008ApJ...676..704L}
of a double coronal SXR configuration in SOL2002-04-30T08:22 (M1.3)
suggesting the presence of a current sheet, as originally discovered
by \cite{2003ApJ...596L.251S}.  The separation between coronal and
loop-top sources decreases as a function of photon energy.  The
separation between the 16-19~keV source centroids ($+$~signs) was
found to be 4.6~$\pm$~0.3$''$ in this case.  } \label{fig:liuw}
\end{center} \end{figure} 
\index{flare (individual)!SOL2002-04-30T08:22 (M1.3)!illustration}

In a series of flares that occurred during April 2002, a coronal
X-ray source was observed above the flare loops and was detectable
to about 20~keV \citep{2003ApJ...596L.251S,2004ApJ...612..546S}.
This source was initially stationary before moving outwards at
around 300~km s$^{-1}$ \citep{2003ApJ...596L.251S}.  \textit{RHESSI} images
typically show a height dependence on energy in the flare loops,
with the higher energy X-ray sources located above the lower energy
sources, whereas here the lower-energy sources are located above
the higher-energy sources  (e.g., Figure~\ref{fig:liuw}).  These
observations matched the theoretical expectations for signatures
of a current sheet\index{magnetic structures!current sheet} 
formed between the top of the flare loops and
the coronal source, although the velocity of the coronal source may
be somewhat too low to be consistent with Alfv{\' e}nic ejection speeds
from the upper end of a large-scale current sheet (the same comment
applies to the speed of supra-arcade
downflows).\index{downflows!supra-arcade}\index{flare models!CSHKP}\index{reconnection!standard flare model}\index{CSHKP}
 For one of these events EUV spectroscopic observations were available
 which showed high-speed, high-temperature plasma flows near the
 inferred current sheet; these observations were interpreted as
 reconnection outflows \citep{2007ApJ...661L.207W}\index{reconnection!outflow}.
Similar but less prominent double coronal X-ray sources were reported
by \cite{2006A&A...446..675V} in SOL2003-11-0T309:55 (X3.9) and
by \cite{2007AdSpR..39.1389L} in the occulted flare SOL2002-11-02T06:07 (C3.9)\index{flare (individual)!SOL2003-11-0T309:55 (X3.9)!current sheet}\index{flare (individual)!SOL2002-11-02T06:07 (C3.9)!current sheet}.

\begin{figure}      
 \begin{center}
 \includegraphics[width=0.8\textwidth]{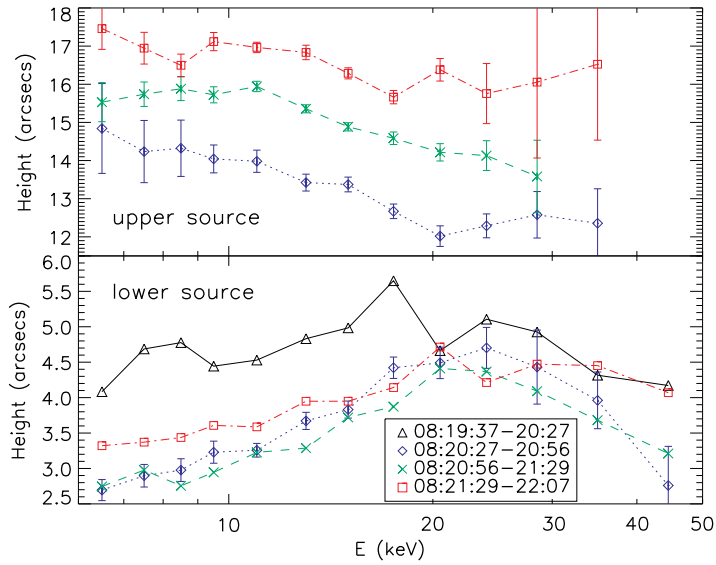}
 \end{center}
 \caption[]{ 
 Height above the limb of the centroids for the upper and lower 
coronal sources in SOL2002-04-30T08:22 (M1.3) 
 plotted as a function of energy for four consecutive time intervals.
 Note the reversal of the trend from low-energy thermal to high-energy non-thermal regime
 \citep[from][]{2008ApJ...676..704L}.
 }  \label{fig:LiuW2008_2LT_fig4}
 \end{figure}
 \index{flare (individual)!SOL2002-04-30T08:22 (M1.3)!illustration} 
 \index{soft X-rays!double coronal source!illustration} 

Imaging spectroscopy of a double coronal source was carried out for
SOL2002-04-30T08:22 (M1.3) (Figure~\ref{fig:LiuW2008_2LT_fig4}; see
Liu et al. 2004).
\nocite{2008ApJ...676..704L}\index{flare (individual)!SOL2002-04-30T08:22 (M1.3)!current sheet}\index{coronal sources!double}
The HXR footpoints were occulted by the limb,
and this facilitated imaging the otherwise relatively faint coronal
sources at energies up to $\sim$40~keV. 
The two coronal sources,
both visible for about 12~minutes, had similar light curves and
power-law spectra above $\sim$20~keV, suggesting production by
similar populations of non-thermal electrons possibly energized by
a common acceleration mechanism.  
At low energies ($\lesssim$20~keV),
both sources were dominated by thermal emission, and the lower
coronal source had a larger emission measure but a lower temperature,
suggesting that the different magnetic connectivity above and below
the current sheet could lead to different plasma densities.  In
addition, the trend of the energy-dependent source structure
\citep{2003ApJ...596L.251S} visible at thermal energies showed a
reversal above $\sim$25~keV (see Figure~\ref{fig:LiuW2008_2LT_fig4}),
with the two sources being further away from each other at higher,
non-thermal  energies. 
A possible explanation is the larger stopping
distances, from the acceleration site, of the higher energy electrons.
The above two properties were also found in SOL2003-04-24T15:53 (C8.2) 
\index{flare (individual)!SOL2003-04-24T15:53 (C8.2)!current sheet} (Liu
et al. 2009; \nocite{2009ApJ...698..632L} see their Figure~5).

\subsection{Early-phase coronal sources}\label{sec:fletcher_coronal_early}

\index{flares!hard X-ray precursor}\index{precursor!hard X-ray}
In its first $\gamma$-ray flare, SOL2002-07-23T00:35 (X4.8), \textit{RHESSI} also discovered
a remarkable new type of coronal HXR source. 
\index{flare (individual)!SOL2002-07-23T00:35 (X4.8)!hard X-ray precursor}
The observations showed
a coronal HXR predecessor to the main impulsive phase of a flare
\citep{2003ApJ...595L..69L,2006PASJ...58L...1A}. The HXR spectrum
of these early coronal sources extends directly down to low energies
($<$10~keV) with power-law indices of around~5, but the spectra
also show the characteristic Fe emission feature at 6.7~keV \citep[e.g.,][]{2004ApJ...605..921P},
establishing that some high-temperature background plasma also
exists in (or near) the source \citep{2010ApJ...725L.161C}. 
While the time evolution of the
thermal component is gradual, the emission at higher energies shows
time variations of tens of seconds' duration suggesting that the two
components are produced by different emission mechanisms.  
X-ray spectral fitting of the high energy component shows that either a
non-thermal model (i.e., broken power-law spectrum) or a multi-thermal
model (temperatures up to $\sim$100~MK are needed) can
represent the spectra well.  
However the microwave observations
\citep{2006PASJ...58L...1A} favor the non-thermal alternative 
\citep[see][]{Chapter5}. 
This interpretation suggests strong
coronal magnetic fields (around 200~G) at relatively high
altitudes ($>$~2~$\times$~10$^4$~km).  
\index{flares!hard X-ray precursor}\index{superhot component}\index{precursor!superhot}

Despite the ambiguous continuum models, the Fe and Fe/Ni line emission can also be used to constrain the thermal plasma parameters\index{flare (individual)!SOL2002-07-23T00:35 (X4.8)!Fe line complex}\index{Fe lines}.\index{spectrum!SXR continuum!and emission lines}
During the impulsive and decay phases of the flare, the fluxes of the two line complexes and their ratio are correlated with the continuum temperature; by assuming that the same relationship holds during this pre-impulsive phase, the observed line fluxes and ratio (which can be accurately measured by \textit{RHESSI}) thus provide upper and lower limits on the temperature and emission measure of the thermal plasma.  
During the peak of the pre-impulsive phase, the line observations constrain the thermal component temperature to be between $\sim$29 and $\sim$37~MK; a cooler component with temperature between $\sim$21 and $\sim$18~MK (respectively) is also required to fit the SXR spectrum.  
Consequently, the low-energy cutoff of the non-thermal electrons is at least as low as $\sim$20 and $\sim$27~keV, respectively \citep{2010ApJ...725L.161C}\index{flare (individual)!SOL2002-07-23T00:35 (X4.8)!magnetic containment}\index{low-energy cutoff}.

Strong coronal magnetic fields are also supported by observations of the 
thermal plasma later in the flare.
\index{soft X-rays!superhot component}\index{superhot component}\index{magnetic field!coronal!and plasma containment}
At the time of peak temperature of SOL2002-07-23T00:35 (X4.8), the thermal energy density of the super-hot component (T $>$50~MK) was $\sim$4800~erg~cm$^3$, suggesting a coronal field strength exceeding $\sim$350~G to contain it.
\index{flare (individual)!SOL2002-07-23T00:35 (X4.8)!superhot component}
We note that a survey of 37 M- and X-class flares shows that strong coronal fields ($>$200 G) are required in all X-class flares, which were also invariably super-hot \citep{Caspi2010} in nature.  
All three HXR-dominated pre-impulsive sources observed by RHESSI were X-class flares, and this suggests an intimate link between the existence of super-hot plasma, strong coronal fields, and the existence of such pre-impulsive HXR sources.

\subsection{High coronal sources}\label{sec:fletcher_coronal_high}

For flares occurring more than about~20$^{\circ}$ behind the solar
limb, the occultation\index{occulted sources} should normally be deep enough to hide not
only the footpoint sources\index{footpoints!occultation}, but also the main flare loops as well
\citep[e.g.,][]{2001A&A...366..294T}. This opens the possibility
of observing emissions from the high corona ($\sim$200~Mm above
flare site), and in fact the early non-imaging observations showed
that such events really do happen
\citep{1971ApJ...165..655F,1978ApJ...224..235H,1982SoPh...75..245H}, if
rarely.  HXR emissions from a flare occulted by 40$^{\circ}$ as
seen from Earth, corresponding to an occultation height of roughly
a third of a solar radius, have been reported \citep{1992ApJ...390..687K}.
Despite this large occultation height, HXR emissions were observed
up to 80~keV, with a rather hard spectrum ($\gamma < 3.5$). 
Another high coronal event observed by {\it Yohkoh} \citep{2001ApJ...561L.211H}
revealed rapid outward source motions with accompanying microwave
emission.
The event morphology suggested filament eruption and and CME occurrence.  
\index{coronal mass ejections (CMEs)!and coronal sources}
An early HXR
stereo\index{stereoscopic observations!HXRs} observation using multiple spacecraft \citep{1979ApJ...233L.151K}
showed that the HXR emissions from the high corona can occur during
the impulsive phase of the flare simultaneously with the HXR footpoint
emissions, and the source size might at the same time be large (of
order 200$''$).
\index{occulted sources!impulsive phase}

The \textit{RHESSI} observations \citep{2007ApJ...669L..49K} of high coronal
events all tend to have similar time profiles.  
They show a fast rise and a slower exponential decay\index{occulted sources!exponential decay}, as illustrated in Figure~\ref{fig:jan20}.  
\index{flare (individual)!SOL2005-01-20T07:01 (X7.1)!coronal hard X-ray source}
The exponential decay is surprisingly
constant, lasting sometimes several minutes without significant
deviation, and the photon spectrum exhibits progressive spectral
hardening \citep[hence the ``soft-hard-harder''
morphology -- ][]{1986ApJ...305..920C,1995ApJ...453..973K,2008ApJ...683.1180G}.
The decay suggests that collisional losses\index{electrons!collisional losses} -- without further
acceleration -- dominate. 
Density estimates
of the ambient plasma support this; these allow estimates of
collisional loss timescales of 25~keV electrons comparable with
those measured.  
\index{electrons!dominant tail population}
While the flare-accelerated electrons in the high
corona are only a small fraction (0.1\%) of the total number of
accelerated electrons in the flare \citep{1992ApJ...390..687K}, the
relative number of energetic electrons ($>$10~keV) in the high
coronal source may be of order 10\% of the thermal electrons\index{electrons!dominant tail population}.

\index{flare (individual)!SOL2005-01-20T07:01 (X7.1)!illustration}
\index{soft-hard-harder}
\begin{figure}[h]
\begin{center}
          \includegraphics[width=0.8\textwidth]{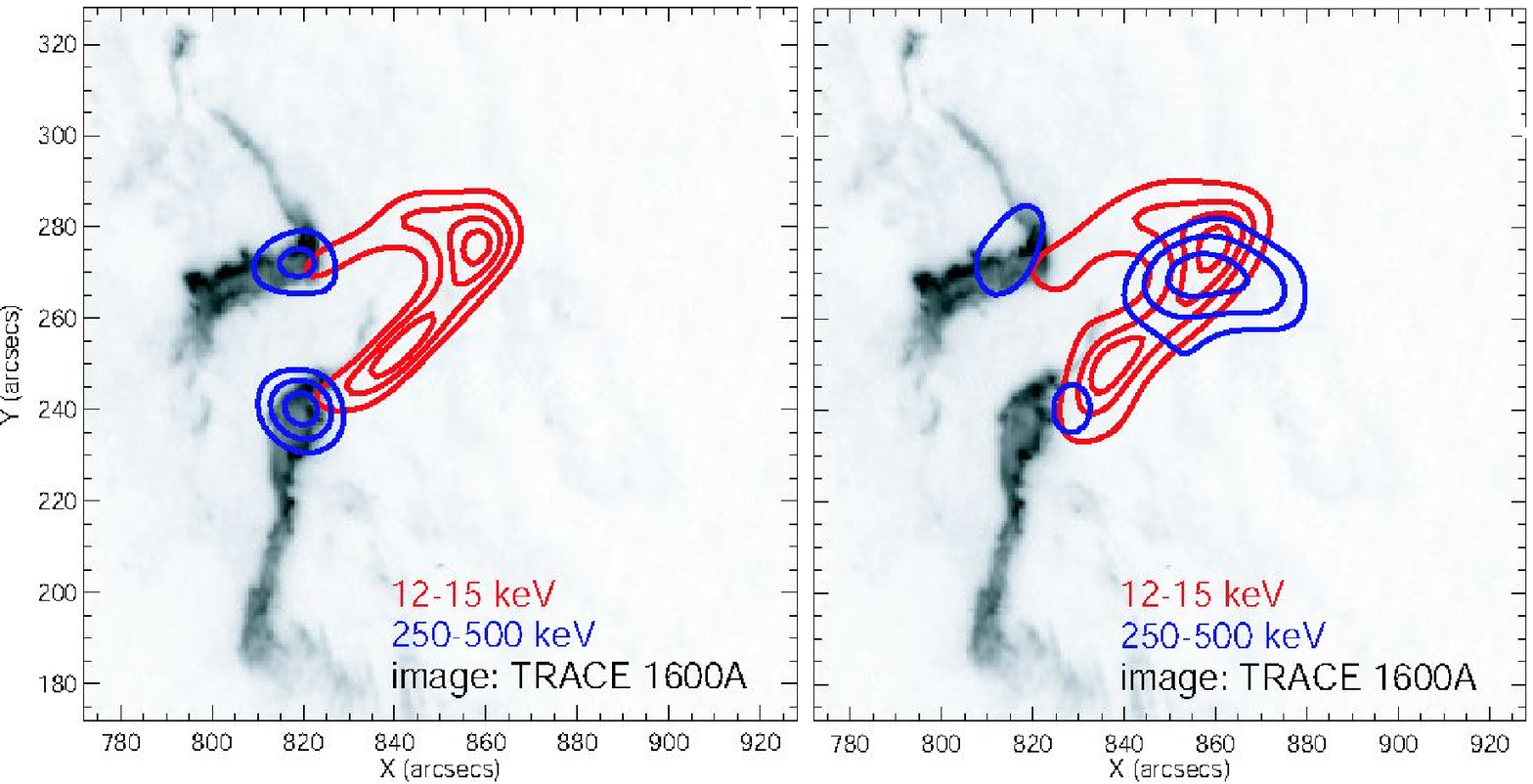}
          \includegraphics[width=0.8\textwidth]{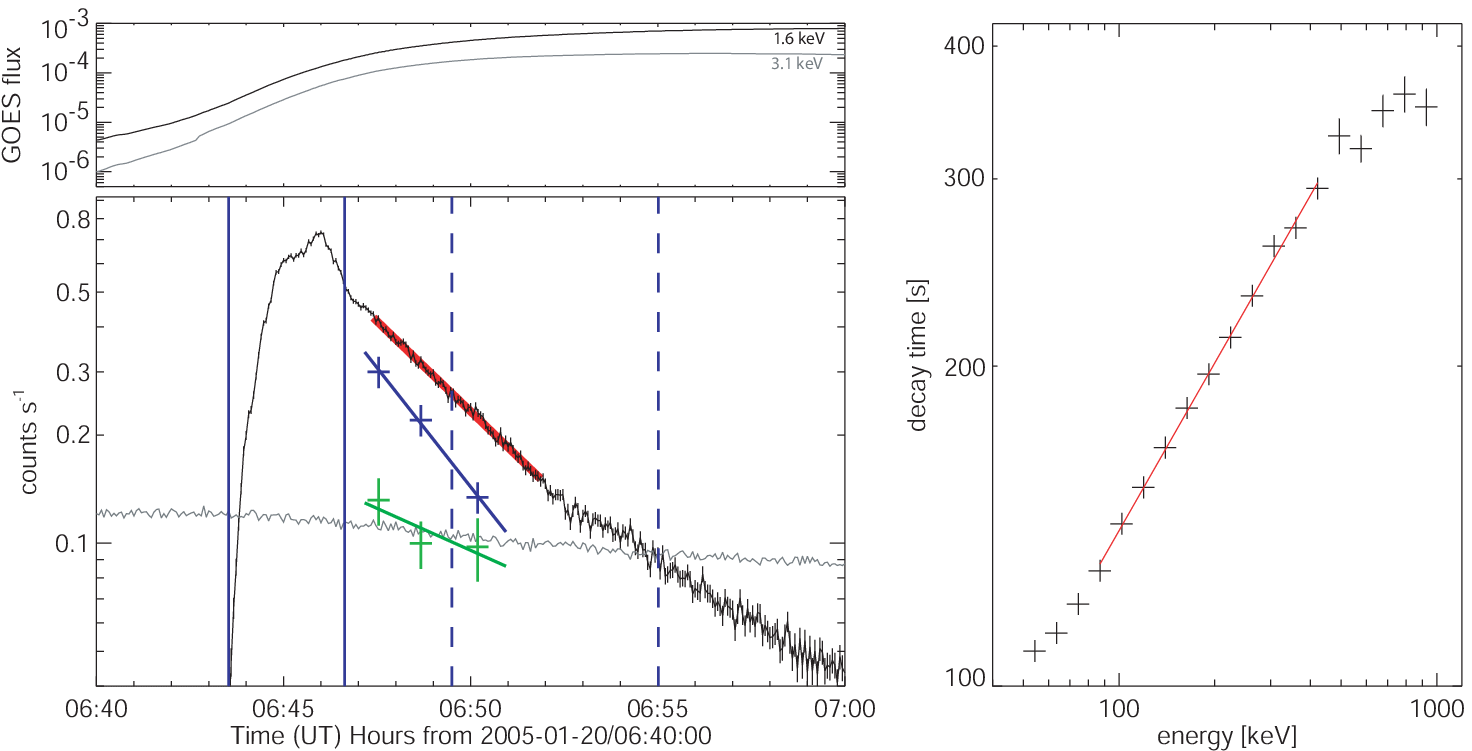}
\caption{\textit{Upper:} \textit{RHESSI} hard X-ray observations at 12-15~keV
(red) and $>$250~keV (blue) from SOL2005-01-20T07:01 (X7.1)
\citep{2008A&ARv..16..155K}, showing a strong high-energy coronal
hard X-ray source.  The background \textit{TRACE} image shows the flare
ribbons as observed in the UV~by \textit{TRACE}, and the contour levels  are
at 30-90\% of the image maxima.  \textit{Lower:} at left, the light
curves, with \textit{GOES} soft X-rays at the top and \textit{RHESSI} hard X-rays at
the bottom.  The latter show three points in the time histories of
the looptop\index{looptop sources} and footpoint sources at  $>$250~keV, while the full
curve shows the total.  The footpoints decay more rapidly.  The
plot at lower right shows the decay time as a function of photon
energy. } 
\label{fig:jan20} 
\end{center} 
\end{figure}
\index{flare (individual)!SOL2005-01-20T07:01 (X7.1)!illustration}
\index{occulted sources!illustration}

\subsection{Gradual late-phase sources}\label{sec:coronal_gradual}

Gradual late-phase sources are characterized by flat HXR spectra
(power-law index $\gamma \approx 2$), gradual time profiles, low
microwave peak frequencies, anomalously weak SXR emission, and
association with coronal radio bursts.
\index{global waves!type II radio burst}
The prototype event SOL1969-03-30T02:47
\index{flare (individual)!SOL1969-03-30T02:47 (pre-\textit{GOES})}
\index{coronal sources}
\index{satellites!OSO-5@\textit{OSO-5}}
\citep{1971ApJ...165..655F} occurred in an active region known
circumstantially to have been some distance behind the solar limb
\citep{1972SoPh...26..460P} so that the HXRs visible from the
Earth-orbiting \textit{OSO-5} spacecraft probably originated from relatively
high in the corona.

A ``soft-hard-harder'' \index{hard X-rays!soft-hard-harder}
pattern of spectral evolution characterizes
many long-duration HXR events
\citep{1986ApJ...305..920C,1995ApJ...453..973K,2008ApJ...683.1180G}\index{flare types!long-duration}.
This pattern differs from the otherwise ubiquitous ``soft-hard-soft''\index{hard X-rays!soft-hard-soft}\index{soft-hard-soft}
pattern associated with the impulsive phase
\citep{1969ApJ...155L.117P,2002svco.conf..261H,2004A&A...426.1093G}.\index{soft-hard-soft}  
From
a non-imaging perspective, the HXR spectrum of such a source consists
of a gradual, continuously hardening component plus a series of
spikes with soft-hard-soft evolution.  
Often these spikes become more gradual as the event develops  
\citep{2008ApJ...673.1169S}; see also 
the early non-imaging observations
from the \textit{TD-1A}\index{satellites!TD1-A@\textit{TD-1A}} spacecraft \citep{1976SoPh...48..197H}.
Figure~\ref{fig:shh} illustrates this\index{hard X-rays!soft-hard-harder}.

The physics of the gradual-phase coronal HXR sources with their
``soft-hard-harder'' temporal development remains to be worked out.
There is every reason at present to suppose that the observed
spectral flattenings and exponential-law time decays can be explained
by some combination of trapping and collisions, but important
theoretical work involving wave-particle interactions, specifically
\index{wave-particle interactions}
\index{plasma instabilities!loss-cone}
loss-cone instabilities, and large-scale magnetic restructurings
remains to be done. A possibly interesting theoretical aspect of
these sources is the idea that the non-thermal particles could
dominate the background plasma component energetically
\index{electrons!dominant tail population}
\citep[e.g.,][]{2001ApJ...561L.211H}.

\begin{figure}    
\begin{center}
          \includegraphics[width=0.95\textwidth]{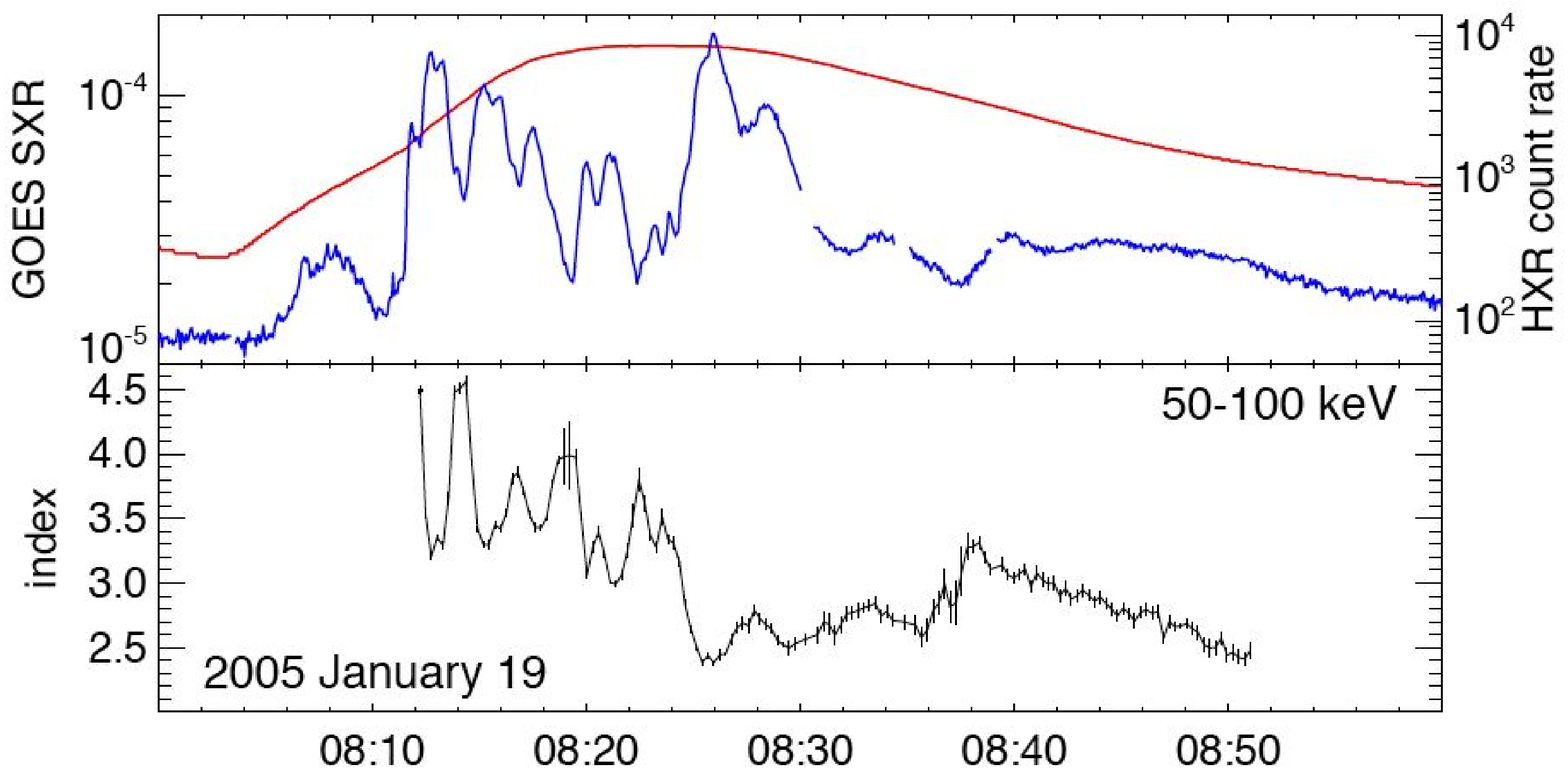}
\caption{Recent illustration of the soft-hard-harder spectral
evolution in the late phase of SOL2005-01-19T08:22 (X1.3)
\citep[adapted from][]{2008ApJ...673.1169S}. Note the anticorrelation
between HXR flux (upper, blue line; the red line shows the \textit{GOES} SXR
flux) and HXR spectral index (lower) at the beginning of the event,
during the more impulsive variations.  Towards the end the variability
has longer time scales and the spectral index systematically
diminishes (gradual hardening).  } 
\label{fig:shh} \end{center}
\end{figure} 
\index{flare (individual)!SOL2005-01-19T08:22 (X1.3)!illustration}
\index{hard X-rays!soft-hard-harder}

Because of their long duration, a thin-target explanation might be
imagined,  such that the HXR emission would come predominantly from
a coronal trap\index{magnetic structures!coronal trap}\index{trapping!magnetic}. 
However it now appears \citep{2004ApJ...603..335Q}
\index{bremsstrahlung!thin-target}
\index{trapping!coronal}
that these late sources emit hard X-rays mainly from footpoints,
at least at energies below 100~keV, as with ordinary flare loops\index{hard X-rays!footpoint sources in gradual events}\index{coronal sources!hard X-ray footpoints}.

\subsection{Looptop source motions}\label{sec:ltmotion}
\index{looptop sources}

\bigskip
\noindent{\bf{Upwards:}} 
A flare arcade gradually develops to larger and larger scales with
time, a process which has long been observed in chromospheric
emission lines such as H$\alpha$~and in SXRs.  
\textit{RHESSI} observes this
phenomenon in a different manner because of its uniquely sensitive
imaging spectroscopy in the 3-20~keV range. Figure~\ref{fig:apr21rise}
shows this graphically for one of the first X-class long-duration
\index{flare types!long-duration}
\index{soft X-rays!long-decay events (LDE)}
flares observed by \textit{RHESSI} \citep{2002SoPh..210..341G}. The ``shrinkage''
\index{magnetic structures!shrinkage}
expected \citep{1987SoPh..108..237S,1996ApJ...459..330F} from the
standard reconnection model is clearly visible, in the sense that
the higher-energy (higher-temperature) X-ray sources, identifiable
with loop tops, lie systematically at higher altitudes than the EUV
sources that they presumably evolve into.  \index{arcade}

\begin{figure}    
\begin{center}
          \includegraphics[angle=0, width=1\textwidth]{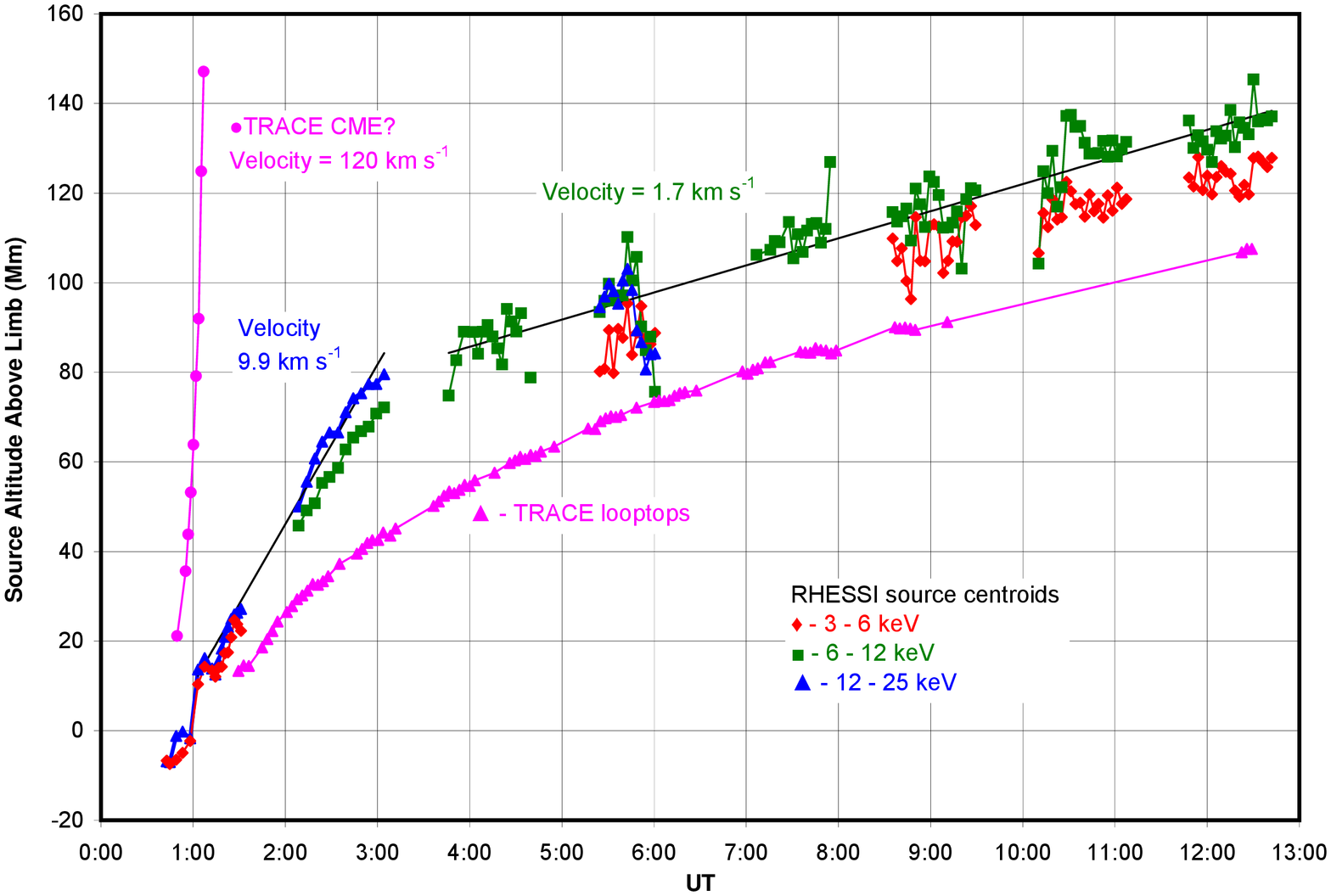}
\caption{Observations of SOL2002-04-21T01:51 (X1.5), an event
at the extreme W~limb \cite[adapted from][]{2002SoPh..210..341G}.
The plot shows apparent altitude above the solar limb as a function
of time
  for features seen in \textit{RHESSI} and \textit{TRACE} images.  The slower motions
  are related to the growth of the arcade, and the \textit{RHESSI} data
  clearly show higher temperatures at larger altitudes.  The rapid
  motions at the beginning are associated with the CME
  \citep{2003ApJ...588L..53G}.  }
\label{fig:apr21rise} \end{center} \end{figure} 
\index{flare (individual)!SOL2002-04-21T01:51 (X1.5)!illustration} 
\index{coronal mass ejections (CMEs)!illustration}
\index{arcade!growth}

\bigskip
\noindent{\bf{Downwards:}}\label{sec:ltdown}
Observations of hot flare loops made by \textit{RHESSI} in X-rays show that
in many cases the loops contract downward during the early, most
explosive part of the flare before the apparent outward expansion
is observed.  The downward motion of the X-ray loop-top centroid
early in the impulsive phase of solar flares has only recently been
recognized.  
\index{magnetic structures!implosion}
\index{coronal sources!contraction}
This new observation may have been missed  with the
\textit{Yohkoh} observations because of coverage biases induced
by the operation of its flare mode, which only initiated hard X-ray
spectral observations at a soft X-ray flux level typically corresponding
to a low C-class flare.  
The first reported observation, during the
rising phase of SOL2002-04-15T00:15 (M3.7)\index{flare (individual)!SOL2002-04-15T00:15 (M3.7)!shrinking loops}, showed shrinking
of the underlying HXR flaring loop at
$\sim$9~km~s$^{-1}$\citep{2003ApJ...596L.251S}. Several further
reported events confirm this pattern
\citep[e.g.,][]{2004ApJ...611L..53L,2004ApJ...612..546S,2006A&A...446..675V,2007SoPh..242..143J}.

In all events, the looptop sources\index{looptop sources} of the flares at higher X-ray
energy bands were located at higher altitudes and showed higher
downward velocities than at lower energies.  For example, in SOL2003-11-03T09:55 (X3.9) \citep{2004ApJ...611L..53L,2006A&A...446..675V}
the mean downward velocities range from 45~km~s$^{-1}$ in the 25-30
keV band to 14~km~s$^{-1}$ in the \textit{RHESSI} 10-15~keV band, and in SXR
observations from the \textit{GOES}  Soft X-ray Imager (SXI)\index{GOES@\textit{GOES}!SXI} 
the looptop
altitude decreased at 12~km~s$^{-1}$, in agreement with the general
trend \citep{2006A&A...446..675V}.  
\index{flare (individual)!SOL2003-11-03T09:55 (X3.9)!illustration} 
This trend of lower speed at lower energies is
also carried through the wavelength range to loops that are both
shrinking and cooling, visible in EUV and H$\alpha$~\citep{2006SoPh..234..273V}.
\index{cooling!and shrinkage}
An interesting phenomenon that follows the energy-dependent looptop
source velocity is the anti-correlation between the HXR flux and
the separation\index{footpoints!separation correlated with HXRs} 
between emission centroids of the looptop sources at
different energies as shown in Figure~\ref{fig:LiuW2004_LT-FP_fig1}
\citep{2004ApJ...611L..53L, 2008ApJ...676..704L}.  
This has been
interpreted as the looptop source being more spatially homogeneous
during the HXR peaks, as a consequence of the rapid energy release.
It is now known that the early downwards motion of coronal HXR
sources is a common characteristic of flares\index{looptop sources!downward motions}.

 \begin{figure}      
 \begin{center}
 \includegraphics[width=0.6\textwidth]{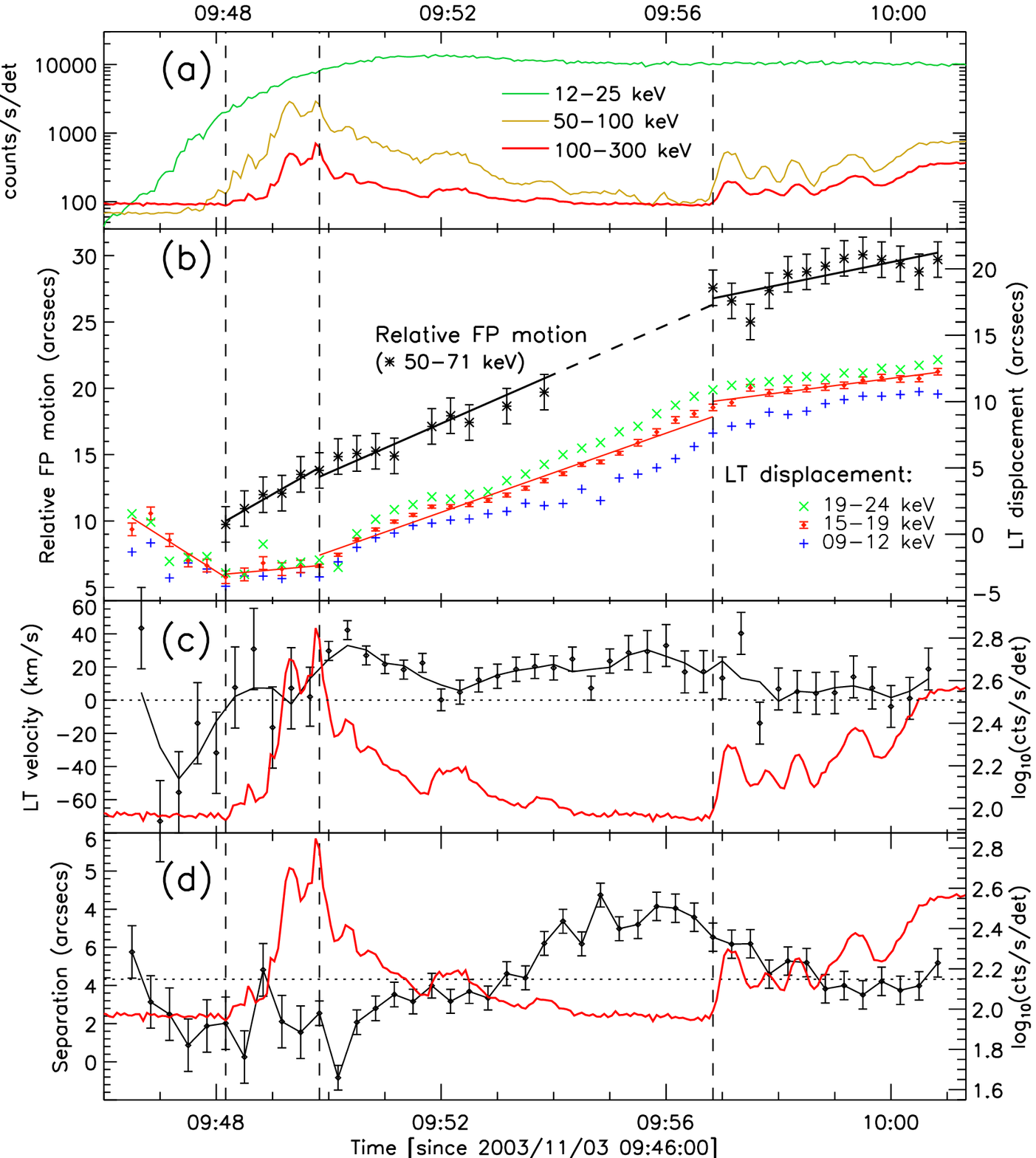}
 \end{center}
 \caption[]{Looptop and footpoint source motions in SOL2003-11-03T09:55 (X3.9) and the energy-dependent source structure.  
 ({\it a})
 {\it RHESSI} light curves.  ({\it b}) Projected height of the
 looptop centroid (right scale) and the separation of the two
 footpoints (left scale), which both increase at comparable speeds.
 Note the early downward looptop motion.  ({\it c}) Velocity of the
 looptop at 15-19~keV, with the values smoothed over 1-minute
 intervals shown as the dark line.  The red curve here and in panel
 {\it d} is the logarithm of the 100-300~keV count rate (right
 scale).  ({\it d}) Separation of the looptop centroids at 19--24~keV 
 and 9-12~keV, which is in anti-correlation with the HXR count
 rate \citep[from][]{2004ApJ...611L..53L}.  
 }
 \index{flare (individual)!SOL2003-11-03T09:55 (X3.9)!illustration}
 \label{fig:LiuW2004_LT-FP_fig1} \end{figure}

The explanation for the converging motion of conjugate footpoints
and the simultaneous descending of looptop sources\index{looptop sources!downward motions} is still in a
preliminary stage of development.  As a basic consequence of the
extraction of excess energy from the magnetic field the contraction
of flaring loops at the initial phase of solar flares should occur
as an ``implosion''
\index{conjugacy}
\index{footpoints!converging motions}
\index{magnetic structures!implosion}
\citep{2000ApJ...531L..75H,2007A&A...472..957J,2009ApJ...696..121L}, a
process consistent with the observations mentioned in the previous
paragraph (see also Section~\ref{sec:fletcher_fpmotions}).
In the framework of a reconnection model with
sheared magnetic field lines, the contraction might be caused by
the relaxation of highly sheared magnetic field after magnetic
reconnection.  
\index{magnetic structures!sheared arcade}
\index{arcade!magnetic structure}
In a  force-free arcade\index{magnetic structures!force-free arcade} in which the magnetic field
strength decreases exponentially with height\index{magnetic field!height dependence}, the dissipation of magnetic energy in a flaring region could lead to a decrease in the
scale height of the magnetic field and thus a shortening of the
field lines  \citep{2007ApJ...660..893J}. 
Therefore, in the initial
phase of flares, the contraction caused by the relaxation of
highly-sheared core magnetic field may dominate over the apparent
expansion of the hot loops which occurs as a result of reconnection
taking place at higher and higher altitudes.  
\index{flares!energy content!magnetic} 
In such a case it is interesting to note that the
remaining shear may indicate stresses in the field remaining even
after the flare has occurred.

\section{Flare relationships to coronal mass ejections, dimmings, and particles}\label{sec:cme}
\index{flares!and coronal dimming}\index{flares!and CMEs}\index{flares!and SEPs}

\subsection{Overview}

The large-scale behavior of the solar corona during a flare or CME\index{coronal mass ejections (CMEs)}
disruption has many observational consequences. Traditionally the
corona itself has only been observable as Thomson-scattered
photospheric light, but well above the photosphere so as to avoid
its glare. Thus most of the low corona, with high density and strong
magnetic field, was unobservable.  Now with SXR and EUV imaging one
can study essentially the same structures via these thermal emissions
(adding to what one can learn from radio techniques; see White et al. 2011,
for further information).
\nocite{Chapter5}

A flare and/or CME disrupts the coronal structure in ways that are
not completely clear yet, and in the process of this large-scale
dynamics there is powerful particle acceleration.  
\index{fables!blind man and elephant}
It is impossible
in this short review to do justice to the vast number of studies
of CMEs both at the Sun and in interplanetary space, and so we focus
here on how the flare, the CME and the related particle acceleration
fit together observationally, a process sometimes likened to the
fable of the blind man and the elephant \citep[e.g.,][where much
more extensive and amusing reviews of the flare/CME relationship
can also be found]{2001JGR...10625199H,2002JASTP..64..231C}.

A flare/CME event marks the conversion of stored magnetic energy
into various other forms that propagate through the solar atmosphere
and into interplanetary space.  Traditionally the brightening is
associated with the ``flare'' and the motions with the ``CME''
\citep{1995SoPh..157..285C}, but the physics of either phenomenon
seems to require both motions and brightenings.  Ultimately, all
of the energy extracted from the magnetic field appears either as
energy associated with the CME, or as enhanced radiative output of
the Sun.  The dominant term of the CME energy appears to be its
kinetic energy, which can be estimated from LASCO images
\citep{2000ApJ...534..456V}. The dominant energy product of a flare,
of course, is the transient excess it makes in the solar luminosity.
We discuss this in detail in Section~\ref{sec:energetics}.  One
immediately evident property of the flare/CME combination is the
comparative sizes of the flaring region and the associated CME,
illustrated quite startlingly in Figure~\ref{fig:flarecme}.

\subsection{Flare energy release and CME dynamics}
\index{flares!energy release and CME dynamics}

\begin{figure}
\begin{center}
\includegraphics[width=0.9\textwidth]{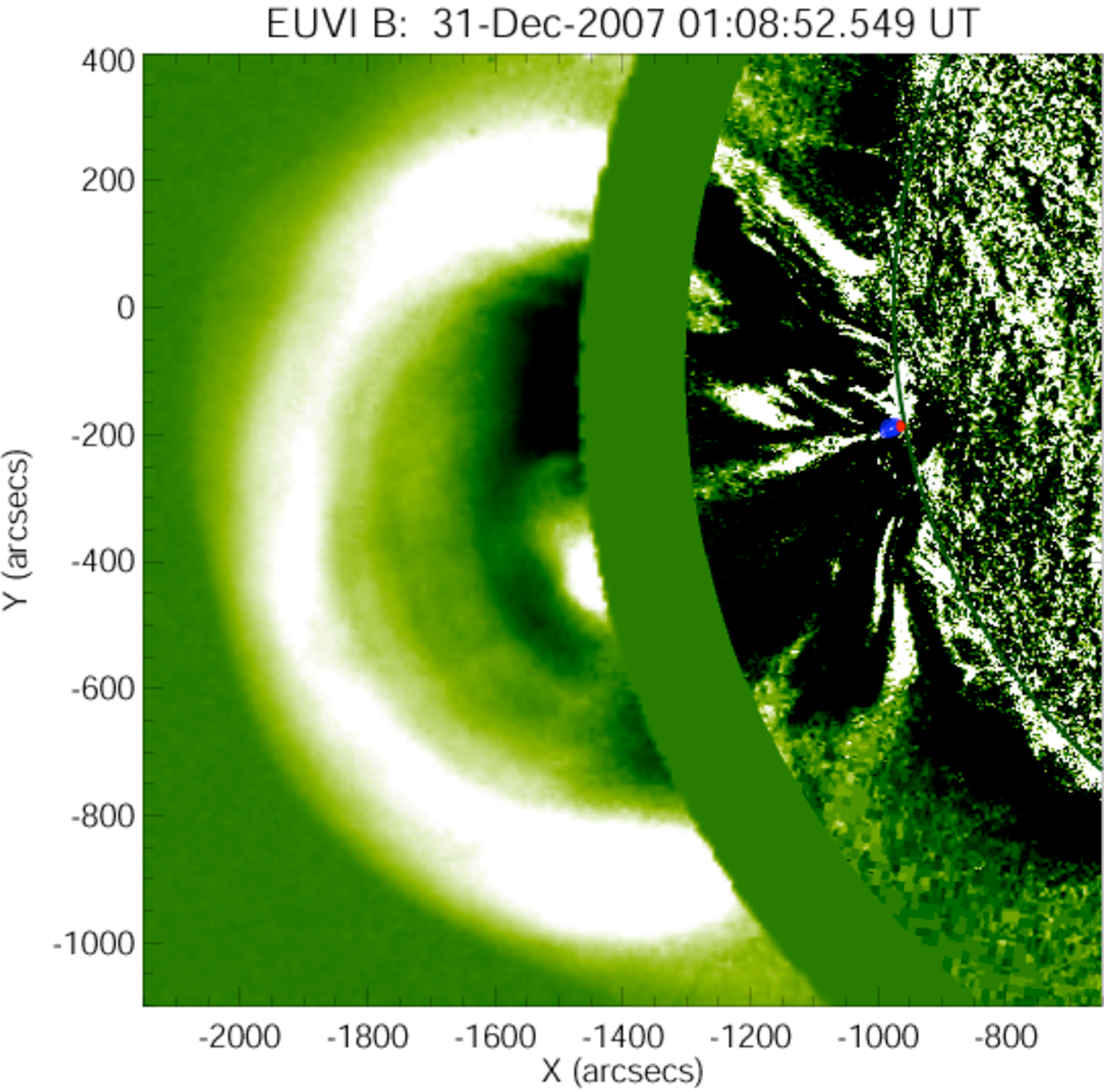}
\caption{\label{fig:flarecme} A combined \textit{RHESSI}/\textit{STEREO} image of SOL2007-12-31T01:11 (C8.3).
\index{flare (individual)!SOL2007-12-31T01:11 (C8.3)!illustration}
It is composed of a difference image of the
solar disk and inner corona, which illustrates well the loss of
coronal plasma referred to as the ``dimming,''  \textit{RHESSI} sources (red
and blue contours) and the CME bright front, dark following cavity
and inner bright core. As is apparent, a CME is a very much more
spatially extensive phenomenon than is the flare (image courtesy S. Krucker).} 
\end{center}
\end{figure}

\index{coronal mass ejections (CMEs)}
From reconnection models it is supposed that the CME kinematics and
the energy release of the associated flare are closely related.
The degree of association between flares and CMEs had always been
problematic, largely  because of the poor coverage of the low corona
provided by coronagraphs.  Almost half of the CMEs, for example,
originate on the far side of the Sun, for which no low-coronal
observations had been available until the advent of \textit{STEREO}.
Nevertheless the X-ray observations of coronal dimmings
\citep{1996ApJ...470..629H} and EUV observations of compact CME
sources \citep{1997SoPh..175..601D} had made it clear that there
was often a very tight relationship between flares and CMEs.
\index{coronal dimmings}

Further studies revealed a close correlation between the CME
acceleration and the derivative of the flare SXR flux, taken as a
proxy for the flare energy release\index{coronal mass ejections (CMEs)!acceleration and flare energy release}
\citep{2001ApJ...559..452Z,2004ApJ...604..420Z,2007SoPh..241...99M}. 
\index{Neupert effect!and CME acceleration phase} 
For
some well-observed CME/flare events, further direct evidence  was
recently provided showing a very close synchronization between the
CME  acceleration and the \textit{RHESSI} HXR flux.  
For example, in the SOL2005-01-17T09:52 (X3.8) flare/CME event  and in SOL2006-07-06T08:36 (M2.5) \citep{2008ApJ...673L..95T},
\index{flare (individual)!SOL2005-01-17T09:52 (X3.8)!CME} 
\index{flare (individual)!SOL2006-07-06T08:36 (M2.5)!CME} 
the use of \textit{GOES}/SXI and \textit{TRACE} running difference images\index{difference images} 
showed that the CME
impulsive acceleration in the low corona and the flare energy release
(deduced from the \textit{RHESSI} HXR flux) are closely synchronized, and
peak simultaneously within $\pm$3~min, i.e., within the CME
measurement uncertainties (see Figure~\ref{fig:veronig_cme}). 
\index{satellites!GOES@\textit{GOES}!SXI}
Such
correlations provide strong evidence that the CME large-scale
acceleration and the flare particle acceleration are intimately
\index{acceleration!and CME acceleration}\index{coronal mass ejections (CMEs)!acceleration}
connected phenomena, reflecting the rapid extraction of energy from
the reconfiguring field both below and above the coronal reconnection
region. The effect of the magnetic boundary conditions (i.e., with
field being line-tied to the photosphere below the coronal reconnection
region, and able to expand relatively freely above) obviously has
a substantial impact on the energetically dominant terms in the
flare versus the CME.

\begin{figure}
  \begin{center}
          \includegraphics[width=0.9\textwidth]{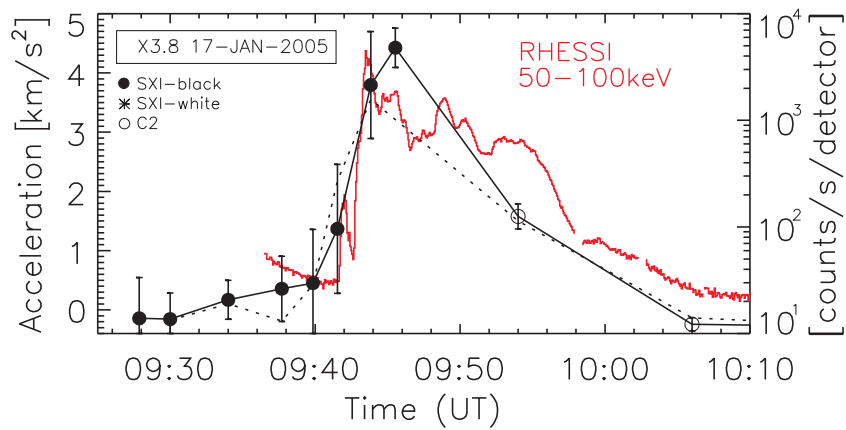}
 \end{center}
  \caption{Flare HXR flux (\textit{RHESSI} 30--100 keV) and CME acceleration
    profile derived from \textit{GOES}/SXI SXR images and \textit{SOHO}/LASCO 
    coronagraph images for SOL2005-01-17T09:52 (X3.8). 
    Note the close synchronization  \citep{2008ApJ...673L..95T}. }\label{fig:veronig_cme}
\end{figure}
\index{flare (individual)!SOL2005-01-17T09:52 (X3.8)!illustration}
\index{satellites!SOHO@\textit{SOHO}!LASCO}

\subsection{Large-scale waves}\label{sec:waves}
\index{global waves}

The restructuring of the large-scale coronal magnetic field implicit
in a flare and a CME can be considered as a magnetic impulse, and
the magnetic field that permeates the surrounding corona and deeper
atmosphere guarantees that large-scale waves will ripple away from
the site~\citep[for a recent review of coronal waves
see][]{2008SoPh..253..215V}.
Because of dispersion, any of
the global waves will shock and dissipate their energy in non-thermal
effects, such as the acceleration of few-keV electrons in a radio
type~II burst\index{global waves!type II radio burst}\index{shocks!type II radio burst}.
The waves may contain large energies, as discussed
in Section~\ref{sec:sep_energy}.  In general an analysis of the
structure and timing of these waves may provide key clues to the
nature of the energy release by helping to define its geometry.

The existence of large-scale coronal shock waves has in fact been
known since the early  interpretation of the ``slow drift'' or
metric type~II radio bursts \citep[e.g.,][]{1963ARA&A...1..291W}.
These were then linked to the Moreton waves, which are observed in
chromospheric H$\alpha$~signatures \citep{1961ApJ...133..935A}, by the
theory of a weak fast-mode MHD shock \citep{1974SoPh...39..431U}.\index{shocks!and Moreton waves}
Figure~\ref{fig:tsunami} shows a well-observed example of a Moreton
wave, which typically travels across the chromosphere at 1000~km~s$^{-1}$\index{radio emission!type II burst}\index{tsunami}\index{Moreton wave}\index{global waves!Moreton}.
Even before this time ``flare ejecta''
or ``driver gas,'' now known as an ICME (for Interplanetary CME),
had been identified as the cause of the ``storm sudden commencement''
geomagnetic effect\index{interplanetary coronal mass ejections (ICMEs)}\index{magnetic storm}\index{storm sudden commencement}.
This is the abrupt onset of a magnetic storm
resulting from the compressive interaction of an interplanetary
shock wave\index{shocks!interplanetary} driven by what we now term an ICME
\citep[e.g.,][]{1964AnAp...27..333C} with the magnetosphere\index{magnetosphere!terrestrial}
\index{global waves!EIT} 
\index{EIT wave} 
\index{Moreton wave} 
\index{radio emission!type II burst}
Now in addition to this well-known coronal and chromospheric evidence
of large-scale waves, we can add the  ``EIT~wave'' observations
\citep{1997SoPh..175..571M,1998GeoRL..25.2465T}.  
These EUV perturbations take the form of an expanding
wave front most clearly visible in difference images. In a small
subset of the EIT~waves one can make an identification with the
Moreton wave/type II phenomenon, but for the most part they have
clearly different properties \citep{2002ApJ...569.1009B}. 
\index{global waves!and X-ray dimming}
Their
interpretation in terms of X-ray, EUV or white light dimming (or
depletion) \cite[e.g.,][]{1996ApJ...470..629H,1997ApJ...491L..55S}
is complicated because of the temperature sensitivity of the EIT
response. The EIT signature is presumably a mixture of true depletion,
simple waves, and large-scale restructurings of the field as required
by the CME. 

\begin{figure}[h]
  \begin{center}
          \includegraphics[width=1\textwidth]{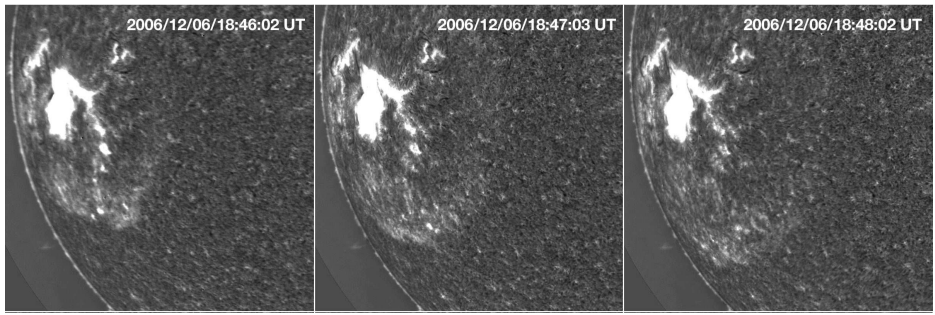}
 \end{center}
  \caption{Moreton wave observed in H$\alpha$~at Sacramento Peak in SOL2006-12-06T18:47 (X6.5).
Courtesy K.~S. Balasubramaniam. }\label{fig:tsunami}
\end{figure}
\index{flare (individual)!SOL2006-12-06T18:47 (X6.5)!illustration} 

The similarity of the radio signatures (metric type~II for the
flare-associated wave, and interplanetary type~II for the CME-driven
wave) have led to much recent  discussion \citep[see][for a recent
comprehensive survey]{2006SSRv..123..341P} regarding the distinction,
if any, between the meter-wave and the interplanetary shock signatures.\index{shocks!spectrogram signature}
The radio spectrograms at long and short wavelengths each have
complex signatures, and have been traditionally made in disjoint
spectral bands. It has thus been common
\citep[e.g.,][]{2004SoPh..225..105C} to speculate that a common
mechanism, specifically the CME bow wave, could explain all of the
large-scale wave observations. Indeed, the coronagraphic observations
show image evidence, in many cases, for CME-driven shocks in the
middle corona \citep{2003ApJ...598.1392V,2009ApJ...693..267O}\index{shocks!CME-driven}.
However there is no clear evidence for continuity in the radio
signatures, using new observations in the 1-14~MHz range that
separates the traditional ground-based and interplanetary observations
\citep{2005ApJ...623.1180C}. The distinction between a flare origin
and a CME origin has also become more difficult to make now that
improved data have established tighter relationships between flares
and CMEs in both point of origin and timing (see
Figure~\ref{fig:veronig_cme}).  The coronagraph data show the
importance of the flanks of the CME-driven wave, both directly and
also indirectly via the excitation of streamers that the flanks
intersect.  
SXR observations may show the earliest signatures of
the large-scale wave disturbance
\citep{2002A&A...383.1018K,2003SoPh..212..121H}.

\subsection{Solar Energetic Particles and particle acceleration}
\index{shocks!particle acceleration}
\index{acceleration}
\index{solar energetic particles (SEPs)}

Shock waves can be efficient accelerators of high-energy particles,
and there is clear (though indirect) evidence for shock acceleration
of SEPs \citep[e.g.,][]{1999SSRv...90..413R}.  Indeed, the energy
ending up in energetic particles can be a substantial fraction of
the CME kinetic energy (see Section~\ref{sec:sep_energy}).  The
shock acceleration of SEPs probably takes place at some distance
from the event origin \citep{1994ApJ...428..837K}.  This would be
consistent with the idea that the shock condition does not develop
immediately, presuming that larger Mach numbers correspond to more
efficient acceleration\index{Alfv{\' e}n speed!middle corona}.
The existence of a minimum in the 
Alfv{\' e}n speed in the middle corona \citep{2003A&A...400..329M} allows
the CME-driven disturbance to attain a higher Mach number even as
its absolute speed may be constant or even decreasing.
Figure~\ref{fig:mann} shows a model view of the coronal Alfv{\' e}n
speed  \citep{2003A&A...400..329M}.

\begin{figure}
  \begin{center}
          \includegraphics[width=0.8\textwidth]{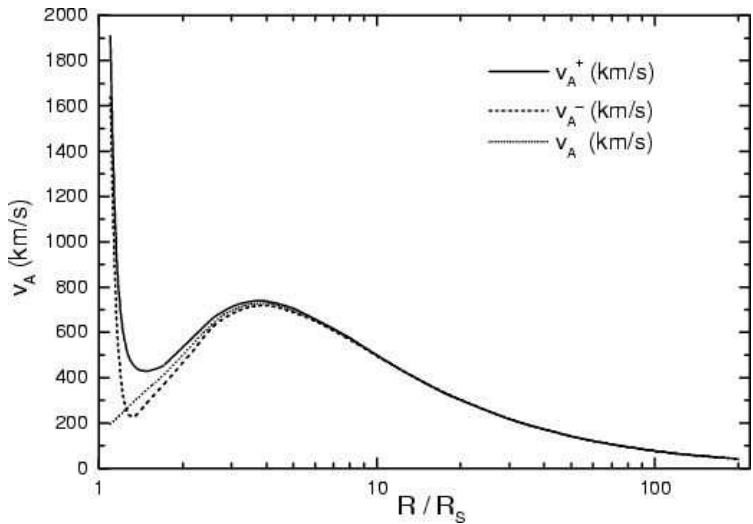}
 \end{center}
  \caption{Model variation of Alfv{\' e}n speed with height in the
  corona, showing the local minimum above an active region
  \citep{2003A&A...400..329M}.  Here the dotted line shows the quiet
  Sun, modeled as a simple dipole at Sun center, and the active
  region as a vertical dipole near the surface. The solid and dashed
  lines show the cases where these two components have parallel or
  antiparallel radial contributions.  } 
\label{fig:mann}
\end{figure}
\index{Alfv{\' e}n speed!variation with height!illustration}

A more directly flare-related acceleration of SEP ions may also
occur, and the fact that SEP occurrence is strongly associated
with a ``soft-hard-harder'' spectral
 evolution \citep{1995ApJ...453..973K,2009ApJ...707.1588G} may
 support this more direct connection.
\index{hard X-rays!soft-hard-harder}
\index{soft-hard-harder}
Solar electron events detected in interplanetary space have a strong
association with metric-decimetric type~III bursts
\citep[e.g.,][]{1970SoPh...15..453L}.  
Here the starting frequency
of the radio emission, taken to be the plasma frequency or its first
harmonic, points to relatively high densities (the lower corona)
consistent with a process physically close to the flare site.
\index{radio emission!type III burst}\index{frequency!starting}

\section{Physical properties of flares}\label{sec:physical}

\subsection{X-ray spectroscopy}
\index{soft X-rays!spectroscopy}

The emission spectra of flares, especially in the SXR range, convey the
most direct information obtainable by remote sensing of the flaring
coronal plasmas.  
\textit{RHESSI} touches on this domain via its capability
for measuring the thermal free-free and free-bound continua, as
well as to detect the K-shell emission lines of highly ionized
Fe~around 6.7~keV\index{soft X-rays!Fe K-shell emission}\index{Fe lines}.
These spectral features appear commonly in a
wide variety of astrophysical sources, such as active galactic
nuclei, stellar flares, and supernova remnants.  
The parameters thus available include the electron temperature $T_e$, the emission
measure $n_e n_i V$, and information about elemental abundances
given adequate models of the plasma physics and the atomic physics.
\textit{RHESSI} data, for example, can determine the abundance ratio Fe/H
from the equivalent width\index{equivalent width}\index{Fe lines!equivalent width} of the Fe-line feature at $\sim$6.7~keV
(a measure of the ratio of the Fe-line flux to the continuum flux
at the Fe-line energy).\index{Fe lines} Note that the Fe~feature
actually consists of many lines in the K-shell energy range, as
described by \cite{2004ApJ...605..921P}.  
The solar Fe~K-shell spectral feature typically does not exhibit the $\sim$6.4~keV
emission of Fe~in low ionization states.\index{Fe lines!low ionization states}\index{ionization state}
This is commonly observed
elsewhere in the Universe, for example, in reflection spectra from
accretion disks, but is largely absent in the solar spectra (in
favor of the $\sim$6.7~keV emission band due to high-temperature
plasmas).

\bigskip
\noindent{\bf{Thermal continua observed by \textit{RHESSI}:}}
\index{continuum!soft X-ray}
The flare thermal spectrum observed by \textit{RHESSI} in the $\sim$3-20~keV energy range consists of free-free (bremsstrahlung) and
free-bound (recombination) continuum emission.
\index{free-free emission}\index{free-bound emission}\index{soft X-rays!continuum}\index{recombination radiation}
The contributions
made by these continua vary with energy and $T_e$. In general,
thermal free-free radiation is predominant at lower energies and
higher temperatures, as was evident in early calculations
\citep{1969MNRAS.144..375C,1978A&AS...32..283G}.
Newer continuum calculations are included in the Chianti\index{Chianti} atomic database and software package \citep{1997A&AS..125..149D, 2003ApJS..144..135Y}, but the
dependence on energy and temperature is nearly the same as the
earlier work.  
There is now, however, the recognition
that the abundances of some elements important for free--bound
emission are enhanced in the corona, giving rise to enhanced
free-bound emission\index{abundances!coronal}.
For flare temperatures between 10 and 20~MK
and with coronal element abundances, the cross-over energy where
free-free and free-bound continua fluxes are equal is at the lower
end of the range that \textit{RHESSI} observes, so both free-free and
free-bound continua are important\index{free-bound emission!thermal} \citep[e.g.,][]{2005SoPh..227..231W}.  
(The two-photon continuum\index{continuum!two-photon}, due to the de-excitation of metastable 
levels in H-like and He-like ions, is much less important.)

\bigskip
\noindent{\bf{The Fe-line and Fe/Ni-line features observed by
\textit{RHESSI}:}
\label{sec:Fe-line}} 
\index{Fe lines} 
As well as the continuous
emission, \textit{RHESSI} observes two line features at $\sim$6.7~keV and
$\sim$8~keV, known as the Fe-line and Fe/Ni-line complexes.  They
are composed of numerous individual spectral lines emitted mainly
by He-like Fe~{\sc xxv} ions and dielectronic satellite lines
emitted by mainly Li-like Fe~{\sc xxiv} and lower Fe ions, with
a small contribution from highly ionized Ni lines to the $\sim
$8~keV feature. 
\index{ionization equilibrium}
In coronal ionization equilibrium, these ions are
expected to be abundant at temperatures above $\sim$10~MK, and in
confirmation of this, the Fe-line feature is evident in \textit{RHESSI}
spectra with $T_e \gapprox 10$~MK.  Both line features are conspicuous
for $T_e~\gapprox~20$~MK.  The spectral resolution of the \textit{RHESSI}
detectors at these energies is $\sim$1~keV FWHM, which is insufficient
for resolving the line structure of the Fe-line and Fe/Ni-line
features.  Nevertheless, \textit{RHESSI} has the advantages of covering a
much broader energy range than previous high-resolution crystal
spectrometers (which have only covered the immediate vicinity of
the $\sim$6.7~keV lines) and by directly observing the continuum
-- some crystal spectrometers have had a strong background due to
crystal fluorescence, which can obscure the true flare continuum.
A valuable diagnostic means for studying the hot component of the
solar flare plasma is thus available \citep{2004ApJ...605..921P}.

Temperature, line equivalent width\index{equivalent width} and abundance\index{abundances} analysis has been carried out for SXRs ($\gapprox $5~keV) spectra taken during 27 flares 
observed by \textit{RHESSI}, with \textit{GOES} class between C3 and X8
\citep{2006ApJ...647.1480P}. 
The measured spectra were fitted with
model spectra consisting of a continuum (isothermal free-free plus
free-bound emission) and lines at $\sim$6.7~keV to characterize
the Fe-line feature and at $\sim$8~keV for the Fe/Ni-line feature.
\index{flare (individual)!SOL2003-04-26T03:06 (M2.1)}
\index{soft X-rays!Fe lines}
Figure~\ref{fig:dennis_spectrum} shows an example of this during
SOL2003-04-26T03:06 (M2.1), at a time when \textit{RHESSI} was in its
single-attenuator state\footnote{\textit{RHESSI} achieves great dynamic range
via two attenuators, which incrementally (in four combinations, but in 
actual practice only three) cut
off the intense low-energy fluxes of major events and also allow
the sensitive detection of weak microflares \citep{2002SoPh..210....3L}.}.\index{microflares}
The temperatures $T_e$ in this analysis
ranges between 11~MK and 29~MK \citep{2006ApJ...647.1480P}.

\begin{figure}
\begin{center}
\includegraphics[bb = 80 50 565 752, height=10cm, angle=90, clip]{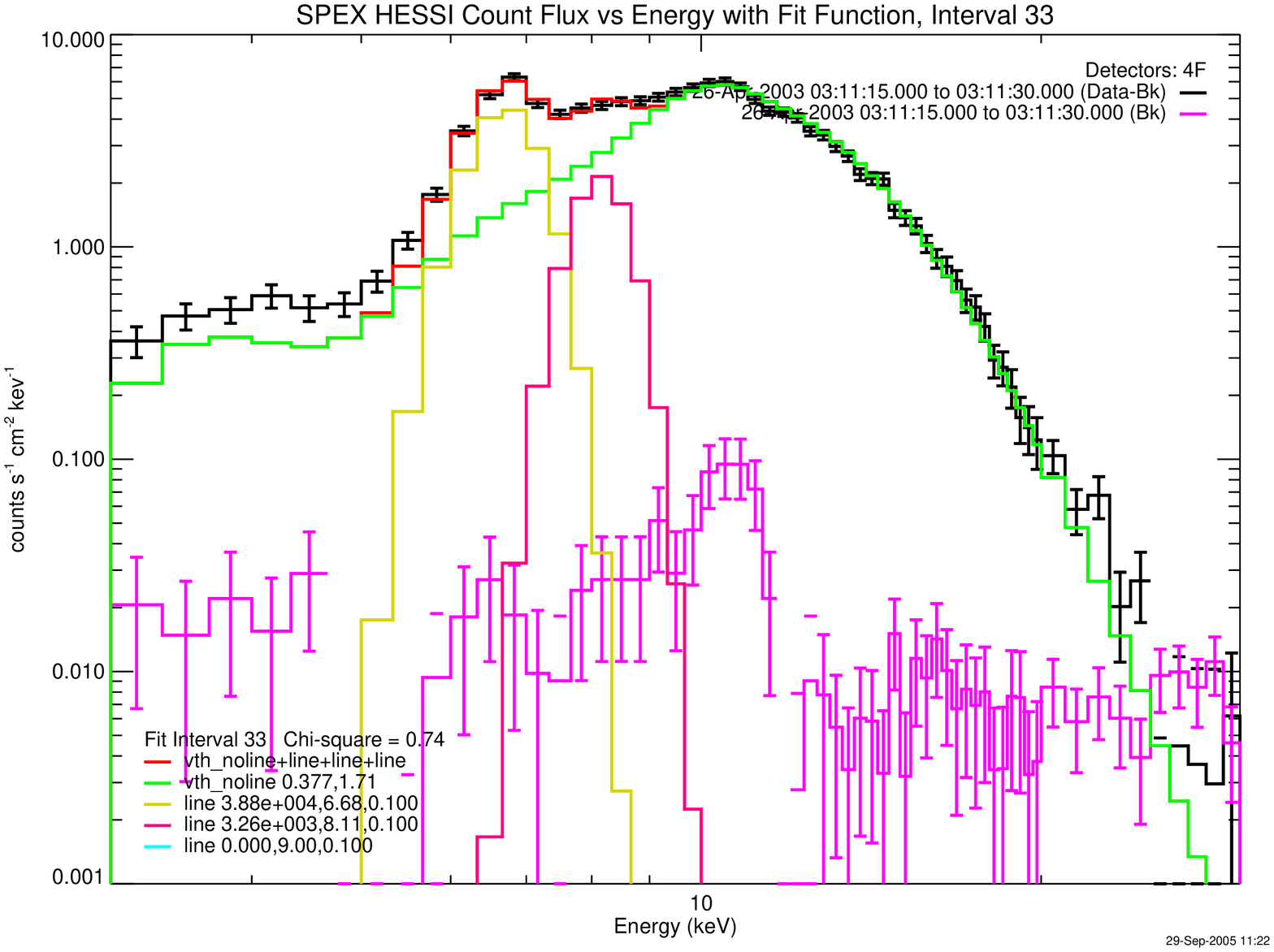}
\includegraphics[width= 10cm]{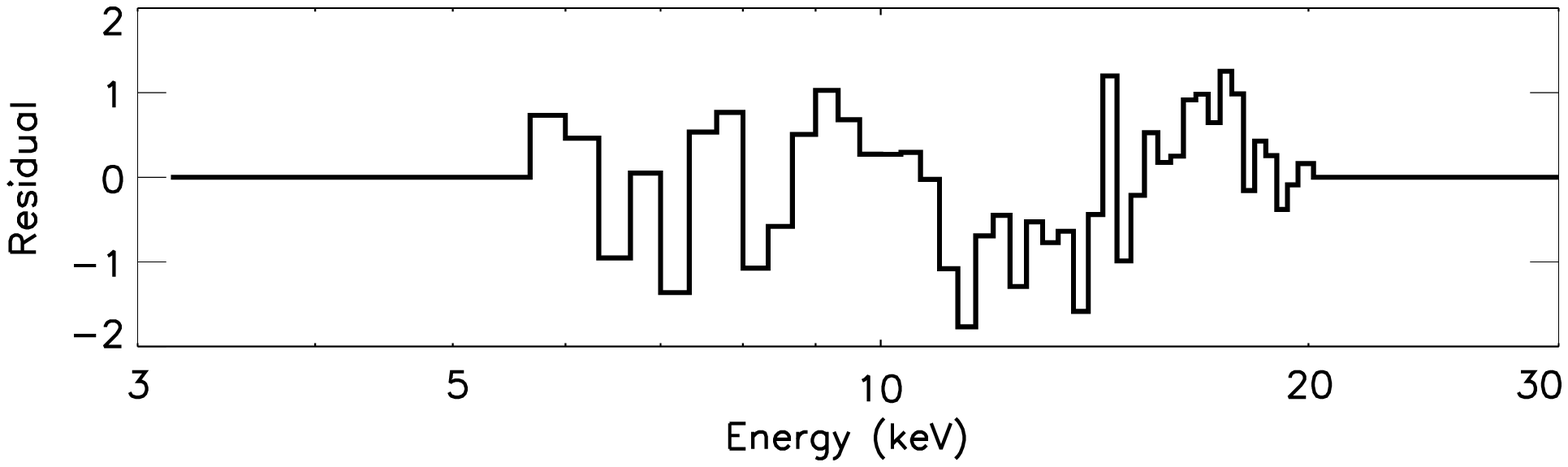}
\caption{\textit{Top:} measured and modeled count rate spectra in
the energy interval 3--30 keV for \textit{RHESSI} detector~4 shortly after
the peak of SOL2003-04-26T03:06 (M2.1).  The measured
background-subtracted spectrum in the interval 03:11:15-03:11:30
UT is the black histogram with $\pm1\sigma$ uncertainties in each
energy bin.  The background spectrum is the purple histogram with
error bars at count rates between $\sim $0.03 and
0.1~counts~(cm$^{2}$~s~keV)$^{-1}$.  The green histogram shows the
thermal continuum calculated with the MEKAL atomic code
\citep{1985ApJS...57..173M} 
and folded through the spectral response matrix
\index{RHESSI@\textit{RHESSI}!spectral response matrix}  of the \textit{RHESSI} instrument.  
The histograms with yellow and magenta lines are
two Gaussian line features representing the Fe-line feature ($\sim
$6.7~keV) and the Fe/Ni-line feature ($\sim $8~keV) respectively,
while the red histogram represents the total model.  
The fit range was 5.7 to 20~keV, and the
reduced $\chi^2$ of the fit was~0.74. 
\textit{Bottom:} residuals
in the fit range plotted as the number of standard deviations.}
\label{fig:dennis_spectrum} \end{center} 
\end{figure} 
\index{flare (individual)!SOL2003-04-26T03:06 (M2.1)!illustration}

The equivalent widths\index{equivalent width} of the Fe-line and Fe/Ni-line features were
also derived.  
Many flares had were long enough to allow repeated
measurements of particularly the Fe-line equivalent width, mostly
during the flare decay phase.  
All the measurements were made with the thin 
\textit{RHESSI} attenuator in place.  
\index{RHESSI@\textit{RHESSI}!attenuating shutters} 
An example
of measured equivalent width variations during SOL2002-05-31T00:16 (M2.4) is shown in Figure~\ref{fig:dennis_eqw}. 
\index{flare (individual)!SOL2002-05-31T00:16 (M2.4)!Fe feature} 
Similar calculations were done for
the $\sim $8~keV Fe/Ni-line feature. 
Both are based on a coronal
Fe/H abundance ratio of $1.26 \times 10^{-4}$ \citep{2000PhyS...61..222F},
or 4~times the photospheric value. 
These are compared with theoretical
calculations of the equivalent width vs. temperature.  
There is a general agreement in the trend for both the Fe-line and Fe/Ni-line
equivalent widths.  
\index{abundances!Fe/H}
This indicates that the coronal Fe abundance
is appropriate for this flare, though there is a systematic
displacement of the points towards higher temperatures.  This may
be due to the multithermal nature of the flare plasma, or to the
presence of non-thermal effects, or to instrumental effects at high
photon count rates.  They may also be due to incorrect atomic rates
used in the calculation of the He-like Fe~{\sc xxv} ion fractions
in ionization equilibrium calculations, since for most of the
temperature range shown the fraction of Fe~{\sc xxv} ions is small
($\lapprox 0.3$) where the uncertainties are greatest.

The measured Fe/H ratios for up to 22 of the 27 flares when in the
first \textit{RHESSI} attenuator state were found to be consistent with a
Fe/H abundance ratio between 0.8 and 1.0 times the coronal value.
No large-scale time variations in the Fe/H abundance are apparent,
as were derived for Ca/H abundance ratios from the BCS instrument
on \textit{SMM}\index{satellites!SMM@\textit{SMM}!BCS} \citep{1984Natur.310..665S,1998ApJ...501..397S}.  
Ratios measured in the thickest \textit{RHESSI} attenuator state were up to nearly
a factor of two higher than the theoretical curve, but for these
there was generally a poor spectral fit.  
\index{RHESSI@\textit{RHESSI}!attenuating shutters}
The best agreement of measured equivalent widths\index{Fe lines!equivalent width!and \textit{RHESSI} attenuator state} are for spectra
taken with \textit{RHESSI} in its first attenuator state during flare decay
stages, for which spectra were apparently more nearly isothermal
than near the flare peak and rise stages.

\begin{figure}
\begin{center}
\includegraphics[width= 10cm,angle=0]{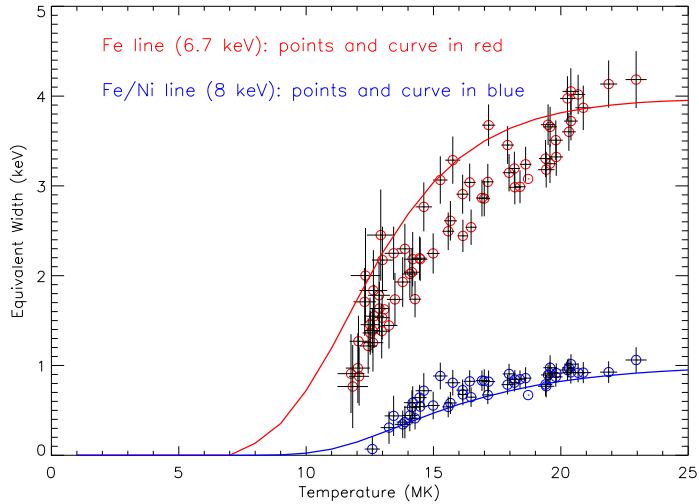}
\caption{Measured and predicted equivalent widths (see text) as a
function of $T_e$ for SOL2002-05-31T00:16 (M2.4).  The red upper
points are for the Fe-line feature with $\pm1\sigma$ error bars
obtained from spectral fits similar to the one shown in
Figure~\ref{fig:dennis_spectrum}.  The blue lower points show the
Fe/Ni-line feature.  The corresponding solid curves show the
theoretical dependence, updated from \cite{2004ApJ...605..921P}.
Atomic data used is from the Chianti\index{Chianti} database (v. 5.1)
\citep{1997A&AS..125..149D, 2003ApJS..144..135Y} (from \cite{2006ApJ...647.1480P}. 
} \label{fig:dennis_eqw} 
\end{center} 
\end{figure} \index{flare
(individual)!SOL2002-05-31T00:16 (M2.4)!illustration}

\bigskip
\noindent{\bf{Ratio of Fe-line to Fe/Ni-line in \textit{RHESSI} spectra:}}
\index{bound-bound emission}
\index{spectrum!SXR emission lines}
The Fe-line feature at $\sim $6.7~keV in \textit{RHESSI} spectra is made up
of Fe~{\sc xxv} lines  and Fe~{\sc xxiv} satellite lines, both
emitted as a result of transitions like $1s-2p$ or $1s-2s$, whereas
most ($\sim $85\%) of the Fe/Ni-line feature at $\sim$8~keV is
made up of Fe~{\sc xxv} lines  and Fe~{\sc xxiv} satellites with
$1s-np$ ($n\geqslant 3$) transitions (the remaining 15\% being due
to highly ionized Ni lines). The flux ratio of the Fe-line to the
Fe/Ni-line features should therefore be sensitive to $T_e$ because
of the different excitation energies of the transitions, and so
$T_e$ could be derived independently of the temperature from the
continuum emission which might be contaminated with non-thermal
emission near the flare impulsive stage. The agreement of the
observed and calculated equivalent widths of both line features
(Figure~\ref{fig:dennis_eqw}) indicates that for most \textit{RHESSI} spectra
the continuum temperature (which is plotted as the abscissa in
Figure~\ref{fig:dennis_eqw}) describes both equivalent widths well,
so that the temperature from the ratio of the line features is
nearly equal to the continuum temperature.

For SOL2002-07-23T00:35 (X4.8), the continuum spectrum is well-fit by two thermal components throughout the impulsive and decay phases.  
\index{flare (individual)!SOL2002-07-23T00:35 (X4.8)!soft X-ray line emissions}
\index{Chianti}
However, the Chianti predictions \citep[cf.][]{2004ApJ...605..921P}
of the Fe~and Fe/Ni line fluxes and ratio based on the observed temperatures and emission measures\index{soft X-rays!emission measure} of the two thermal components are significantly larger than the measured line values -- by, on average, $\sim$55\%, $\sim$20\%, and $\sim$34\%, respectively, with larger deviations at lower continuum temperatures \citep{2010ApJ...725L.161C}.
\index{Chianti!Fe/Fe-Ni ratio}
These discrepancies suggest against the interpretation of non-thermal excitation for the lines (as that would produce fluxes in excess of the predictions); while abundance variations could explain the deviations of the line fluxes, the variation in the ratio is more likely explained by ionization-fraction uncertainties as suggested by \cite{2006ApJ...647.1480P}.
\index{ionization equilibrium!uncertainties}


\bigskip
\noindent{\bf Non-isothermal effects in \textit{RHESSI} spectra:}
\index{non-isothermality} 
As indicated earlier, the analysis of
Fe-line emission in \textit{RHESSI} spectra to derive the flare Fe/H abundance
ratio \citep{2006ApJ...647.1480P}  was based on the assumption that
the emitting plasma was isothermal.  This conflicts with the
appearance of flares with multiple loops, in which each loop may
have a different T$_e$.  In spite of this the isothermal approximation
appears to apply to many  flares, even including long-duration
\index{flare types!long-duration}
\index{soft X-rays!long-decay events (LDE)}
flares observed spectroscopically from \textit{Yohkoh} in S~{\sc xv}
and Ca~{\sc xix} lines \citep{2005ApJ...634..641P}.  Such flares
have obvious multiple-loop structures.  The success of isothermal
fits also applies to \textit{RHESSI} spectra at low energies, for which good
fits to spectra obtained even with the \textit{RHESSI} attenuators in place
are generally achieved.  For the rise phase of many flares, however,
fits to \textit{RHESSI} spectra are not so satisfactory, even for spectra
which appear not to have any non-thermal component in the low-energy
continuum. Inspection of \textit{RHESSI} images and the higher-resolution
images from \textit{TRACE}\index{TRACE@\textit{TRACE}!and DEM} 
and \textit{SOHO}/EIT\index{satellites!SOHO@\textit{SOHO}!EIT} instrument at the developing
stages of flares shows that many individual loop structures contribute
to the total flare emission, perhaps with different temperatures.

Analysis of the differential mission measure (DEM) is therefore required,
or at least \textit{RHESSI} imaging spectroscopy, for these initial flare
stages\index{emission measure!differential}\index{RHESSI@\textit{RHESSI}!imaging spectroscopy}
Although the parameters of the DEM are difficult to determine
from \textit{RHESSI} spectra alone, some progress can be made using simple
forms for the temperature structure of the developing flare plasma,
such as DEM $ = T_e^{-\alpha}$ or ${\rm exp}\,(-T_e/T_0)$ where
$\alpha$ or $T_0$ characterize the emitting plasma at any particular
time. 
\index{cooling!equilibrium}
A more physically based description would involve a set
of nearly isothermal components reflecting the multiple loops each
in its own cooling equilibrium, so that the DEM parameters $\alpha$
or $T_0$ would characterize the distribution of the components
needed.
However, \cite{2010ApJ...725L.161C} showed that for at least one flare (SOL2002-07-23T00:35), the spectrum is well represented by two isothermal components, and a DEM analysis is consistent with a bimodal DEM.
\index{flare (individual)!SOL2002-07-23T00:35 (X4.8)!bimodal DEM distribution}
Cursory analysis of other X-class flares reveals a similar bimodal structure.  
A simple exponential or power-law DEM is therefore not necessarily a reasonable approximation.


\bigskip
\noindent{\bf{Non-thermal excitation of continuum and line emission
in \textit{RHESSI} spectra:}
\label{non-thermal_excitation}} 
It has been commonly assumed that the non-thermal electrons accelerated during
the flare impulsive stage produce bremsstrahlung (free-free
radiation) as they interact with the chromospheric or coronal
material. 
The free-free emission is basically due to hydrogen,
with small contributions from He atoms or ions, reflecting the solar
composition.  
Free-bound radiation from the non-thermal electron
continuum\index{free-bound emission!non-thermal} 
\index{continuum!non-thermal electron}
has until recently been neglected, on the grounds that
high-energy electrons are less likely to recombine on ions than
they are to emit free-free radiation.\index{recombination radiation (non-thermal)} 
This has now been questioned
\citep{2008A&A...481..507B, 2009ApJ...697L...6B}, who argue
that for Fe~particularly the free-bound radiation may be very
important for hot flare sources such as limb flares in which the
footpoints are occulted  
\index{footpoints!occultation}.

This little-studied effect could considerably alter the emission
for such flares, and should be taken into account where it is
detectable.  
This mechanism is fundamentally different from free-free
emission in that the emitted X-ray energy maps one-to-one with the
energy of the parent electron, rather than as an integral over the
entire distribution.  This means that its spectral features in
principle can be interpreted much more directly.

As well as non-thermal effects on the flare continuum, non-thermal
electrons may give rise to excitation of lines such as those making
up the Fe-line feature at $\sim $6.7~keV. Excitation could occur
by the ionization of K-shell electrons in near-neutral Fe, with
re-arrangement of the Fe atoms and emission of Auger electrons
(67\%) or photons (33\%), resulting in the K$\alpha$ or K$\beta$
lines (inner-shell transitions $1s-2p$ and $1s-3p$ respectively).
An energy of at least 7.1~keV is required for the removal of the
K-shell electron in each case. Observations of these lines (at
6.4~keV for K$\alpha$ and 7.1~keV for K$\beta$) could provide a
diagnostic for a non-thermal electron distribution that has sharp
cut-off energy $E_0$, since the lines would not be observed for
$E_0 > 7.1$~keV but would if $E_0 < 7.1$~keV \citep{1973SoPh...32..209P}.
In practice a sharp cut-off would be quickly smoothed out by
interaction of the lower-energy electrons in the distribution with
ambient plasma.  Most of the observed Fe K$\alpha$ line emission
is due to fluorescence of neutral Fe in the photosphere for disk
flares \citep{1979SoPh...62..113B, 1984ApJ...279..866P} and not to
K-shell ionization by electrons. However, an intriguingly marginal
case was observed with the BCS instrument on \textit{SMM} in which excitation
by non-thermal electrons might have been significant
\citep{1986SoPh..103...89E}.  
\index{acceleration!low-energy cutoff}
\index{low-energy cutoff}

\bigskip
\noindent{\bf{Summary:}}
\index{soft X-rays!summary of RHESSI results}
\textit{RHESSI} was intended primarily as a probe of the non-thermal emission
spectra of flares at high energies, but observation of low-energy
($\sim $3-20~keV) flare spectra has yielded important information.
Results include the derivation of electron temperature evolution
during the peak and decay stages of flares from the thermal continuum
based on an isothermal assumption, though simple approximations to
the temperature distribution have also been used. The continuum is
theoretically due to free-free and free-bound radiation in
comparable amounts in the \textit{RHESSI} energy range. 
The two line features,
the Fe-line (at $\sim $6.7~keV) and Fe/Ni-line ($\sim $8~keV)
features, enable the abundance of Fe relative to H~to be determined
from their fluxes relative to nearby continuum emission. 
\index{abundances!Fe/H}
Analyses
of spectra during the peak and decay phases of flares suggest a
coronal value of Fe/H, i.e., one that is larger than the photospheric
value by a factor 2-4. 
This is confirmed by measurements from the
SOXS instrument on the \textit{GSAT-2} spacecraft \citep{2006SoPh..239..217J}. 
\index{satellites!GSAT2@\textit{GSAT-2}!SOXS}
Additional temperature
information is offered by the flux ratio of the two line features:
generally the temperatures derived are similar to those obtained from the
energy dependence of the thermal continuum. Nonthermal effects are
currently being investigated particularly for the Fe-line and
Fe/Ni-line features, and could be a sensitive probe of the low-energy
cut-off energy in the non-thermal electron distribution if one exists.

\subsection{Flare energetics \label{sec:energetics}}
\index{flare energetics}\index{flares!energetics}

\index{flare (individual)!SOL2002-04-21T01:51 (X1.5)!energetics}
\index{flare (individual)!SOL2002-07-23T00:35 (X4.8)!energetics}

In the matter of constructing the overall energetics of a flare and
its associated mass ejection, the requirement for multi-wavelength
observations is clear. Direct measurements of the total radiative
output are available for only a few flares, notably SOL2003-10-28T11:10 (X17.2) from the Total Irradiance Monitor on the \textit{SORCE} spacecraft
\index{satellites!SORCE@\textit{SORCE}!TIM}\index{TSI}
\index{flare (individual)!SOL2003-10-28T11:10 (X17.2)!TSI detection}
\citep{2004GeoRL..3110802W}. 
While such measurements are the most
accurate for estimating the total energy released in an event,
information about the nature of the energy release process itself
can be acquired only through analysis of the partition of energy
amongst the various components such as energetic particles and
thermal plasma that are present as the flare proceeds.

The first attempts to estimate the total irradiance excess from a
flare were made with the Active Cavity Radiometer Irradiance Monitor
(ACRIM) 
\index{ACRIM} 
\index{satellites!SMM@\textit{SMM}!ACRIM}
on board the \textit{Solar Maximum Mission}. 
These
observations, unfortunately, yielded only upper limits \citep{1983SoPh...86..123H}.  
The TIM instrument currently flying on 
\textit{SORCE}\index{satellites!SORCE@\textit{SORCE}!TIM}
\index{Total Irradiance Monitor (TIM)} has now made definite observations
\citep{2004GeoRL..3110802W,2006JGRA..11110S14W}, and these have
proven to be a key factor in our new ability to characterize the
partition of energy since they provide a direct measure of the total
flare radiation.  
These results may improve with time if filtering can be developed to reduce the TIM background fluctuations. 
However the best way to measure this important parameter sensitively could be to
have imaging bolometric measurements\index{flares!bolometric measurements}, which would avoid the large background fluctuations due to p-modes and convective motions in
the rest of the photosphere.

Considerable constraints on the energy release processes follow
from a consideration of the partition \index{flares!energy
content!partition} of the released energy between accelerated
particles, radiation,  heated plasma, and ejected solar material.
But any exercise of this sort must be done rather carefully in order
not to ``double count''\index{flares!energetics!double counting}\index{double counting}
energy terms that are directly related to
each other, e.g., energy in accelerated electrons that is used to
produce thermal plasma  \citep{2005JGRA..11011103E} or radiation.
This requires distinguishing amongst ``primary'' components of
energy (e.g., the magnetic field), ``intermediate'' components
(e.g., accelerated particles and thermal plasma), and ``final''
components (e.g., kinetic energy of ejecta, radiant energy in various
wavebands), and recognizing the overlap of these components.

In the \textit{RHESSI} era,\index{eras!RHESSI@\textit{RHESSI}} the partitioning of energy in two well-observed solar flare/CME events was carried out using data from a variety
of missions including \textit{RHESSI, ACE, SOHO} and \textit{GOES}\index{satellites!ACE@\textit{ACE}}\index{satellites!SOHO@\textit{SOHO}}\index{satellites!GOES@\textit{GOES}}.
This study yielded the result that ``flare radiant energy and CME mechanical energy
are the same order of magnitude.'' 
The SXR flare (from 1.5~keV) appears to contain substantially less than about 10\% of the total radiant energy \citep{2005JGRA..11011103E}.
The impulsive-phase radiation appears to dominate the flare luminosity.
Both the SEPs and the impulsive phase acceleration contain a
substantial fraction of the total energy \citep{2005JGRA..11011103E},
as described below.

\subsection{Energetics of two large \textit{RHESSI} flares}\label{sec:emslie}

A number of previous studies have examined the energy budget of a
limited number of energy components in certain flares.  The radiative
energy budget of SOL1973-02-05  \index{flare
(individual)!SOL1973-02-05 (pre-\textit{GOES})!energetics} was evaluated by
\cite{1980sfsl.work..451C}, but in the absence of HXRs or $\gamma$-ray
observations for this event the role of energetic particles in the
event could not be assessed.  
The X-ray and $\gamma$-ray observations
of several flares, including the major $\gamma$-ray flare SOL1972-08-04, \index{flare (individual)!SOL1972-08-04 (pre-\textit{GOES})!energetics}
were used to show that the kinetic energy of the accelerated electrons
constituted a surprisingly large fraction of the total flare energy,
perhaps as high as 10~to 50\% of the $\sim$10$^{32}$~erg released
during the flare \citep{1976SoPh...50..153L}.  Two flares within
the same active region on 1980 August~31 
\index{flare (individual)!SOL1980-08-31T12:49 (M2.8)!energetics} 
provided the energy content
in thermal plasma, non-thermal electrons, and hydrodynamic mass
motions of non-ejected material \citep{1984SoPh...91..325S},  while
the energy content in radiative, thermal, non-thermal electron, and
non-CME associated plasma ejected was calculated for SOL2002-02-26T10:27 (C9.6)
\citep{2002SoPh..210..287S}.
\textit{RHESSI} X-ray observations
\index{flare (individual)!SOL2002-02-26T10:27 (C9.6)!energetics}
were used determine the energy in accelerated electrons and in the
hot plasma for nine medium-sized flares (\textit{GOES} class C6 to M8), with
the conclusion that despite the large uncertainties, the energies
in these two components were of the same magnitude in each case
\citep{2005A&A...435..743S}.

The energetics of two X-class flares (SOL2002-04-21T01:51 
\index{flare (individual)!SOL2002-04-21T01:51 (X1.5)!energetics} 
and SOL2002-07-23T00:35)
\index{flare (individual)!SOL2002-07-23T00:35 (X4.8)!energetics} 
have been analyzed in a very comprehensive study made possible by
overlapping observations at a variety of wavelengths
\citep{2004JGRA..10910104E,2005JGRA..11011103E}.  
SOL2002-04-21T01:51 was a long-lived SXR event which occurred near the west limb; SOL2002-07-23T00:35 was much more impulsive, a strong emitter of HXRs and
$\gamma$-rays \citep[see][]{2003ApJ...595L..69L}, and was located
near the east limb at S13E72.  
Observations were used from instruments
on the \textit{ACE}\index{satellites!ACE@\textit{ACE}}, 
\textit{SOHO},\index{satellites!SOHO@\textit{SOHO}}\index{Solar and Heliospheric Observatory@\textit{Solar and Heliospheric Observatory (SOHO)}}\index{Advanced Composition Explorer@\textit{Advanced Composition Explorer (ACE)}}
and \textit{RHESSI} to provide quantitative estimates of the energy
contents of (1) the coronal mass ejection, (2) the thermal plasma
at the Sun, (3) the accelerated electrons producing hard
X-rays, (4)
the accelerated ions producing gamma rays, and (5) the solar energetic particles
accelerated by the outward eruptive disturbance/CME.  The detailed
energy budget for these two events, including the CME kinetic and
potential energies and the energy in the 
SEPs at 1~AU, is reproduced in Table~\ref{table:dennis_energetics}
\index{flares!energy content!overall}.

\begin{table}
\caption{Flare and CME Energy Budgets$^*$}
\begin{tabular}{l@{\extracolsep{1cm}}c@{\extracolsep{1cm}}c}
\hline
&\textbf{21 April 2002}&\textbf{23 July 2002}\\
\hline
\textbf{Primary Energy}&&\\
\qquad Magnetic & $32.3 \pm0.3$&$32.3 \pm0.3$\\
\textbf{Flare}&&\\
\qquad \textbf{Intermediate Energies}&&\\
\qquad \qquad Electrons $(> E_{\rm min})$& $31.3 \pm 0.5$ &$31.3 \pm 0.5 $\\
\qquad \qquad Ions ($>1$~MeV~nucleon$^{-1}$)& $<31.6$ &$31.9 \pm 0.5$\\
\qquad \qquad Thermal Plasma ($T > 5$~MK)& $31.1^{+0.4}_{-1.0}$&$30.4^{+0.4}_{-1.0}$\\
\qquad \textbf{Radiant Energy}&&\\
\qquad \qquad From \textit{GOES} plasma & 31.3 $\pm0.3$ &31.0 $\pm0.3$\\
\qquad \qquad Assuming $L_{total}/L_X = 100$ &32.2 $\pm{0.3}$&32.2 $\pm{0.3}$\\
\textbf{CME}&&\\
\qquad Kinetic & $32.3 \pm 0.3$& $32.3 \pm 0.3$\\
\qquad Gravitational Potential &$30.7 \pm 0.3$ & $31.1 \pm 0.3$ \\
\textbf{Energetic Particles at 1 AU} & $31.5 \pm 0.6$& $<30$\\
\hline
\end{tabular}
\par
\smallskip
$^*$Tabulated values are $\log_{10}$ of the energies in erg of the different
components given in table 1 of \cite{2005JGRA..11011103E}.

\label{table:dennis_energetics}
\end{table}
\index{flares!energetics!table}

\bigskip
\noindent{\bf{Magnetic energy:}} 
\index{flares!energy content!magnetic}
The total magnetic energy available for conversion into other forms
(flare and CME) can be estimated in principle from the extrapolation
of the field observed at the photosphere.  There are many uncertainties
with such a procedure.  The flares providing the information in
Table~\ref{table:dennis_energetics} were each at the limb and
therefore not amenable to the extrapolation in any case.  Thus we
view the uncertainty of 0.3~dex quoted in the table as highly
optimistic\index{magnetic structures!extrapolations from photosphere!caveats}\index{caveats!photospheric field extrapolation}.

\bigskip
\noindent{\bf{Energy in the CME:}}
\index{flares!energy content!CME}
To derive the energy in the CME one must first derive a coronal
density distribution from  the excess brightness (due to Thomson
scattering\index{scattering!Thomson} of photospheric light) in coronagraphic white-light
images \citep[e.g., from the LASCO~C2 and 
C3~coronagraphs\index{satellites!SOHO@\textit{SOHO}!LASCO} on
\textit{SOHO};][]{1995SoPh..162..357B}.  
These observations permit an estimate
of the mass distribution of the CME \citep{1981SoPh...69..169P,
2000ApJ...534..456V, 2002ESASP.506...91V}.  
Then, the flow patterns
of the plasma during the ejection must be determined from the
projected images as a function of time to derive the velocity 
field\index{coronal mass ejections (CMEs)!velocity field}
of the CME material. 
Finally these properties can be used to find
the potential ($U_{\Phi}$) energy, kinetic ($U_{\rm K})$ energy,
and  enthalpy associated with the CME\index{coronal mass ejections (CMEs)!energy partition}. 
For SOL2002-04-21T01:51 and SOL2002-07-23T00:35, the potential energies obtained were $U_\Phi = 10^{30.7}$
and $10^{31.1}$~erg, respectively, and kinetic energies $U_{\rm
K} = 10^{32.3}$ and $10^{32.0}$~erg.  
These are unusual CME events.
Their large kinetic energies 
\index{flare (individual)!SOL2002-04-21T01:51 (X1.5)!CME energetics} 
\index{flare (individual)!SOL2002-07-23T00:35 (X4.8)!CME energetics} 
place both of them in the top~1\% of all
observed CMEs for the period 1996-2000 \citep{2002ESASP.506...91V}.
In both cases the gravitational potential energy is 
\lapprox10\% of the total energy contained in the CME.  
Note that the magnetic energy of a CME, thought to be its dominant term, is
almost impossible to assess observationally.

\bigskip
\noindent{\bf{Thermal energy:}} 
\index{flares!energy content!thermal plasma}
The thermal energy of the heated plasma is obtained from the
temperature $T_0$~(K) and emission measure\index{emission measure!definition} 
$EM = \int_V n_e^2 \,
dV$ (cm$^{-3}$) for the thermal portion of the overall spectral fit
to the HXR data \citep[see, e.g.,][]{2003ApJ...595L..97H}. Here
$n_e$ is the electron density (cm$^{-3}$) and $V$ is the emitting
volume (cm$^3$). 
Account must be taken of the filling factor
\index{filling factor} 
$f$ equal to the ratio of the emitting volume to the apparent
volume ($V_{\rm ap}$) as determined with an imaging instrument
having limited spatial resolution. The thermal energy content of
the plasma is then given by

\begin{eqnarray}
\nonumber U_{\rm th} & = & 3 \, n_e \, kT_0 \, f \, V_{\rm ap}
\simeq
3 \, kT_0 \, {\sqrt{EM \times f \, V_{\rm ap}}} \\
& = & 4.14 \times 10^{-16} \, T_0 \, {\sqrt{EM f \,V_{\rm ap}}} \, \ {\rm erg}, \label{eqn:emslie_u_thermal}
\end{eqnarray}
where $k$ is the Boltzmann constant and $V_{\rm ap}$~the source volume,
estimated from the area information contained in the \textit{RHESSI}
observations assuming $V_{\rm ap}=A^{3/2}$.  For the common assumption
of  a filling factor of unity ($f=1$), this is an upper limit for
\index{filling factor}
the instantaneous thermal energy.  
\index{cooling!and energy estimates}
It also can be taken as a lower
limit to the total thermal energy since it does not account for the
cooling of the plasma prior to a given time, nor for any heating
at later times.  Each of these contributions could add perhaps a
factor of two to the total thermal energy.  
Application of equation~(\ref{eqn:emslie_u_thermal}) to SOL2002-04-21T01:51 and 
SOL2002-07-23T00:35  yielded values of 
\index{flare (individual)!SOL2002-04-21T01:51 (X1.5)!thermal plasma} 
\index{flare (individual)!SOL2002-07-23T00:35 (X4.8)!thermal energetics} 
$U_{\rm th} = 10^{31.3}$~erg and $10^{31.1}$~erg, respectively. 
An estimate
of the total radiated energy can be rather straightforwardly obtained
from the \textit{GOES}\index{GOES@\textit{GOES}!energetics} SXR data, simply by integrating
the product of the emission measure\index{emission measure} and the optically thin radiative
loss function\index{radiative loss function}\index{cooling!radiative} \citep{1969ApJ...157.1157C} (for coronal abundances)
over the duration of the flare \citep{2005JGRA..11011103E}.
\index{Chianti!energetics} 
This exercise gives values of $U_R \sim$10$^{31.3}$~erg for the SOL2002-04-21T01:51
flare and $10^{31.0}$ erg for SOL2002-07-23T00:35. 
Note that no knowledge of the source
volume, density, or filling factor is required to make this
calculation; hence the good agreement between these values and those
obtained immediately above suggests that the simplifying  assumption
of unity filling factor is not unreasonable. In particular, the
\index{filling factor}
volume filling factor for the soft-X-ray-emitting plasma cannot be
too small ($<0.01$), otherwise the plasma energy calculated using
the \textit{RHESSI} source areas would be significantly below the estimate
obtained from the \textit{GOES} data.

A separate physical argument also suggests that the filling factor cannot be too small.
At the time of the peak temperature, the energy density in the thermal plasma, 
assuming $f = 1$, already requires coronal field strengths exceeding $\sim$350~G to 
contain the plasma; this number increases as $f^{-1/2}$ so the filling factor must be no smaller than $\sim$0.01 unless the coronal field can significantly exceed $\sim$1000~G.
\index{magnetic field!coronal!and plasma containment}
\cite{Caspi2010} found that this same argument holds for essentially all X-class flares, as they all require coronal field strengths exceeding $\sim$220-460 G.
This again suggests that a filling factor in the range 0.1-1 is a reasonable assumption for super-hot, X-class flares.

\bigskip
\noindent{\bf{Accelerated Electrons:}} 
\index{flares!energy content!accelerated electrons}
The energy in accelerated electrons can be determined from applying
a thick-target model see \citep[see][]{Chapter3} to the measured HXR spectrum
in order to obtain the injected spectrum $F_0(E_0)$ (electrons
s$^{-1}$~keV$^{-1}$) and calculating the corresponding injected
power from
\begin{equation}
U_{\rm e} = A_i \, \int_{E_{\rm min}}^\infty E_0 \, F_0(E_0) \,
dE_0, \label{eqn:emslie_ue}
\end{equation}
where $E_{\rm min}$ is the lowest particle energy in the non-thermal
component of the electron distribution and $A_i$ is the injection
area (which, however, cancels in further  manipulation). The
accumulated energy in non-thermal electrons is then obtained by
integrating the injected electron power over time. The thermal
spectrum is typically dominant at low energies, so the largest value
of $E_{\rm min}$ 
\index{acceleration!low-energy cutoff}
consistent with an acceptable fit to the spatially integrated
spectral data, is chosen; the energies obtained are necessarily
lower limits.  The values of $U_e$ thus determined were $U_{\rm e}
= 10^{31.5}$~erg for SOL2002-07-23T00:35 and $10^{31.3}$~erg for
SOL2002-04-21T01:51.  These results are higher than the
corresponding values of $10^{31.3}$ erg and $10^{31.1}$ erg for the
energy contained in the thermal plasma $U_{\rm th}$.  This result
is reinforced by the wide lower error bar on $U_{\rm th}$ caused
by the uncertain filling factor $f$ and the fact that $U_e$ may be
an {\it under}estimate.  This suggests that much of the electron
power is radiated in other wavelengths, such as optical and EUV
(see below).
\index{filling factor}

\bigskip
\noindent{\bf{Accelerated Ions:}} 
\index{flares!energy content!accelerated ions}
As explained in \cite{Chapter4}, accelerated ions
\index{flares!energy content!accelerated ions} 
are also energetically important in energetic solar flares with
significant emission above $\sim $300~keV \citep{2000IAUS..195..123R}.\index{spectrum!gamma@$\gamma$-ray lines}
The primary ions undergo nuclear collisions and thereby produce
$\gamma$-ray lines and continua of various kinds, for example, by
direct de-excitation following inelastic scattering to produce lines
mainly in the  $\sim $1-10~MeV range
\citep[e.g.,][]{1979ApJS...40..487R}.  These are broad enough for
\textit{RHESSI} to resolve. 
There is also a highly-broadened set of lines\index{continuum!gamma@$\gamma$-ray pseudo-continuum}
(a ``pseudo-continuum'' because of overlaps) resulting from
$\alpha$-particles and higher-Z ions striking ambient nuclei.\index{spectrum!gamma@$\gamma$-ray pseudo-continuum}
The
flux in the highly-broadened component is typically $>$3 times that
in the moderately-broadened component.
 The threshold energies for producing all of these nuclear lines
 are $\gapprox$2.5~MeV, and so the spectrum below that energy is
 unknown observationally. Assuming a flat spectrum below 2.5~MeV,
 a lower limit of $(1.0 - 4.0) \times 10^{30}$~erg of energy  and
 an upper limit of $ (1.2 - 120) \times 10^{32}$~erg (assuming a
 power-law spectrum down to 0.1~MeV) was found for protons in
SOL2002-07-23T00:35 (X4.8) \citep{2003ApJ...595L..69L}.  Protons and heavier
 ions together range from $U_{\rm i} \simeq (6 - 24) \times
 10^{30}$~erg to $(7 - 700) \times 10^{32}$~erg. 
 For SOL2002-04-21T01:51 (X1.5), for which no significant $\gamma$-ray line emission
 was produced, the upper limit is in the range $U_{\rm i} \simeq
 4.0 \times 10^{30}$~erg to $1.2 \times 10^{34}$~erg, depending on the spectral
model used\index{flare (individual)!SOL2002-04-21T01:51 (X1.5)!$\gamma$-ray upper limits}.

\bigskip
\noindent{\bf{Solar Energetic Particles:}} 
\label{sec:sep_energy}
\index{flares!energy content!solar energetic particles}
SEPs accelerated at the flare site
and/or at shocks driven by the CME represent another significant
contribution to the global energy budget.\index{shocks!particle acceleration}
SOL2002-07-23T00:35 (X4.8)\index{flare (individual)!SOL2002-07-23T00:35 (X4.8)!SEPs}
occurred near the East limb of the Sun (S13E72) and, as is typical
for east-limb events, was apparently not magnetically well-connected
to Earth. 
As a result, near-Earth spacecraft such as \textit{ACE} and \textit{GOES}
did not observe significant SEP fluxes that could be traced to this
event. 
On the other hand, SOL2002-04-21T01:51 at  S14W84
\index{flare (individual)!SOL2002-04-21T01:51 (X1.5)!SEPs} was
relatively well connected to Earth, and indeed, a strong interplanetary
shock (Mach number M$_A = 3.7$) was observed some two days later,
at $\sim$04:15~UT on April~23. That the intensities of the SEPs
coincided well with the time of shock arrival indicated that
acceleration was taking place locally at the shocks. 
Integrating the energy spectra over energy and solid angle covered
by the shock (typically $\sim$$\pi$~sr) gives the total particle
energy incident at 1~AU as $U_{\rm  p} \approx 2.8 \times 10^{31}$~erg),
which is a significant fraction ($\sim$15\%) of the CME kinetic
energy ($\sim$$1.8 \times 10^{32}$~erg).  This implies that shock
acceleration must be relatively efficient. Indeed, a broader survey
of SEP events shows that this result holds commonly, though not
invariably \citep{2008AIPC.1039..111M}. Figure~\ref{fig:mewaldt}
illustrates this\index{solar energetic particles (SEPs)!total energy content}\index{acceleration!SEPs}.

\begin{figure}
\begin{center}
\includegraphics[width=0.8\textwidth]{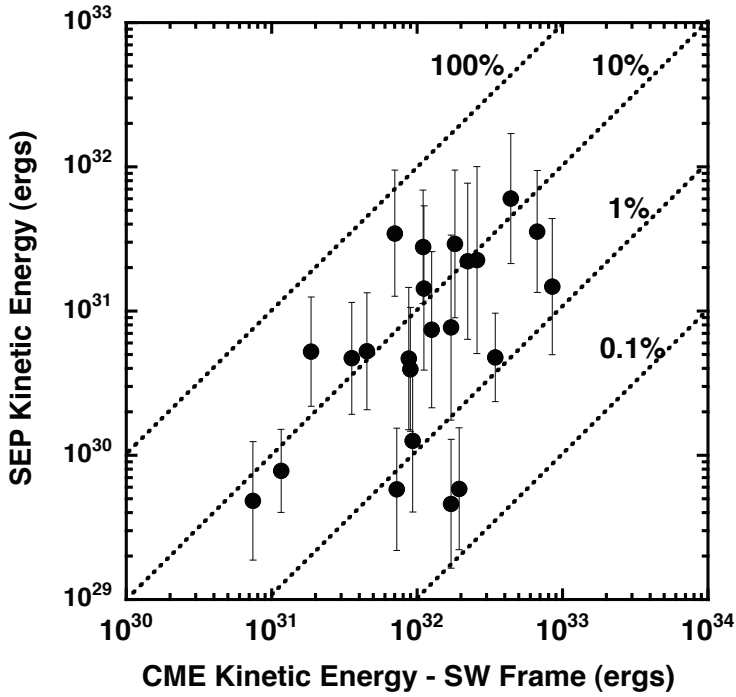}
\end{center}
\caption[]{Comparison of SEP energy with CME kinetic energy for a
sample of events, adapted from \cite{2008AIPC.1039..111M}.
There is no suggestion of a correlation, but these estimates do imply that the
SEPs contain a large fraction of the total CME kinetic energy.
}
\label{fig:mewaldt}
\end{figure}
\index{coronal mass ejections (CMEs)!and SEP energy!illustration}

\bigskip
\noindent{\bf{Future improvements:}} 
The measurements in
Table~\ref{table:dennis_energetics} all have sizeable error bars.
How can these measurements be refined? We tackle this line-by-line
in the table.

(i) Estimating the primary magnetic energy reliably
is a difficult problem, but its solution will certainly involve
measurements of the vector magnetic field at the ``top'' of the
chromosphere, with spatial resolution adequate to capture the field
close to the magnetic polarity inversion line -- scales below one 
arcsecond. 
This must be accompanied by fast and robust mathematical
methods for the non-linear force-free field extrapolations 
and/or by direct measurements of the vector field in the corona, e.g., by CoMP or similar instruments.\index{CoMP}

(ii) Table~6.1 identifies ``intermediate energies,'' reservoirs between
the fundamental magnetic field and the true losses (radiation and
ejecta).
These table entries are themselves model-dependent, for
example, in the cooling time scale of the thermal plasma.
Detailed EUV-SXR imaging spectroscopy of flare arcades can sharpen these
estimates substantially\index{flare models!thick-target}.
The current uncertainties on the non-thermal
electron energy budget derived from the collisional thick-target
model (which is the model requiring the smallest amount of non-thermal
particle energy) lie wholly in a lack of knowledge of the ``low energy
cutoff'' of the electron spectrum, if there is one\index{low-energy cutoff}.
One theoretical possibility for deducing this information, is to 
find edges in the X-ray spectrum produced by free-bound
radiation \citep{2008A&A...481..507B,2010A&A...515C...1B}\index{free-bound emission!non-thermal}.
No hard X-ray edge structure has yet been reported.
The ion (or proton) component is also very poorly constrained, and improving
this will require very much better $\gamma$-ray spectroscopy, with
much improved signal-to-noise ratio. 
We also have no knowledge at all of lower energy protons (of
a few tens or hundreds of~keV), though diagnostics such as 
\index{Lyman-$\alpha$!Doppler shifts}
Doppler-shifted Lyman-$\alpha$ (emitted
by charge-exchanging\index{charge exchange}\index{reactions!charge-exchange} low-energy protons \citep{1976ApJ...208..618O}
and the (disputed) observations of H$\alpha$\index{Fraunhofer lines!H$\alpha$!impact polarization} impact polarization
\index{Balmer-$\alpha$!impact polarization}
\citep{1999A&A...349..283V} might help us in this direction.\index{polarization!H$\alpha$}

(iii) Substantial improvement in CME energetics is now being
obtained from the \textit{STEREO}\index{satellites!STEREO@\textit{STEREO}} observations.
The major uncertainty now, as
before, is the lack of knowledge of the magnetic field and its
dynamics.
Table~6.1 does not even list this item though it is the dominant one.

(iv) The sparse sampling of SEP fluxes, and the difficulty of bookkeeping their distribution in the heliosphere, are some of the issues.
Data from nearer the Sun will help, as will
improvements in our understanding of their sources. 
The observation of energetic neutral atoms via charge 
exchange \citep{2009ApJ...693L..11M}
and neutrons, again from inner-heliospheric vantage points if possible,
will be a major step forward\index{energetic neutral atoms}\index{neutron emission}.
Because of the sampling issues, though,
this will likely remain inthe domain of statistics rather than
precise measurement.

\section{Summary: models and observations}
\index{flare models!summary}

\subsection{Flare model constraints}\label{sec:pm}
\index{flare models!observational constraints}

Where within this mass of data can we identify the decisive
observations that will choose one model over another? It is a
daunting task to find ``a flare model''  that is capable of explaining
all observed phenomena in all events. As we have seen a flare can
be geometrically complicated and have many associated temporal
components, and its effects appear across a large dynamic range of
physical parameters.
The extend from the solar interior to beyond one AU.  
Furthermore, not all flares exhibit the same behavior and
many deviate sharply from the ``standard''  two-ribbon eruptive flare
scenario that has framed much of our thinking for decades. 
For example, some flares do not have associated eruptions, and many
have multiple ribbons instead of just two.\index{ribbons!multiple}
Some flares continue
accelerating electrons well into their X-ray decay phase, some have
associated $\gamma$ rays but many -- perhaps most -- do not.

Despite having orders of magnitude more data than was available at
the time that the ``standard''\index{flare models!standard} scenario was first formulated, our
knowledge is still incomplete, and this incompleteness forces us
to invoke cartoons\index{cartoons} (see
\url{http://solarmuri.ssl.berkeley.edu/~hhudson/cartoons/} for a
compilation) to relate one observable to another, and to suggest
cause and effect. Analytical or computational models with some
degree of refinement tend to deal separately with different aspects
of the flare; e.g., the magnetic configuration and how it evolves,
electron acceleration, expansion of the chromosphere, production
of radiation of one type or another. These various parts are then
stitched together into cartoons\index{cartoons}\index{flare models!and cartoons} -- also loosely called models --  but
without the details of any individual aspect yet being fully worked
out.

So, what are the ``top-level''  components to a flare model? Firstly,
we must understand how a particular coronal magnetic configuration
can become loaded with or emerge with sufficient stored magnetic
energy. We must understand how, after a period of stability, the
configuration becomes unstable in such a way as to produce a dramatic
energy release. The conversion of stored energy into the various
forms which we infer observationally is a -- perhaps the -- central
aspect of flare models, which still defies detailed explanation. A
fourth model element -- undoubtedly also related to the initial
magnetic configuration -- must address the relationship between the
localized radiation burst that is the flare, and the coronal mass
ejection. Each of these top-level model components has many
sub-components. For example,  considering one element of the energy
conversion problem, even after 150 years we do not know the origin
of flare optical emission. Several alternative theoretical scenarios
exist for this part alone, and likewise for the tens of other
observed and inferred flare phenomena. 
Where within this mass of models\index{flare models!predictive power} can we find the decisive predictions?

Let us confine further discussion to the model elements identified
as top-level above, and try and identify some ways in which
observations support or refute the various options. This is necessarily
a very abbreviated discussion as a full comparison between the
multifarious models and the observations would require many more pages.

\bigskip \noindent{\bf{Energy storage:}} 
The storage of energy\index{flares!energy storage}
presents an interesting problem. 
If magnetic reconnection, releasing
stored energy, can readily happen in a flare, or indeed in coronal
heating\index{coronal heating!and reconnection}, why does it not happen right away\index{magnetic field!energy storage}\index{reconnection!non-occurrence}?
This may come down to
the configuration of the magnetic field. Observationally, some
progress has been made in understanding energy storage using field
extrapolations from observed photospheric vector magnetic fields
but these remain problematic, not least because the photosphere is
not force-free, in contradiction to the basic mathematical assumptions
of the extrapolations. The suggestion from the existing extrapolations,
at least of newer active regions, is that free energy storage occurs
low down in the atmosphere, close to the polarity inversion line
\citep{2007A&A...468..701R,2008ApJ...675.1637S}\index{magnetic field!free energy}\index{magnetic field!extrapolation}.
Unfortunately,
present observations are barely able to resolve the photospheric
vector magnetic field, let alone the chromospheric field,  at the
arcsecond scales required, and the reconstruction techniques are
temperamental, so these conclusions should be treated with caution.
Nonetheless, many flares also start with their footpoints close to
a magnetic neutral line.
This often involves the activation of low-lying active
region filaments overlying strong sheared polarity inversion lines.
\index{magnetic structures!polarity inversion line}
\index{magnetic structures!core fields}
This suggests that the properties of what is known as the ``core
field''\index{magnetic structures!core fields} in filament models -- the strong, twisted or sheared field
supporting filament material --  are also core to understanding the
early phase of flares, and models capable of producing such
configurations, whether by shearing or by emergence of structures
with concentrated twist, are very relevant. To that end, we will
profit from paying close attention to active region filament magnetic
field observations, such as those of \cite{1991A&A...247..379W} or
\cite{2009A&A...501.1113K}, as well as anything that can be learned
from coronal field diagnostics in the microwave, IR and UV regimes
(Section \ref{sec:fletcher_fpmagnetic}).  
Observations diagnosing
the typical properties in or near the reconnection region, such as
the field strength and connectivity, number density, temperature
and velocity, are necessary input to impulsive phase reconnection
and acceleration or heating models.
Unfortunately the bulk properties readily observable with spectroscopic remote sensing may not be
sufficient to understand the conditions that lead to what is in
large part a plasma kinetic process (i.e., the decoupling of particles
from the magnetic field and from each other).
We should be prepared to learn what we can, in terms of the microphysics,
 from {\emph{in situ}} observations in the solar wind or terrestrial environment.\index{in situ@\textit{in situ}!comparisons with remote sensing} 

\bigskip
\noindent{\bf{Instability and field rearrangement:}} The evolution
of a stable, energy-loaded configuration towards an unstable one,
and the release of the stored energy, are related problems. Both
require an understanding of the conditions under which magnetic
reconnection will set in -- or not --  and this is a vast field of
study in itself\index{reconnection!microphysics vs global physics}.
One view is that the rate of magnetic reconnection
is determined primarily by the plasma microphysical conditions.
Another is that reconnection is determined by the magnetohydrodynamic
state, and the microphysics somehow adapts to keep up with what the
field dictates. But with neither field measurements nor detailed
information on non-equilibrium plasma conditions in the corona where
flare reconnection is thought to take place, we know little for
sure about the reconnection physics. 
Observations of plasma inflow\index{reconnection!inflow}
in the late phase of flares allows estimates to be made of the
reconnection rate, which turns out to be compatible with fast
reconnection \citep{2001ApJ...546L..69Y,2006ApJ...637.1122N}. 
But the rate alone is not very informative, and the late phase of a
flare is a less challenging environment in terms of the reconnection
and rate of energy conversion required.

The standard flare reconnection model has basically been in development
since \cite{1948MNRAS.108..163G}\index{flare models!standard}.
Note that Giovanelli's\index{cartoons!Giovanelli}\index{Giovanelli, R. G.} sketches
of the magnetic scenarios for flares mainly emphasized the behavior
of current systems, rather than flux transfer; as late as 1963
\cite{1963ApJS....8..177P}\index{Parker, E. N.} could describe magnetic-field
``annihilation'' as a ``presently popular belief'' to explain solar
flares, but that ``There is very little in the observations to
support such views.'' 
\index{magnetic field!flux annihilation}
With \cite{1964psf..conf..425P}\index{Petschek, H. E.} augmenting
the earlier work of Giovanelli, Cowling\index{Cowling T. G.}, Dungey\index{Dungey, J. W.}, Sweet\index{Sweet, P. A.}, and Parker
himself, the case became much stronger\index{reconnection!history}.
Nowadays it would be difficult to discuss flare phenomenology without
appealing to magnetic reconnection in one form or another, although
all of the abundant evidence for it is necessarily indirect. But
once again, observational constraints on model geometry are few.
On the disk, observationally-grounded extrapolations of the magnetic
field are suggestive of one topology or another, sometimes backed
up by the shape of the field outlined in EUV. For example, there
is evidence for coronal nulls in flaring active regions
\citep{2000ApJ...540.1126A,2001ApJ...554..451F,2009ApJ...700..559M}\index{coronal sources!current-sheet morphology}\index{soft X-rays!double coronal sources}.
On the limb, double coronal sources are sometimes observed with a
temperature structure consistent with outflows from a reconnecting
structure, and a vertical displacement between the two sources
suggestive of a vertically-extended current sheet
\index{current sheets}
\index{magnetic structures!coronal current sheets}
\citep{2003ApJ...596L.251S}. 
Though not strictly flare activity,
the magnetic and EUV coronal evolution during a flux emergence event
is consistent with a model involving separator reconnection
\citep{2005ApJ...630..596L}. 
During the impulsive phase, the portion
of the field outlined by hot plasma can look very asymmetric and
disordered, but later on in the EUV/SXR arcade it looks rather
symmetric and quasi-2-D, like the standard cartoon\index{cartoons!standard}. 
So all topologies
are possible, and perhaps flare magnetic systems evolve through
different topologies.  
However, returning to Parker's discussion
of ``annihilation,''\index{reconnection!vs. annihilation} until we know whether we are dealing with
reconnection or flux transfer at a neutral point, a neutral line,
a separator, a current sheet or indeed a volume filled with many
small current sheets it is not possible to say whether the energy
released is dominated by field annihilation (i.e., dissipation of
antiparallel components of {\bf B} within the reconnecting structure) or
field relaxation following reconnection (``dipolarization'')\index{dipolarization}\index{magnetic structures!dipolarization}.
For example, reconnection at a null dissipates
essentially no magnetic field because the reconnection volume is
so small -- energy conversion happens elsewhere in the system -- but
this would not necessarily apply to a volume filled with many small
current-sheet structures.

\bigskip
\noindent{\bf{Energy conversion:}} The sudden transformation of
energy contained in the solar magnetic field is the heart of the
flare problem. As mentioned above, magnetic reconnection itself
does not transform much energy, but may facilitate large-scale
restructurings that do.  The common ingredient in all flares, by
any definition of their sudden appearance, is the impulsive phase.
We identify this as the timing of the energetically fundamental
non-thermal process of particle acceleration as originally observed
at radio wavelengths, and then in X-rays and $\gamma$-rays as well.
It is this impulsive phase, which we can associate with all of the
dominant energetic processes of a flare (the white-light/UV continuum
emission, large-scale wave generation, CME eruption, and powerful
particle acceleration), that is not captured in the many cartoons\index{cartoons!omissions}
showing different versions of the standard flare model.

The dominant models hold that particle acceleration is an entirely
coronal process, with HXR footpoints produced by accelerated coronal
electrons which precipitate, possibly having escaped a magnetic
trap formed by the expansion of magnetic field into the corona. The
presence of coronal hard X-ray sources \citep{2008A&ARv..16..155K}
as well as radio bursts provides ample evidence that non-thermal
electron populations are present in the corona during the flare
impulsive phase, with high fractions of all electrons present being
accelerated. 
In one case in particular, the data suggest that
essentially all coronal electrons present in the source are
non-thermal \citep{2010ApJ...714.1108K}, with mimimal remaining thermal
distribution. 
Details of the acceleration mechanism are obscure,
and observations in the optical to SXR regime which deal primarily
with thermal (i.e., ``processed'' energy) are of limited help beyond
suggesting the geometry and evolution of the environment in which
acceleration occurs. Observational suggestions of a current sheet
provide one possible environment, but if one is to produce the
electron fluxes necessary to explain the HXR footpoints the required
current sheet dimensions are rather unfeasible, unless the Alfv{\' e}n
speed is high (the number of electrons accelerated per second being
limited to what can be advected into the sheet each second). We
have evidence of magnetic field relaxation \citep{2003ApJ...596L.251S}
consistent with shrinkage beneath a reconnection region, in which
betatron acceleration would occur \citep{1997ApJ...485..859S}. But
though this process can accelerate a good number of electrons, the
energy increases achieved are modest, requiring a suprathermal
population\index{suprathermal populations} 
to start with \citep{2005ApJ...635..636G,2006ApJ...647.1472K}.
Non-thermal line broadening in soft X-ray \citep[e.g.,][]
{2000A&A...364..859R} and EUV lines
\citep{2008ApJ...679L.155I,2008PASJ...60..275H} may  be interpreted
as evidence for plasma turbulence\index{plasma turbulence}, a central ingredient in many
electron acceleration models, though our observations are at a
spatial scale far larger than that at which plasma wave energy can
be effectively dissipated by electron acceleration.

At some level, coronal electron acceleration is straightforward
even though the details are unknown. 
A certain amount of energy
dumped into the coronal plasma in the form of plasma turbulence,
for example, must be shared between the coronal particles resulting
in a mean energy per particle. 
If that mean energy is high enough
that the electrons are collisionless on timescales of interest, a
non-thermal distribution of some description must result. Even in
the exceptional coronal source studied by \cite{2010ApJ...714.1108K}
the numbers and energies are plausible. More challenging are the
chromospheric HXR sources interpreted as due to escaping accelerated
coronal electrons. The standard collisional thick-target electron
beam model requires electron fluxes up to a few $\times 10^{35}$
electrons per second \citep{2003ApJ...595L..97H} leaving the corona.
It is well known that this places strong demands on the coronal
electron population, which amounts to perhaps $10^{36}$ in the
volume above a flare region, and which would require replenishing
during the course of the flare. 
Replenishing of the coronal
acceleration volume by a counterstreaming return current generated
by the plasma in which the electrons propagate (of equal and opposite
electron flux) has been postulated, but it was pointed out relatively
early on there would be problems with beam stability
\citep{1977SoPh...52..117B}, and this remains the case.\index{return current!and beam stability}
White-light footpoint areas, which we might take as a proxy for the electron
beam areas (HXR footpoint areas being hard to measure reliably)
imply a beam flux and a return current speed in excess of what can
propagate stably through the corona. At least according to our
present theoretical understanding, the beam can travel stably only
if the density of the loop in which it moves is large - on the order
of $10^{11}\ {\rm cm}^{-3}$ \citep[see, e.g.,][for
details]{1990A&A...234..496V}. 
Though this is possible later in the flare (once evaporation has started) there is little imaging or
spectroscopic evidence for such loop densities before the flare
\citep[except in coronal thick target loop flares of][which do not
show footpoints]{2004ApJ...603L.117V}.  
Other exceptions might be found in the pre-flare phase of SOL2002-07-23T00:35 (X4.8), as discussed by \cite{linetal} and \cite{2010ApJ...725L.161C}.
Despite this uncertainty
the coronal electron beam model has been accepted for decades, but
it is clear that renewed theoretical effort must be dedicated to
understanding the propagation of a dense electron beam through the
corona. Some alternatives to this model have also recently been
proposed.  \cite{2008ApJ...675.1645F} have introduced a model in
which electrons producing the HXR footpoints are wholly accelerated
in the chromosphere, and \cite{2009A&A...508..993B} discuss the
chromospheric re-acceleration of a small number of originally coronal
electrons so that their photon yield per electron is increased,
reducing the electron number and flux requirement. 
Overall, an instantaneous emission measure\index{emission measure!non-thermal} 
for the non-thermal electrons of as much as
$10^{46}\ {\rm cm}^{-3}$ is required to explain chromospheric HXR
footpoints\index{footpoints!non-thermal emission measure} 
\citep{1976SoPh...48..197H} which is achievable in the
chromosphere, though still demanding. For example, if the flare
footpoint area were around $10^{17}\ {\rm cm}^{-2}$, then assuming a
chromospheric slab of thickness $10^8$~cm would require on average
$3\times 10^{10}$ electrons~$\rm{cm}^{-3}$ to be accelerated -- on
average about a tenth of the total electron population (bound or
unbound) in the top 1000~km of the VAL-C chromospheric model
\citep{1981ApJS...45..635V}.

\bigskip
\noindent{\bf{Flares and CMEs:}} At present it seems clear that the
standard flare scenario of large-scale reconnection in magnetic
fields stretched by an eruption must explain a great deal of flare
phenomenology. Probably \cite{1974SoPh...34..323H} provided the
first clear 3-D~visualization of how this might feasibly happen,
although the ideas certainly had been available long before this
seminal paper appeared. In fact, the Hirayama work described a
filament eruption (we would associate it with a CME nowadays),
rather than a flare as such. Furthermore, we recognize that it is
not the mass of the filament that is important in the overall
dynamics, rather the evolution of the magnetic field which carries
the filament mass along. The magnetic free energy released as the
outwards ejection of mass is therefore at least as large as the CME
kinetic energy estimated from the visible material. The CME is of
course just another manifestation of the re-arrangement of coronal
magnetic fields, but in an environment in which the perturbation
can expand relatively freely into the corona above the reconnection
region. This provides an interesting contrast to the situation below
the reconnection region, where the magnetic rearrangement of strong
fields releases a similar amount of energy into a small, confined
volume of very low beta plasma, resulting primarily in non-thermal
particles and heat, rather than mass motions. A basic calculation
shows that $10^{31}$~erg released into a volume of $10^{27}\ {\rm
cm}^{3}$ at a mean density of say $10^{11}\ {\rm cm}^{3}$  results
in a mean energy per particle of $60$~keV. 
Of course radiative and
conductive losses reduce the instantaneous value, but it does make
reasonably clear that flares are almost certain to be efficient
producers of non-thermal populations.

The rise of the active region filament heralding the imminent onset
of a flare identifies the importance of the MHD instability - which
will go on to produce the CME -- in the whole flare/CME combination.
But the extremely good correspondence in time between HXR bursts
and both filament lift-off \citep{2007SoPh..241...99M,2008ApJ...673L..95T}
and coronal ``supra-arcade downflows''\index{downflows!supra-arcade} 
\citep{2004ApJ...605L..77A}
(interpreted as retracting flux tubes formed in reconnection behind
a CME) does not really permit us to say whether one is the ``cause''
and the other the ``effect.''

A further pioneering paper by \cite{1960MNRAS.120...89G} had already
described a large-scale reconnection scenario that did not require
a prior eruption, and many flares, even including some as powerful
as the low X~class \citep{2007ApJ...665.1428W} events\index{flares!CMEless}, do not result
from the open fields stretched out by a CME. 
Therefore models which
permit the release of stored magnetic energy without field opening
are also necessary - such as those proposing internal reconnections
in active region core field \citep{2006ApJ...637L..65G}, or
reconnection without full field opening \citep{2008ApJ...680..740D,
2009ApJ...700..559M}.

A final observational link which deserves modeling attention is the
association of solar energetic particles with ``soft-hard-harder''
X-ray spectral evolution and fast CMEs \citep{2009ApJ...707.1588G}.
The SHH evolution is clearly a property of a solar acceleration
process which operates long after the CME has left, and long after
the flare impulsive phase. It is apparently unique to flares
exhibiting CMEs. Perhaps slow reconnection behind the departing
CME, or dipolarization of reconnected fields which have been greatly
stretched by the process, plays a part in the ongoing acceleration.

\subsection{Future observational progress}\label{sec:missing}

Here we list a few important areas in which observations should be improved.

\begin{enumerate}

\item {\bf UV/EUV imaging spectroscopy.}
\index{Lyman-$\alpha$!need for observations}
It is a major embarrassment to solar physics that we often turn to
stellar observations to learn how to fill in missing ``details''
from the solar data. One such ``detail'' is the spectroscopy and
morphology of Ly-$\alpha$ in flares  \citep{2009A&A...507.1005R}.
Ly-$\alpha$ is a primary radiating component, rich in diagnostic
information about the chromosphere.  In general the visible/UV
continuum contains the majority of flare radiated energy and yet
we have few good observations of it \citep[e.g.,][]{1989SoPh..121..261N}.

\item {\bf Sensitive high-energy observations.}
\index{hard X-rays!need for sensitive observations}
\textit{RHESSI} has made it abundantly clear that the key non-thermal processes
involved in the disruption of coronal plasmas (i.e., flares and CMEs)
can readily be detected even in the tenuous middle corona.  There
is a vast parameter space awaiting sensitive instruments.

\item {\bf Microwave/meter-wave imaging spectroscopy.}
\index{radio emission!need for imaging spectroscopy}
Solar radio astronomy has not had the benefit yet of broadband
observations in this key domain, or of radio imaging at more than
a few  frequencies. We know it to contain emission and absorption
features of great diagnostic significance, as well as giving insight
into the 3-D structure of the coronal magnetic field.

\item {\bf Neutral particle emissions.} The detection of neutrons
and energetic neutral atoms from solar flares is in its infancy but
holds great promise for understanding the behavior of accelerated
ions in the virtually unknown domain below a few~MeV. Neutron
detectors placed at a few tenths of an AU will be of great value\index{neutrons!need for low-energy observations}\index{neutrons}.

\item {\bf Coronal seismology.} 
\index{magnetic structures!and coronal seismology}
The wave population -- background and transient -- is another means
whereby the coronal magnetic structure can be probed. This can be
observed by high-resolution imaging and imaging spectroscopy of the
global corona.  The first steps in this area are now emerging from
ground-based and \textit{Hinode} observations, and the Solar Dynamics
Observatory will provide a comprehensive imaging view because of
its large telemetry bandwidth\index{coronal seismology}.

\end{enumerate}

\section{Conclusions}

We have reviewed flare observations in a broad sense, touching on
related phenomena and models that attempt to describe the overall
process. The multifarious observations across the broad spectrum
of phenomena each help us to characterize the equilibrium change
in the corona and chromosphere that we call a flare, and it should
be clear that the multiwavelength approach is crucial in flare
studies.  It tells us where the flare energy starts and where it
ends up, and something about the intermediate steps.  It also
provides some geometrical and diagnostic information about the flare
magnetic environment, at different levels in the atmosphere, and
how and when this changes as the flare proceeds.  This big picture
cannot be reached using one spectral region on its own. The
multiwavelength observations have many detailed applications as we
try to understand specific mechanisms that are at work in various
phases and regions of the flare development.  Some of the mechanisms
are at the stage of recent discovery and have only the sketchiest
understanding at present. The coming decade will see a flood of
multi-wavelength data, mastery of which which will require the
development of new analysis techniques, such as fast image processing
and feature recognition. It will be clear to the reader that much
of the observational evidence presented here is based on the detailed
analysis of small numbers of flares, and even basic statistical
studies are rather few. But a comprehensive understanding of the
flare phenomenon will require a blend of both approaches -- i.e., the
collection, sifting, comparison and assimilation of detailed
properties of large samples of events. We look forward to the
challenge.

\begin{acknowledgements}
The authors would collectively like to acknowledge the work by the many instrument teams and software teams, whose sustained efforts over the years have made this kind of multiwavelength analysis possible. 
The chief architects of and major contributors to the Solarsoft library deserve particular thanks (Sam Freeland, Richard Schwartz, William Thompson, Kim Tolbert and Dominic Zarro).
L.F. would like to acknowledge financial support from the UK STFC (Rolling grant ST/F0026371), the EU's SOLAIRE Research and Training Network (MTRN-CT-2006-035484) and L.F. and M.B. also acknowledge the support of the  Leverhulme Foundation (Grant F/00 179/AY). 
The work of H. Ji was supported by NSFC 10833007W. 
W.~Liu was partly supported by an appointment to the NASA Postdoctoral Program at Goddard Space Flight Center, administered by Oak Ridge Associated Universities through a contract with NASA. 
Partial support to W.~Liu's work was also provided by \textit{Hinode} SOT contract NNM07AA01C.
A. Caspi, H. Hudson, and S. Krucker were supported by NASA under contract NAS5-98033 and grants NAG5-12878 and NNX08AJ18G.
A.V. and M.T. acknowledge the Austrian Science Fund (FWF) projects no. P20867-N16 and P20145-N16.

\end{acknowledgements}
\bibliographystyle{ssrv}

\bibliography{ch2}

\printindex

\end{document}